\documentclass[preprint,showpacs,preprintnumbers,amsmath,amssymb]{revtex4-2}
\usepackage{bm}
\usepackage{amsmath}
\usepackage{amssymb}
\usepackage{amsfonts}
\usepackage{graphicx}
\begin{document}

\title{ \textbf{
    On the Realization of Quantum State Teleportation in Proton Systems
} }

\author{H.~Wita{\l}a}
\affiliation{M. Smoluchowski Institute of Physics, 
Faculty of Physics, Astronomy and Applied Computer Science,
Jagiellonian University, PL-30348 Krak\'ow, Poland}

\date{\today}

\begin{abstract}
  We discuss how to generate entangled Bell states of two nucleons using
  unpolarized nucleon–nucleon scattering or the exclusive deuteron breakup
  reaction. We follow the approach of
  Z. X. Shen et al., arXiv:2510.24325v1  [nucl-th],
  where Bell states were identified in unpolarized proton–proton
  elastic scattering. We confirm these results and show that, in the
  unpolarized proton–deuteron breakup reaction, it is also possible to
  generate proton–proton entangled Bell states in kinematically complete
  proton–proton quasi-free scattering (QFS) and final-state interaction (FSI)
  configurations.

  We also discuss an experimental setup that, by exploiting such
  entangled states, could enable the teleportation of quantum mechanical
  states in a three-proton system. Such an experiment requires triple 
  coincidences among the outgoing nucleons, which precludes the use of
  entangled Bell states generated with extremely polarized incoming particles.
  Since counting rates for unpolarized reactions are much higher than for
  polarized ones, the present results open a pathway toward searching
  for signatures of quantum state teleportation in hadronic systems.
\end{abstract}


\maketitle \setcounter{page}{1}

\section{Introduction}
\label{intro}

Recently, entanglement properties of the nucleon–nucleon (NN) states formed
in nuclear processes have attracted the attention of the nuclear physics
community. The generation of entanglement in NN scattering has been
extensively investigated in numerous studies; see, for example,
Refs. \cite{beane2019,beane2021,bai2022,bai2023,liu2023,kirch2,bai2024,miller1,miller2,low,witnp2024,shen_2025,wit_2025,cavallin_2025}
and the references therein.

One of the important problems considered was how to efficiently generate
highly entangled Bell-like states of two nucleons that could subsequently
be employed in various applications. In the first studies, the possibility
of producing such states using polarized incoming particles was
investigated \cite{witnp2024,wit_2025}. In Ref. \cite{witnp2024}, the final
spin density matrix in polarized neutron–proton (np) scattering was analyzed,
and entangled np pairs of all possible Bell types were found at higher
energies of np elastic scattering; however, this occurred only in the
case of extreme polarizations of both incoming nucleons.

It was also shown that scattering maximally polarized protons off
maximally polarized deuterons can lead to entangled np 
and proton–proton (pp) pairs in specific kinematically complete
configurations of the exclusive deuteron breakup \cite{wit_2025}.
Although such methods of generating entangled nucleon pairs are theoretically
possible, they are presently impractical due to the lack of polarization
sources capable of producing protons or deuterons in a single magnetic
substate. Even if such sources were to become available, the applicability
of entangled states formed in polarized reactions would be severely limited
by their much lower counting rates compared to those of unpolarized processes.

It thus seems natural to examine the possibility of entanglement formation
in unpolarized pp elastic scattering and in the proton-deuteron (pd)
 breakup reaction, despite
the puzzlement such an attempt might initially evoke. However, it should
not be surprising that, using an unpolarized proton beam and target, or
an unpolarized deuteron breakup reaction, one can produce entangled
pp pairs in a pure Bell state.

In fact, in Ref.~\cite{lamehi}, a test of Bell’s inequalities was
performed using unpolarized pp scattering at $E_{lab} \approx 13$~MeV.
A similar test was carried out in Ref.~\cite{sakai2006} using
the unpolarized breakup reaction $p(d,{}^2H)n$ with an unpolarized
deuteron beam energy of $E_{lab} = 270$~MeV. Therefore, in both experiments
the two outgoing protons must have been in an entangled Bell state.

It has indeed been shown in a recent study \cite{shen_2025} that entangled
proton–proton pairs can be produced in pp elastic scattering
with an unpolarized beam and target. The availability of high-intensity
entangled pp states, made possible by these breakthrough results, opens
the way to a wide range of applications, including, among others,
the teleportation of a quantum mechanical state between two protons
in a three-proton system.

Such an experiment, proposed in \cite{teleport}, is essentially based
on two elements. The first is the formation of entangled pp pairs, and
the second is the scattering of one of the entangled protons off a polarized
proton target. The latter process is in fact responsible for
the teleportation of the quantum state of the target proton to
the second entangled proton, leaving the two remaining protons in
a new entangled state.

For this experiment to be successful, in addition to the requirement of
producing a specific Bell-type entangled state of two protons, the actual
occurrence of teleportation requires a specific single-term structure of
the pp scattering transition matrix, with the dominant contribution
originating from one particular Bell component.

In the present paper, we study the energy dependence of these two
ingredients of the teleportation experiment in order to identify
the optimal conditions for its realization.

We investigate the extent to which unpolarized pd 
breakup can be considered a useful tool for generating entangled pp pairs.
Before turning to the breakup process, we verify whether our approach is
able to reproduce the results of Ref. \cite{shen_2025} for pp elastic
scattering. We study the energy dependence of the transition matrix
in pp elastic scattering, with particular emphasis on revealing its
structure in the Bell-state basis.

In Sec.~\ref{formNN}, we outline our approach and present results for
NN elastic scattering. An analogous study for the exclusive
nucleon-deuteron (Nd) breakup
reaction is presented in Sec.~\ref{formbr}. Details of the teleportation
experiment and its optimal realization at low and higher energies are
discussed in Sec.~\ref{teleport}. Finally, Sec.~\ref{sumary} contains
a summary and conclusions.

\section{Unpolarized NN elastic scattering}
\label{formNN}

A representative case of maximal entanglement is provided by
the Bell-state basis \cite{bookqinf}, which, following the
convention of Ref.~\cite{shen_2025}, is given by:
\begin{eqnarray}
  | \psi_1\rangle  &\equiv& | \Phi^+ \rangle = \frac {1} {\sqrt{2}}
  (| +\frac {1} {2} +\frac {1} {2}\rangle  +
   | -\frac {1} {2} -\frac {1} {2}\rangle ) \equiv
 \frac {1} {\sqrt{2}} (| ++ \rangle  + | -- \rangle )  \cr
  | \psi_{2}\rangle  &\equiv&  | \Phi^- \rangle = \frac {1} {\sqrt{2}}
  ( | +\frac {1} {2} +\frac {1} {2}\rangle  -
    |-\frac {1} {2} -\frac {1} {2} \rangle  ) \equiv
 \frac {1} {\sqrt{2}} (| ++ \rangle  - | -- \rangle )  \cr
  | \psi_{3}\rangle  &\equiv&  | \psi^+ \rangle = \frac {1} {\sqrt{2}}
  ( | +\frac {1} {2} -\frac {1} {2}\rangle  +
    |-\frac {1} {2} +\frac {1} {2} \rangle  )  \equiv
 \frac {1} {\sqrt{2}} (| +- \rangle  + | -+ \rangle )  \cr
  | \psi_{4}\rangle  &\equiv&  | \psi^- \rangle = \frac {1} {\sqrt{2}} 
  ( | +\frac {1} {2} -\frac {1} {2}\rangle  -
    |-\frac {1} {2} +\frac {1} {2} \rangle  )  \equiv
 \frac {1} {\sqrt{2}} (| +- \rangle  - | -+ \rangle ) ~.
\label{eq_1}
\end{eqnarray}

The elastic scattering of a nucleon beam off a nucleon target is described
by a transition operator $M$ \cite{book}, which can be expressed in the Bell
basis (\ref{eq_1}) in terms of sixteen coefficients $C_{i'i}$:
\begin{eqnarray}
  M &=&  \sum_{i,i'=1}^4  \langle \psi_{i'} | M | \psi_{i} \rangle ~
  | \psi_{i'} \rangle \langle \psi_i | \equiv 
   \sum_{i,i'=1}^4  C_{i'i} ~
  | \psi_{i'} \rangle \langle \psi_i |
   ~.
 \label{eq_6}
\end{eqnarray}

By solving the Lippmann–Schwinger (LS) equation \cite{book} with a realistic
NN potential (including the pp Coulomb force in the pp case), the matrix
elements $\langle m_1' m_2' \lvert M \rvert m_1 m_2 \rangle$ can be computed.
From these, the sixteen coefficients $C_{i'i}$ of Eq.~(\ref{eq_6}) are
determined:
\begin{eqnarray}
  C_{11} &=&  = \frac {1} {2}
    (~~\langle --|M|--\rangle + \langle --|M|++\rangle
  + \langle ++|M|--\rangle + \langle ++|M|++\rangle~ ) \cr
  C_{12} &=&  = \frac {1} {2}
    (-\langle --|M|--\rangle + \langle --|M|++\rangle
  - \langle ++|M|--\rangle + \langle ++|M|++\rangle~ ) \cr
  C_{13} &=&  = \frac {1} {2}
    (~~\langle --|M|-+\rangle + \langle --|M|+-\rangle
  + \langle ++|M|-+\rangle + \langle ++|M|+-\rangle~ ) \cr
  C_{14} &=&  = \frac {1} {2}
    (-\langle --|M|-+\rangle + \langle --|M|+-\rangle
  - \langle ++|M|-+\rangle + \langle ++|M|+-\rangle~ ) \cr
  C_{21} &=&  = \frac {1} {2}
    (-\langle --|M|--\rangle - \langle --|M|++\rangle
  + \langle ++|M|--\rangle + \langle ++|M|++\rangle~ ) \cr
  C_{22} &=&  = \frac {1} {2}
    (~~\langle --|M|--\rangle - \langle --|M|++\rangle
  - \langle ++|M|--\rangle + \langle ++|M|++\rangle~ ) \cr
  C_{23} &=&  = \frac {1} {2}
    (-\langle --|M|-+\rangle - \langle --|M|+-\rangle
  + \langle ++|M|-+\rangle + \langle ++|M|+-\rangle~ ) \cr
  C_{24} &=&  = \frac {1} {2}
    (~~\langle --|M|-+\rangle - \langle --|M|+-\rangle
  - \langle ++|M|-+\rangle + \langle ++|M|+-\rangle~ ) \cr  
  C_{31} &=&  = \frac {1} {2}
    (~~\langle -+|M|--\rangle + \langle -+|M|++\rangle
  + \langle +-|M|--\rangle + \langle +-|M|++\rangle~ ) \cr
  C_{32} &=&  = \frac {1} {2}
    (-\langle -+|M|--\rangle + \langle -+|M|++\rangle
  - \langle +-|M|--\rangle + \langle +-|M|++\rangle~ ) \cr
  C_{33} &=&  = \frac {1} {2}
    (~~\langle -+|M|-+\rangle + \langle -+|M|+-\rangle
  + \langle +-|M|-+\rangle + \langle +-|M|+-\rangle~ ) \cr
  C_{34} &=&  = \frac {1} {2}
    (-\langle -+|M|-+\rangle + \langle -+|M|+-\rangle
  - \langle +-|M|-+\rangle + \langle +-|M|+-\rangle~ ) \cr
  C_{41} &=&  = \frac {1} {2}
    (-\langle -+|M|--\rangle - \langle -+|M|++\rangle
  + \langle +-|M|--\rangle + \langle +-|M|++\rangle~ ) \cr
  C_{42} &=&  = \frac {1} {2}
    (~~\langle -+|M|--\rangle - \langle -+|M|++\rangle
  - \langle +-|M|--\rangle + \langle +-|M|++\rangle~ ) \cr
  C_{43} &=&  = \frac {1} {2}
    (-\langle -+|M|-+\rangle - \langle -+|M|+-\rangle
  + \langle +-|M|-+\rangle + \langle +-|M|+-\rangle~ ) \cr
  C_{44} &=&  = \frac {1} {2}
    (~~\langle -+|M|-+\rangle - \langle -+|M|+-\rangle
  - \langle +-|M|-+\rangle + \langle +-|M|+-\rangle~ ) 
  ~.
 \label{eq_7}
\end{eqnarray}

The spin density matrix of the  outgoing nucleons is given by \cite{ohlsen1972}:
\begin{eqnarray}
  \rho_f &=& M ~ \rho_{in} ~ M^{\dagger}
   ~,
 \label{eq_8}
\end{eqnarray}
where the spin state of the incoming nucleons is described by the density matrix
$\rho_{\text{in}}$. In the case of an unpolarized nucleon beam and an unpolarized
nucleon target, the initial density matrix is
$\rho_{\text{in}} = \frac{1}{4} I_b \otimes I_t$ \cite{ohlsen1972}. The final spin
density matrix, expressed in the Bell basis (\ref{eq_1}), is then characterized
by sixteen coefficients $\bar C_{i'i}$:
\begin{eqnarray}
  \rho_f &\propto&  \sum_{i,i'=1}^4 ~
 (\sum_{i''=1}^4 C_{i'i''} C_{ii''}^*) ~
  | \psi_{i'} \rangle \langle \psi_i | \equiv 
   \sum_{i,i'=1}^4  \bar C_{i'i} ~
  | \psi_{i'} \rangle \langle \psi_i |
   ~.
 \label{eq_9}
\end{eqnarray}

From the properties of the M matrix \cite{book}, it follows that for any
 NN system, at a given energy and center-of-mass angle, only
six of the sixteen coefficients $C_{i'i}$ (or $\bar C_{i'i}$) are nonzero.
These are $C_{ii}$, for $i=1,2,3,4$, as well as $C_{23}$ and $C_{32}$. 
    The corresponding nonvanishing coefficients in the final-state
    density matrix are
    $$ \bar C_{11}=|C_{11}|^2, ~ \bar C_{22}=|C_{22}|^2+|C_{23}|^2, ~
    \bar C_{33}=|C_{33}|^2+|C_{32}|^2, $$
$$\bar C_{44}=|C_{44}|^2, ~ \bar C_{23}=C_{22}C_{32}^* + C_{23}C_{33}^*, ~
\bar C_{32}=C_{32}C_{22}^* + C_{33}C_{23}^*~. $$

    This leads to the following general six-term decomposition of the M  matrix
    in the Bell basis:
\begin{eqnarray}
  M &=&  C_{11}   | \Phi^+ \rangle \langle \Phi^+ |  + 
         C_{22}   | \Phi^- \rangle \langle \Phi^- |  + 
         C_{33}   | \psi^+ \rangle \langle \psi^+ |  + 
         C_{44}   | \psi^- \rangle \langle \psi^- |  + \cr
   &~&   C_{23}   | \Phi^- \rangle \langle \psi^+ |  +         
         C_{32}   | \psi^+ \rangle \langle \Phi^- |   
   ~,
 \label{eq_6aa}
\end{eqnarray}
and, similarly, for the spin density matrix $\rho_f$ in unpolarized NN
scattering, with $C_{i'i}$ replaced by  $\bar C_{i'i}$.

For identical nucleons (the nn or pp system) at the center-of-mass
angle $\Theta_{c.m.}=90^o$,  three additional coefficients
vanish: $C_{11}$, $C_{22}$ and $C_{33}$. This is a consequence of the
fact that, at this specific angle for identical nucleons, the spin-triplet
($s=1$) component vanishes and only the spin-singlet ($s=0$) component
contributes to the M matrix. Therefore, at this angle the M matrix is
fully specified by only three terms:
\begin{eqnarray}
  M &=&  C_{44}   | \psi^- \rangle \langle \psi^- |  + 
         C_{23}   | \Phi^- \rangle \langle \psi^+ |  +         
         C_{32}   | \psi^+ \rangle \langle \Phi^- |   
   ~,
\label{eq_6bb}
\end{eqnarray}
and $\rho_f$:
\begin{eqnarray}
  \rho_f &=&  \bar C_{22}   | \Phi^- \rangle \langle \Phi^- |  + 
              \bar C_{33}   | \psi^+ \rangle \langle \psi^+ |  + 
              \bar C_{44}   | \psi^- \rangle \langle \psi^- |  
   ~.
 \label{eq_6cc}
\end{eqnarray}

In Ref.~\cite{shen_2025}, the properties of the M matrix and
the final-state
 spin density matrix $\rho_f$ were investigated for the pp system.
 We repeated this investigation using a modified approach and extended it to
 np scattering. In Figs.~\ref{fig1aa} and \ref{fig1bb}, we
 present the energy dependence of the nonvanishing coefficients $C_{i'i}$ 
 and $\bar C_{i'i}$ at the center-of-mass angle 
    $\Theta_{c.m.}=90^o$, calculated using the AV18 \cite{av18} NN potential.
    A pronounced contrast between the pp and np systems is observed.

    Whereas in the pp case only three terms contribute to the M matrix and
    to $\rho_f$ at this angle, in the np system six terms of comparable
    magnitude determine the transition matrix and the spin density matrix.
    Even more striking than the reduced number of contributing terms in the
    pp system is the strong dominance of a single term at low energies,
    around $10$~MeV, and at higher energies, around
    $150$~MeV, both for the M matrix and for $\rho_f$.

    The dominance of a single term in $\rho_f$ suggests that the formation of
    strongly entangled Bell states in unpolarized pp scattering is possible.
    Since complete quasi-free scattering configurations in the exclusive
    pd breakup reaction exhibit strong similarities to NN scattering, this
    suggests that the formation of such states may also be possible in
    unpolarized pd breakup.

In Figs.~\ref{fig1}–\ref{fig8}, we present, at two laboratory energies,
$E_{lab}=10$~MeV and
$E_{lab}=151$~MeV 
(identified in Ref.~\cite{shen_2025} as the most promising energies for the
production of entangled states), the angular distributions of the absolute
values of the expansion coefficients $|C_{i'i}|$ and $|\bar C_{i'i}|$, 
calculated using the AV18 NN potential~\cite{av18}.

The results for the pp system (Figs.~\ref{fig1}–\ref{fig4}) confirm those of
Ref.~\cite{shen_2025}, where it was found that at lower energies,
$E_{lab} \approx 10$~MeV, and for angles around
$\Theta_{c.m.} \approx 90^o$, the transition matrix is well
approximated by
$M \approx | \psi^- \rangle \langle \psi^- |$, and the corresponding final
spin density matrix by
$\rho_f \approx | \psi^- \rangle \langle \psi^- |$.
At higher energies, $E_{lab} \approx 150$~MeV,
the transition matrix is instead given by
$M \approx | \psi^+ \rangle \langle \Phi^- |$, and the corresponding final
spin density matrix by $\rho_f \approx | \psi^+ \rangle \langle \psi^+ |$.

    Indeed, Fig.~\ref{fig1} shows that at $E_{lab} = 10$~MeV and
    for center-of-mass angles in the range
    $\Theta_{c.m.} \in (55^o,125^o)$, 
    the dominant contribution to the M matrix comes from $C_{44}$, which is
    more than one order of magnitude larger than $C_{32}$ and $C_{23}$,
    while
    the remaining coefficients are even smaller or vanish altogether. This leads
    to the approximation $M \approx | \psi^- \rangle \langle \psi^- |$.

    For the final-state spin density matrix $\rho_f$, this selectivity is even
    more pronounced: in the same angular region, the coefficient $\bar C_{44}$ 
    exceeds the other coefficients by nearly two orders of magnitude, and the
    outgoing pp pair is therefore produced in a strongly entangled state
   $|\psi_{4}\rangle =  | \psi^- \rangle$.

    At $E_{lab} = 151$~MeV, the largest contribution to M, driven by
    a single coefficient, again occurs around $90^o$ and comes from $C_{32}$,
    which is about one order of magnitude larger than the remaining
    coefficients. At this energy, however, this contribution appears in
    a more restricted angular region, $\Theta_{c.m.} \in (80^o,100^o)$,
    than at lower energies (see Fig.~\ref{fig3}),
    and $M \approx | \psi^+ \rangle \langle \Phi^- |$.

    For the spin density matrix (see Fig.~\ref{fig4}), a similar picture
    is observed, with the dominance of $C_{33}$; however, this occurs
    in an even more restricted angular region,
    $\Theta_{c.m.} \in (85^o,95^o)$. Thus, the outgoing pp pair is in
    an entangled state, $|\psi_{3}\rangle =  | \psi^+ \rangle$.

    In the np system (Figs.~\ref{fig7}–\ref{fig8}), the behavior of
    the $C_{i'i}$ and $\bar C_{i'i}$  coefficients is completely different.
    At $E_{lab}=10$~MeV (not shown), the selectivity observed in the pp system is
    entirely lost, and all diagonal coefficients $C_{ii}$ and $\bar C_{ii}$
    contribute comparably to M and $\rho_f$, respectively. This prevents
    the formation of any type of entangled Bell state.

    At $E_{lab}=151$~MeV, a weak selectivity appears at specific angles.
    At $\Theta_{c.m.} \approx 30^o$ and $\Theta_{c.m.} \approx 140^o$,
    the largest contributions to M and $\rho_f$ come from
    $ | \Phi^- \rangle \langle \Phi^- |$, while at very backward angles
    the dominant contribution is $ | \psi^- \rangle \langle \psi^- |$.
    However, in each case the dominance of the leading contribution is
    too weak to expect the formation of any sufficiently pure Bell state.

    At this point, it is interesting to highlight the difference in the
    formation of Bell-like states in unpolarized and maximally polarized
    NN scattering. In Ref.~\cite{witnp2024} (Table~III), it was shown that
    polarized np scattering with extremely polarized neutrons and protons
    yields all four types of Bell states at $E_{lab}=100$~MeV
    and at specific angles.

    We have verified that, at this energy, the coefficients
    $C_{i'i}$ and $\bar C_{i'i}$ in unpolarized np scattering behave
    similarly to those shown in Figs.~\ref{fig7} and \ref{fig8},
    although with weaker selectivity. This indicates that polarization
    of the incoming neutron and proton allows for the production of
    purer Bell states and a richer spectrum of Bell-state types than
    in unpolarized np scattering.

    Before turning to the breakup reaction, for which we will mostly
    employ solutions of the three-nucleon (3N) Faddeev equations
    with the pp Coulomb force omitted, we first examine the consequences
    of switching off the Coulomb interaction in pp scattering for
    conclusions concerning the formation of entangled states and
    the structure of the M matrix.

    In Figs.~\ref{fig1a} and \ref{fig2a}, we present results for pp
    elastic scattering at $E_{lab} = 10$~MeV, analogous
    to those shown in Figs.~\ref{fig1}–\ref{fig2}, with the only
    difference being that the Coulomb force is switched off. As expected,
    large effects on the magnitudes of the coefficients appear
    at the low energy $E_{lab} = 10$~MeV, predominantly in the forward and
    backward angular regions. These effects diminish with increasing energy
    and, at $E_{lab} = 151$~MeV (not shown),
    are confined to more forward and backward angles.

    However, the pattern of the relative magnitudes of the contributing
    coefficients $C_{i',i}$ and $\bar C_{i',i}$ is very similar whether
    the Coulomb force is included or not. In the mid-angle region,
    where Coulomb effects are much smaller, the magnitudes of the
    contributing coefficients are nearly identical, especially
    at $E_{lab} = 151$~MeV. At $E_{lab} = 10$~MeV, switching off
    the pp Coulomb force
    significantly broadens the angular range in which the formation
    of entangled states appears to be possible.

    It follows that the conclusions regarding the possibility of
    entangled-state formation drawn for the pp system can be directly
    transferred to unpolarized neutron–neutron (nn) elastic scattering.

\section{Unpolarized Nd breakup reaction}
\label{formbr}

In the next step, we perform a similar investigation of the
neutron–deuteron (nd) exclusive breakup reaction $d(n,N_1N_2)N_3$, 
 using in this case the high-precision CD Bonn potential \cite{cdb}.
The matrix elements of the breakup transition operator 
$\langle m_1' m_2' m_3' \lvert U_0 \rvert m_n m_d \rangle$, 
determined from solutions of the three-nucleon (3N) Faddeev equations
\cite{book,glo96,wit88,hub97}, allow us to express the operator $U_0$ itself in the Bell-state basis
$| \psi_{i} \rangle $ for nucleons 1 and 2, together with the spin state of the third nucleon
$| m_3  \rangle$ and the incoming neutron and deuteron spin states
$\rvert m_n m_d \rangle$, as follows: 
\begin{eqnarray}
  U_{0} &=& \sum_{m_n,m_d,m_3'} \sum_{i,i'=1}^4
  \langle \psi_{i'} m_3' | U_0 | m_n m_d \rangle ~
  | \psi_{i'} \rangle | m_3' \rangle \langle m_n m_d | 
   ~.
 \label{eq_20}
\end{eqnarray}

The spin density matrix of the three outgoing free nucleons is given by
\begin{eqnarray}
  \rho_f &=& U_0 ~ \rho_{in} ~ U_0^{\dagger}
   ~,
 \label{eq_21}
\end{eqnarray}
here the spin state of the incoming neutron and deuteron is described by the
density matrix  $\rho_{in}$.

In Ref. \cite{wit_2025}, the possible formation of entangled states for the outgoing nucleon pair
$1-2$ in the breakup reaction was investigated for quasi-free scattering (QFS) and final-state
interaction (FSI) complete configurations, using fully polarized incoming neutrons and deuterons.
Here, we investigate the case of unpolarized incoming neutrons and deuterons, a reaction that
exhibits significantly higher counting rates than the polarized breakup reaction.
Assuming an unpolarized incoming state, $\rho_{in}= \frac {1} {2 \times 3} I_n \otimes I_d$,
the spin density matrix $\rho_f^{12}$, describing the spin state of the outgoing nucleon pair $1-2$,
is given by:
\begin{eqnarray}
  \rho_f^{12} = Tr_3 ~ \rho_f \propto
  \sum_{m_3} \langle m_3 | \rho_f | m_3 \rangle &=&
  \sum_{i,i'=1}^4  | \psi_{i'} \rangle  \langle \psi_{i} | (\sum_{m_nm_dm_3}
  \langle \psi_{i'} m_3 | U_0 | m_n m_d \rangle 
  \langle \psi_{i}  m_3 | U_0 | m_n m_d \rangle^* ) \cr
  &\equiv& \sum_{i,i'=1}^4  | \psi_{i'} \rangle  \langle \psi_{i} |~
  \bar C_{i'i}^{br} 
   ~.
 \label{eq_22}
\end{eqnarray}

We investigate the breakup reaction at two laboratory energies,
$E_{lab}=13$~MeV and
$E_{lab}=156$~MeV. At these energies, the resulting QFS configurations
exhibit strong similarities to the previously studied free NN scattering at
$E_{lab}=10$~MeV and $151$~MeV, respectively \cite{glo96,wit1989}.

We begin with a discussion of the FSI(nn) configuration in the exclusive
$d(n,nn)p$  breakup reaction at an incoming neutron laboratory energy of
$E_{lab}=156$~MeV. In Fig.~\ref{fig9}, we present the expansion coefficients
$\bar C_{i'i}^{br}$  for all such FSI(nn) geometries as a function of
the laboratory angle
$\Theta_1^{lab}$  of the first outgoing neutron.

We would like to point out that, in order to investigate all
kinematically complete
FSI and QFS configurations in the
$d(n,N_1N_2)N_3$ breakup reaction, for both $np$ and $nn$ FSI or
QFS pairs, we consider
their dependence on the laboratory angle of the first nucleon,
$\Theta_1^{\rm lab}$.
We assume that, at this angle, the FSI($N_1N_2$) condition
(where $N_1$ and $N_2$  have equal momenta) or the QFS($N_1N_2$) condition
(where nucleon $N_3$  remains at rest in the laboratory frame) is
exactly fulfilled.
This corresponds, when viewed along the S-curve, to the location S
at which the cross-section
maximum occurs.

At the energies considered here, for some values of $\Theta_1^{\rm lab}$,
two solutions
satisfying the FSI(12) or QFS(12) conditions occur. The second solutions,
which are difficult
to access experimentally, form a branch that does not extend to small
values of
$\Theta_1^{\rm lab}$ in the figures presented below. This branch
corresponds, in the FSI case,
to small energies of the interacting nucleons, and in the QFS case,
to small values of the
laboratory angle  $\Theta_2^{\rm lab}$  and the energy $E_1^{\rm lab}$. 

In all figures showing FSI and QFS configurations as functions
of $\Theta_1^{\rm lab}$,
we present results for both solutions; however, the discussion
and conclusions are restricted
to the experimentally accessible first solutions.

As seen in Fig.~\ref{fig9}, the only nonvanishing coefficient
in any FSI(nn) configuration
is $\bar C_{44}^{br}$. As a consequence, the spin density matrix of
the two outgoing neutrons
in the FSI(nn) complete geometry  at $E_{lab}=156$~MeV is
$ \rho_f =| \psi^- \rangle \langle \psi^- |$, 
and the neutron–neutron pair is therefore in a pure Bell
state $ | \psi^- \rangle$.

For the final-state–interacting neutron–proton pair at this energy
(see Fig.~\ref{fig11}),
the situation changes and additional coefficients contribute
to $\rho_f$.
However, also in this case the dominant contribution to $\rho_f$
at each laboratory angle
$\Theta_1^{\rm lab}$ comes from $\bar C_{44}^{br}$, which in the angular region
$ \Theta_1^{\rm lab} \in (25^o,55^o)$ is about two orders of
magnitude larger than the other
    coefficients. It follows that the final-state–interacting np pair at
    $E_{lab}=156$~MeV is also very likely to be in the Bell
    state $ | \psi^- \rangle $.

    An interesting case is provided by the QFS(nn)
    configuration (see Fig.~\ref{fig13}).
Similarly to the FSI(np) case, a larger number of coefficients
$\bar C_{i'i}^{br}$ contribute to $\rho_f$—six in total in both
cases—which results in ten  
$\bar C_{i'i}^{br}$  coefficients vanishing. The dominant contribution
to $\rho_f$ in the
    angular range $\Theta_1^{\rm lab} \in (30^\circ, 60^\circ)$ 
    comes from   $\bar C_{33}^{br}$, which in this region is approximately
    an order of magnitude
    larger than the other coefficients. In a narrow angular interval around
    $\Theta_1^{\rm lab} \approx 45^\circ$,   $\bar C_{33}^{br}$  exceeds
    the remaining coefficients
    by more than three orders of magnitude. It thus follows that
    the QFS(nn) pair
    at $E_{lab} = 156$~MeV is very likely to be in the Bell state
    $|\psi^+\rangle$,
    at least in the vicinity of $\Theta_1^{\rm lab} \approx 45^\circ$.

For the QFS(np) configuration (see Figs.~\ref{fig15}), the behavior of the
$\bar C_{i'i}^{br}$ coefficients resembles that of their counterparts
in unpolarized np
elastic scattering (see Fig.~\ref{fig7}). Again, six $\bar C_{i'i}^{br}$
coefficients are nonzero,
while ten vanish.
 At $\Theta_1^{\rm lab} \approx 20^\circ$ and $\Theta_1^{\rm lab} \approx 70^\circ$,
the dominant contribution to $\rho_f$ comes from $\bar C_{22}^{br}$, which
in the 
vicinity of these angles is approximately one order of magnitude larger
than the other
coefficients. This suggests that the np pair is in the Bell state
$|\psi^-\rangle$,
although the resulting state is likely not sufficiently pure.

At the lower energy  $E_{lab} = 13$~MeV, similarly to the case at
$E_{lab} = 156$~MeV,
all nn pairs emerging under the FSI(nn) condition are in a pure
Bell state $|\psi^-\rangle$
(see Fig.~\ref{fig17}). 
Similarly, all np pairs under the FSI(np) condition are very likely
also in the Bell state
$|\psi^-\rangle$  at this energy (see Fig.~\ref{fig19}).
Here, the leading coefficient $\bar C_{44}^{br}$ is about three orders
of magnitude larger
than the next-largest coefficients, $\bar C_{11}^{br}$, $\bar C_{22}^{br}$,
and $\bar C_{33}^{br}$,
which ensures that this Bell state is considerably purer than the
corresponding state
at $E_{lab} = 156$~MeV.

Similarly, the nn pairs under the QFS(nn) condition at $E_{lab} = 13$~MeV, as at
$E_{lab} = 156$~MeV, are very likely in a pure Bell state. The main
difference is that the
state is now $|\psi^-\rangle$ rather than $|\psi^+\rangle$. 
This occurs for angles $\Theta_1^{\rm lab} \in (25^\circ, 50^\circ)$,
where  $\bar C_{44}^{br}$ is approximately two orders of magnitude
larger than the other
coefficients. In a narrow angular region around
$\Theta_1^{\rm lab} \approx 40^\circ$, $\bar C_{44}^{br}$
exceeds the remaining coefficients by more than four orders of magnitude,
likely yielding
an even purer Bell state $|\psi^-\rangle$.

We have seen that at $E_{lab} = 156$~MeV it was difficult to find entangled
np pairs in
QFS(np) geometries. This situation is even more pronounced at lower
energies, where
the formation of entangled np pairs is not possible at all.
As shown in Fig.~\ref{fig23}, at $E_{lab} = 13$~MeV at least two coefficients,
$\bar C_{11}^{br}$ and $\bar C_{22}^{br}$, contribute equally to  $\rho_f$
in all QFS(np)
configurations. Moreover, for $\Theta_1^{\rm lab} \leq 20^\circ$, $\bar C_{44}^{br}$
is of comparable magnitude.


In Figs.~\ref{fig9}–\ref{fig23} we show all FSI($N_1N_2$) and QFS($N_1N_2$)
configurations as a function of $\Theta_1^{\rm lab}$. For each angle
$\Theta_1^{\rm lab}$, the corresponding FSI or QFS kinematical condition
is exactly fulfilled. When examining the cross section as a function
of the arc length $S$ for a breakup configuration defined by a specific
$\Theta_1^{\rm lab}$ together with the corresponding
$\Theta_2^{\rm lab}$ and $\Phi_{12}=0^\circ$ for FSI($N_1N_2$)
or $\Phi_{12}=180^\circ$ for QFS($N_1N_2$), a characteristic peak appears.
The cross section reaches its maximum at the value of $S$ where
the corresponding FSI or QFS condition is exactly satisfied.
In the case of the $pd$ breakup, the maximum at FSI($pp$)
is suppressed due to penetration through the $pp$ Coulomb barrier.
An interesting question is how rapidly the entanglement
is lost when moving to geometries corresponding to values
of $S$ away from this maximum.

In Figs.~\ref{fig25}–\ref{fig31} we show the absolute values of
the expansion coefficients $\bar C_{i'i}^{\rm br}$ at both energies as a
function of $S$ for the FSI(nn) and QFS(nn) configurations at selected
values of $\Theta_1^{\rm lab}$ and the corresponding $\Theta_2^{\rm lab}$.
It is seen that, for both configurations and energies, strong
entanglement as well as the character of the Bell state are preserved
over a wide range of $S$ values around the FSI and QFS peaks. This
justifies averaging over these cross-section maxima in actual experiments.


Since in practice one is restricted to measurements of the pd breakup,
we discuss the influence of the pp Coulomb force on the nd breakup results.
As a first step, we compare cross sections in nn or pp QFS and FSI geometries
for the reactions $d(n,N_1N_2)N_3$ and $d(p,N_1N_2)N_3$.
In the latter case, the pp Coulomb force is active and the
three-nucleon Faddeev equations including this interaction must
be solved \cite{wit_coul}.

In Figs.~\ref{fig33} and \ref{fig34} we show the influence of the pp
Coulomb force on all FSI and QFS cross sections. For the FSI case
(Fig.~\ref{fig33}), the effect is huge: the FSI(nn) cross section is
reduced by at least a factor of $\approx 10$  at $13$~MeV
and $156$~MeV. This drastic reduction is caused by penetration through
the pp Coulomb barrier. For the QFS case (Fig.~\ref{fig34}), the strong
Coulomb effect is confined to very forward values of $\Theta_1^{\rm lab}$
and to the second branch of solutions. Otherwise, only a slight reduction
or enhancement of the QFS(nn) cross section is observed at $E_{lab}=13$~MeV.
This effect further decreases with increasing incident energy, and at
 $E_{lab}=65$~MeV only a very weak influence on the cross section remains.

In Fig.~\ref{fig35} we show, for two energies ($E_{lab}=13$~MeV and
$E_{lab}=65$~MeV), the effects of the Coulomb force on QFS and FSI peaks
in kinematically complete configurations defined by specific values of
the outgoing nucleon angles $\Theta_1^{\rm lab}$, $\Theta_2^{\rm lab}$, and
$\Phi_{12}$. The effect on the QFS(nn) peak is relatively small and decreases
with increasing energy, whereas the FSI(nn) peak is completely suppressed
by the Coulomb barrier.

In view of the drastic reduction of the FSI(pp) peak, it is mandatory
to verify whether our conclusions concerning entanglement formation
in the FSI and QFS regions for the nd breakup reaction also remain valid for
the pd breakup. In Figs.~\ref{fig36}–\ref{fig38} we present the absolute
values of the expansion coefficients  $|\bar C_{i^,i}^{br}|$ for all FSI
configurations of two neutrons or two protons, calculated at $E_{lab}=65$~MeV
using solutions of the three-nucleon Faddeev equations without inclusion
of the pp Coulomb force (upper panels) and with the pp Coulomb interaction
included \cite{wit_coul} (lower panels).

The Coulomb interaction between the two protons does modify the values
of the coefficients and allows additional components—beyond the leading one,
which is the only nonvanishing component in the case of FSI(nn)—to contribute.
However, it does not affect the conclusions concerning the entanglement of
the produced Bell states. The difference of at least six orders of magnitude
between the leading coefficient and all remaining ones for FSI(pp)
at any production angle demonstrates that the conclusions regarding
the formation of entangled states drawn for the nd breakup remain unchanged.
This leaves no doubt that Bell states are also produced in these
 configurations in the pd breakup reaction.

 Figure~\ref{fig38} further shows that the entanglement is preserved
 over a relatively wide region of the S-curve around the FSI(pp) peak.

 That the action of the pp Coulomb force does not change the pattern of
 relative contributions from different terms is demonstrated for the
 QFS nn and pp configurations at $E_{lab}=65$~MeV in
 Figs.~\ref{fig40}–\ref{fig42}. Admittedly, at $E_{lab}=65$~MeV
 no entanglement is produced under QFS conditions. However, the clear
 similarity of the expansion coefficients in the nd and pd cases allows
 us to conclude that, also in the QFS case, the pp Coulomb force does not
 influence the conclusions concerning entanglement formation.


 Summarizing the results of the present section and Sec.~\ref{formNN}, based
 on the magnitudes of the contributing components to the final spin density
 matrix, we find numerous cases of Bell-state formation in unpolarized
 pp elastic scattering, as well as in FSI and QFS configurations of
 the exclusive pd breakup. A summary of these results is presented
 in Table~\ref{tab1}. The produced Bell states are predominantly
 of the $|\psi^{-}\rangle$ type. Only at high energies in $p(p,p)p$
 elastic scattering and in QFS(pp) configurations does the Bell-state
 type change to $|\psi^{+}\rangle$.


\begin{table}[ht!]
\centering
\caption{The incoming proton laboratory energy $E_{lab}$ and the angular
  regions in which different types of Bell states  $|\psi_i\rangle$
  of Eq.~(\ref{eq_1}) were identified in unpolarized elastic $p(p,p)p$
  scattering, as well as in the two complete geometries, QFS and FSI, of
  the unpolarized exclusive deuteron breakup reaction $d(p,N_1N_2)N_3$.
  The angular regions shown in the fourth column are based on considerations
  of the magnitudes of contributions from different terms, whereas those
  in the sixth column are based on a comparison of induced polarizations
  and spin correlations.
}
\label{tab1}
\begin{tabular}{ |c|c|c|c|c|c|}
\hline
~No~ & ~reaction~ & ~$E_{lab}$ [MeV]~   & ~region of angles~ & ~Bell state type~ &
~region of angles~  \\
    &             &                    &  (magnitude of terms)      &                   &
 (induced $P_y$ and $<\sigma_i^1 \sigma_j^2>$ \\
 \hline
 1 & $p(p,p)p$ &  10 &  $\Theta_{c.m.} \in (55^o,125^o)$ & $| \psi^{-} \rangle$ &
$\Theta_{c.m.} \in (55^o,125^o)$ \\
\hline
2 & $p(p,p)p$ & 151 &~ $\Theta_{c.m.} \in (85^o,95^o)~ $ & $| \psi^{+} \rangle$ &
 $\Theta_{c.m.} \in (85^o,95^o)$ \\
 \hline
 3 & FSI(pp) &  13  &  all $\Theta_1^{lab}$             & $| \psi^{-} \rangle$  &
all $\Theta_1^{lab}$  \\
\hline
4 & FSI(pp) &  156 &  all $\Theta_1^{lab}$             & $|\psi^{-} \rangle$   &
 all $\Theta_1^{lab}$    \\ 
\hline
5 & FSI(np) &  13  &  all $\Theta_1^{lab}$             & $| \psi^{-} \rangle$  &
 - \\
\hline
6 & FSI(np) &  156 & $\Theta_1^{lab} \in (25^o,55^o)$   & $|\psi^{-} \rangle$ &
 - \\
\hline
7 & QFS(pp) &  13 &   $\Theta_1^{lab} \in (25^o,50^o)$  & $| \psi^{-} \rangle$ &
$\Theta_1^{lab} \in (35^o,45^o)$  \\
\hline
8 &~ QFS(pp)~ &  156 & $\Theta_1^{lab} \in (30^o,60^o) $  & $|\psi^{+} \rangle$ &
$\Theta_1^{lab} \in (40^o,50^o)$ \\
\hline
\end{tabular}
\end{table}


As noted in Sec.~\ref{intro}, the formation of entangled pp pairs in
the FSI(pp) configuration of the exclusive pd breakup is not surprising,
as measurements reported in Ref.~\cite{sakai2006} support this possibility.
Any apparent surprise likely comes from seemingly contradictory considerations.

In a Bell state, the two protons must have vanishing polarizations. Yet,
unpolarized reactions generally induce polarization in the outgoing
nucleons \cite{ohlsen1972}, which must therefore disappear when a Bell state
forms. This provides a natural consistency check for the pp states listed
in Table~\ref{tab1}, showing how closely they approximate pure Bell states
and how their entanglement remains intact despite nonzero components
in the final spin density matrix.

In Figs.~\ref{fig44}–\ref{fig54} we show the induced polarizations $P_y$
and all nonvanishing induced spin correlations
$< \sigma_i^1 \sigma_j^2 >$ for the states listed in Table~\ref{tab1},
calculated using the AV18 NN potential. All other components of the
induced polarization vector, as well as the remaining induced
spin correlations, must vanish due to parity conservation~\cite{ohlsen1972}.
The observables presented here should be compared with their
counterparts for the entangled Bell states, summarized in Table~I
of Ref.~\cite{witnp2024}.
At this point, we emphasize the importance of the choice of coordinate
system when calculating spin observables, which should always be taken
as the right-handed Cartesian coordinate system defined according to
the Madison Convention~\cite{Madison1971}. Here, the coordinate system
is specified by taking the z-axis along the momentum of the incoming nucleon.
For the Bell states, the polarization $P_y$ vanishes, and the spin
correlations satisfy $< \sigma_i^1 \sigma_i^2 > = \pm 1$,
depending on the type of Bell state, while
$< \sigma_x^1 \sigma_z^2 > = < \sigma_z^1 \sigma_x^2 > = 0 $.

In Fig.~\ref{fig44}(c)–(f) we show the angular distributions of the
induced polarizations and spin correlations in unpolarized pp scattering
at two energies, $E_{lab}=10$~MeV and $E_{lab}=151$~MeV.
In panels (a) and (b), we also display the angular distributions of
the cross sections at these two energies.
The identity of the two protons ensures that all observables are
symmetric about $\Theta_{c.m.}=90^o$, and that the induced polarization $P_y$
as well as the spin correlations
$<\sigma_x^1 \sigma_z^2>$ and $<\sigma_z^1 \sigma_x^2>$
must vanish for any energy at $\Theta_{c.m.}=90^o$. At angles different
from $90^o$,  $P_y$, $<\sigma_x^1 \sigma_z^2>$, and $<\sigma_z^1 \sigma_x^2>$
do not vanish. They are only slightly different from zero
at  $E_{lab}=10$~MeV; however, at $E_{lab}=151$~MeV they can reach large values
of approximately $\pm 0.25$ for $P_y$ and $\pm 0.5$ for the spin correlations.
The induced spin correlations
$<\sigma_x^1 \sigma_x^2>$ and $<\sigma_y^1 \sigma_y^2>$
take values of approximately $+1$ at and around $\Theta_{c.m.}=90^o$
at both energies, while $<\sigma_z^1 \sigma_z^2>$ takes the value $+1$
at the lower energy and $-1$ at the higher energy, as expected for
the states $| \psi^{-} \rangle$ and $| \psi^{+} \rangle$, respectively.
This comparison shows that the state of the pp pair produced at
$\Theta_{c.m.}=90^o$ is indeed a Bell-like state of the type shown
in Table~\ref{tab1}, and that this conclusion also holds in the angular
ranges indicated in the fourth column of that table.

The same is also true for QFS pp pairs produced in the pd breakup reaction
under QFS conditions at $E_{lab}=13$~MeV and $\Theta_1^{\rm lab} \approx 39^o$
(see Fig.~\ref{fig45}), as well as at $E_{lab}=156$~MeV and
$\Theta_1^{\rm lab} \approx 45^o$ (see Fig.~\ref{fig46}). At these
particular laboratory angles, the induced polarization $P_y$ and
the induced spin correlations $<\sigma_x^1 \sigma_z^2>$
and $<\sigma_z^1 \sigma_x^2>$ vanish, while the diagonal spin correlations
approach values of approximately $\pm 1$, as expected for strongly
entangled Bell states.
However, the angular regions shown in the fourth column of Table~\ref{tab1},
which are based solely on the magnitudes of the different contributing terms,
must be narrowed to the values indicated in the sixth column of that table.
This refinement is required by the angular dependence of the induced
polarizations and the induced diagonal spin correlations shown in
Figs.~\ref{fig45} and \ref{fig46}.

The results (not shown) for np pairs in the QFS(np) configuration,
in particular the small values of the diagonal
spin correlations, which never approach values close to $1$
(they lie at $E_{lab}=13$~MeV in the range of approximately
$-0.1$ to $0.3$, and at $E_{lab}=156$~MeV they reach values of up to
approximately $\pm 0.8$), show that
under these conditions no Bell-like states are formed. This finding
supports the conclusions based on the magnitudes of the contributing
terms for this case.

Also np pairs formed under FSI conditions at both energies do not reveal
any resemblance to entangled Bell states (also not shown). 
 The diagonal spin correlations take values that depend
on  $\Theta_1^{lab}$: at $E_{lab}=13$~MeV they are approximately $0.8$, while
at $E_{lab}=156$~MeV they are even smaller, ranging between $-0.2$
and $-0.8$. These values are far from those characteristic
of Bell-like states. It therefore appears that, in this case, the
difference between $\bar C_{44}^{br}$ and the other coefficients is not
sufficient to ensure the formation of Bell states.

In Figs.~\ref{fig51} and \ref{fig52} we show, at both energies,
the S dependence of the induced polarizations and spin correlations
for kinematically complete geometries containing a location at which
the exact QFS(nn) condition is fulfilled and an nn pair is formed
in an entangled Bell state. It is clear that pairs formed at different
S locations around this value of S are also in such a Bell state. At the
lower energy, the corresponding region $ S \in (7.5,9.5)$~MeV
is relatively narrow compared with that at the higher energy,
$ S \in (100,130)$~MeV.

In Figs.~\ref{fig53} and \ref{fig54} we illustrate the same behavior
for the FSI(nn) configuration. We note that FSI(nn) (and similarly FSI(pp))
is exceptional in that, for any $\Theta_1^{\rm lab}$,  whenever
the FSI(12) condition is fulfilled, the two neutrons (protons)
form an exact Bell state. The figures show how far one can depart from
the exact FSI condition without spoiling this state. It is seen that
the corresponding ranges of S are
$ S \in (4,7)$~MeV at $E_{lab}=13$~MeV  and $ S \in (50,75)$~MeV
at $E_{lab}=156$~MeV.

In view of the drastic reduction of the FSI(pp) cross section due to
Coulomb barrier penetration, we considered it necessary to verify that
the inclusion of the pp Coulomb force has no significant influence on
polarizations and spin correlations under these conditions.
In Fig.~\ref{exp_scheme}, we present a comparison of induced polarizations
and induced spin correlations at $E_{lab}=65$~MeV, calculated with and
without the Coulomb force included, for a specific breakup geometry
containing the location S, at which the exact FSI(pp) (and FSI(nn))
condition is fulfilled. Indeed, only small Coulomb-force effects are
observed, and these occur far from the FSI(pp) region.

We have added a sixth column to Table~\ref{tab1}, which summarizes
the conclusions regarding the formation of Bell states based on a
comparison of the induced polarizations and spin correlations
with those of the Bell states.

At this point, we would like to stress the fundamental difference between
the formation of strongly entangled pp pairs through pp elastic scattering
or the pd breakup reaction in QFS(pp) configurations, on the one hand, and
the use of complete FSI(pp) configurations in the pd breakup, on the other.
This difference is essential for the experiment discussed in the next section.

In the latter case, the pp pairs are always produced in an exact FSI
Bell state, and the energy of the entangled protons—which depends only
on the incoming proton energy and the FSI production angle—can be
freely adjusted. In contrast, in pp scattering or in QFS geometries of
the pd breakup, entangled pp pairs can be formed only in relatively
pure but nevertheless approximate Bell states. Moreover, the energies
at which such states can be produced are restricted either to a
low-energy region ($\approx 10$~MeV) or to a rather narrow region
around $\approx 150$~MeV.

The requirement of the dominance of a single term in the pp transition
matrix M for subsequent pp scattering of one of the produced entangled
protons—whose energy is approximately half of the incoming proton
energy—renders both of these processes unsuitable for performing 
the experiment in question at higher energies.


\section{Teleportation experiment}
\label{teleport}

Equipped with a method for preparing two protons in an entangled Bell state,
and with knowledge of the structure of the pp scattering matrix M
in the Bell basis, we discuss the feasibility of an experiment—based on
ideas proposed in \cite{teleport}—aimed at teleporting a quantum
mechanical state between two protons. The experimental
setup is shown in Fig.~\ref{exp_scheme}.

The system consists of three protons initially prepared in the state
$| \tilde \psi_i \rangle$. Proton 1 is in the state
$| \psi \rangle_1 = \alpha | - \rangle_1 + \beta | + \rangle_1$, 
while protons 2 and 3 are in an entangled Bell state
$| \psi_i \rangle_{23}$, $i=1,2,3,4$. 
Consequently, the polarizations $P_y$ of protons 2 and 3 vanish, and
the polarization of proton 1 is given by:
\begin{eqnarray}
 P_y &=&  ~_1\langle \psi | \sigma_y | \psi  \rangle_1   = 2 \mathfrak{Im} (\alpha \beta^*) 
   \label{eq_0} ~, 
 \end{eqnarray}
where the normalization condition $| \alpha |^2 + | \beta |^2 = 1$ holds.

The initial  state of the system may 
be expressed in the Bell basis as:
\begin{eqnarray}
  | \tilde \psi_1 \rangle  = | \psi \rangle_1 \otimes | \Phi^+ \rangle_{23}=
  \frac {1} {2} &[&
 + | \Phi^+ \rangle_{12} \otimes ( + \alpha | - \rangle_3  +  \beta | + \rangle_3)
 \cr  
&~& + | \Phi^- \rangle_{12} \otimes ( + \alpha | - \rangle_3  -  \beta | + \rangle_3)
 \cr 
&~& + | \psi^+ \rangle_{12} \otimes ( + \alpha | + \rangle_3  +  \beta | - \rangle_3)
 \cr
&~& + | \psi^- \rangle_{12} \otimes ( + \alpha | + \rangle_3  -  \beta | - \rangle_3) ]
\label{eq_2} \\ 
  | \tilde \psi_2 \rangle  = | \psi \rangle_1 \otimes | \Phi^- \rangle_{23}=
  \frac {1} {2} &[&
 +  | \Phi^+ \rangle_{12} \otimes ( + \alpha | - \rangle_3  -  \beta | + \rangle_3)
 \cr  
&~& + | \Phi^- \rangle_{12} \otimes ( + \alpha | - \rangle_3  +  \beta | + \rangle_3)
 \cr 
&~& + | \psi^+ \rangle_{12} \otimes ( - \alpha | + \rangle_3  +  \beta | - \rangle_3)
 \cr
&~& + | \psi^- \rangle_{12} \otimes ( - \alpha | + \rangle_3  -  \beta | - \rangle_3) ]
\label{eq_3} \\  
  | \tilde \psi_3 \rangle  = | \psi \rangle_1 \otimes | \psi^+ \rangle_{23}=
  \frac {1} {2} &[&
  +  | \Phi^+ \rangle_{12} \otimes ( + \alpha | + \rangle_3  +  \beta | - \rangle_3)
\cr  
&~& + | \Phi^- \rangle_{12} \otimes ( + \alpha | + \rangle_3  -  \beta | - \rangle_3)
\cr 
&~& + | \psi^+ \rangle_{12} \otimes ( + \alpha | - \rangle_3  +  \beta | + \rangle_3)
\cr
&~& + | \psi^- \rangle_{12} \otimes ( + \alpha | - \rangle_3  -  \beta | + \rangle_3) ]
\label{eq_4} \\  
  | \tilde \psi_4 \rangle  = | \psi \rangle_1 \otimes | \psi^- \rangle_{23}=
  \frac {1} {2} &[&
  + | \Phi^+ \rangle_{12} \otimes ( + \alpha | + \rangle_3  -  \beta | - \rangle_3)
\cr  
&~& + | \Phi^- \rangle_{12} \otimes ( + \alpha | + \rangle_3  +  \beta | - \rangle_3)
\cr 
&~& + | \psi^+ \rangle_{12} \otimes (- \alpha | - \rangle_3  +  \beta | + \rangle_3)
\cr
&~& + | \psi^- \rangle_{12} \otimes ( - \alpha | - \rangle_3  -  \beta | + \rangle_3)
  ]  ~.
\label{eq_5}
\end{eqnarray}

We now let proton 2 scatter off proton 1. After the scattering process,
the state of the system becomes:
\begin{eqnarray}
  M_{12} \otimes I_3 ~ | \tilde \psi_i \rangle &=&  M_{12} \otimes I_3
(  | \psi \rangle_1 \otimes  | \psi_i \rangle_{23} ) ~.
 \label{eq_5a}
\end{eqnarray}

Using the form
$M_{12} = | \psi^- \rangle_{12}  ~_{12}\langle \psi^- |$, 
valid at $E_{lab} \approx 10$~MeV, we obtain for each $| \tilde \psi_i \rangle$: 
\begin{eqnarray}
  M_{12} \otimes I_3 ~  | \tilde \psi_1 \rangle &=&
 + \frac {1} {2} 
  | \psi^- \rangle_{12} \otimes ( + \alpha | + \rangle_3  -  \beta | - \rangle_3)
\equiv  + \frac {1} {2} 
   | \psi^- \rangle_{12}  \otimes |  \psi_3 \rangle
  \label{eq_6a}    \\
  M_{12} \otimes I_3 ~  | \tilde \psi_2 \rangle  &=&
 + \frac {1} {2} 
  | \psi^- \rangle_{12} \otimes ( - \alpha | + \rangle_3  -  \beta | - \rangle_3)
\equiv  + \frac {1} {2} 
   | \psi^- \rangle_{12}  \otimes |  \psi_3 \rangle
  \label{eq_6b} \\
  M_{12} \otimes I_3 ~  | \tilde \psi_3 \rangle  &=&
 + \frac {1} {2} 
  | \psi^- \rangle_{12} \otimes ( + \alpha | - \rangle_3  -  \beta | + \rangle_3)
\equiv  + \frac {1} {2} 
   | \psi^- \rangle_{12}  \otimes |  \psi_3 \rangle
  \label{eq_6c} \\  
  M_{12} \otimes I_3 ~  | \tilde \psi_4 \rangle  &=&
 - \frac {1} {2} 
  | \psi^- \rangle_{12} \otimes ( + \alpha | - \rangle_3  +  \beta | + \rangle_3)
\equiv  - \frac {1} {2} 
   | \psi^- \rangle_{12}  \otimes |  \psi_3 \rangle
  \label{eq_6d} ~. 
\end{eqnarray}

In a similar manner, using
$M_{12} = | \psi^+ \rangle_{12}  ~_{12}\langle \Phi^- |$, 
valid at $E_{lab} \approx 151$~MeV, we obtain:
\begin{eqnarray}
  M_{12} \otimes I_3 ~  | \tilde \psi_1 \rangle  &=&
 + \frac {1} {2} 
  | \psi^+ \rangle_{12} \otimes ( + \alpha | - \rangle_3  -  \beta | + \rangle_3)
\equiv  + \frac {1} {2} 
   | \psi^+ \rangle_{12}  \otimes |  \psi_3 \rangle
  \label{eq_7a} \\
  M_{12} \otimes I_3 ~  | \tilde \psi_2 \rangle  &=&
 + \frac {1} {2} 
  | \psi^+ \rangle_{12} \otimes ( + \alpha | - \rangle_3  +  \beta | + \rangle_3)
\equiv  + \frac {1} {2} 
   | \psi^+ \rangle_{12}  \otimes |  \psi_3 \rangle
 \label{eq_7b} \\
  M_{12} \otimes I_3 ~  | \tilde \psi_3 \rangle  &=&
 + \frac {1} {2} 
  | \psi^+ \rangle_{12} \otimes (+ \alpha | + \rangle_3  -  \beta | - \rangle_3)
\equiv  + \frac {1} {2} 
   | \psi^+ \rangle_{12}  \otimes | \psi_3 \rangle
  \label{eq_7c} \\  
  M_{12} \otimes I_3 ~  | \tilde \psi_4 \rangle  &=&
 + \frac {1} {2} 
  | \psi^+ \rangle_{12} \otimes ( + \alpha | + \rangle_3  +  \beta | - \rangle_3)
\equiv  + \frac {1} {2} 
   | \psi^+ \rangle_{12}  \otimes |  \psi_3 \rangle
  \label{eq_7d} ~. 
\end{eqnarray}

Thus, after the scattering, the system evolves into a new state. Protons 1
and 2 are now in an entangled Bell state, while proton 3 ends up in a pure
state $| \psi \rangle_3$,  
which is very similar to the initial state
of proton 1,  $| \psi \rangle_1$. The final state of proton 3 depends
on the initial entanglement between protons 2 and 3. 
When protons 2 and 3 are entangled in the state $| \psi^- \rangle_{23}$,
each with an energy of $\approx 10$~MeV, or when they are in the state
$| \Phi^- \rangle_{23}$, each with an energy of $\approx 151$~MeV, the state of
proton 3 is exactly $| \psi \rangle_1$. This is precisely what is meant
by teleportation.

We have shown that, using unpolarized pp scattering or the
unpolarized pd breakup reaction, it is not possible to produce the entangled
Bell state  $| \Phi^- \rangle_{23}$ of two protons. Instead, the protons
can be produced only in the state  $| \psi^- \rangle_{23}$, and,
at $E_{lab} \approx 151$~MeV,  also in the state $| \psi^+ \rangle_{23}$.

When protons 2 and 3 are entangled,  
each with an energy of $\approx 151$~MeV, teleportation of the state
$| \psi \rangle_1$
results in proton 3 being in the state
$| \psi \rangle_3 =  \alpha | + \rangle_3  +  \beta | - \rangle_3$ 
for the entangled state $| \psi^- \rangle_{23}$, 
with the polarization $P_y$  of proton 3
opposite to that of proton 1, and in the state
$| \psi \rangle_3 =  \alpha | + \rangle_3  -  \beta | - \rangle_3$ 
for  $| \psi^+ \rangle_{23}$, with the polarization of proton 3
 the same as that of proton 1. 
In Appendix~\ref{a1}, we show that these conclusions concerning polarization
remain unchanged when the pure state of proton 1 is replaced by a statistical
mixture of states.

The second possibility is practically not feasible. Producing such an
entangled pair would require pp elastic scattering or QFS(pp) kinematics
in the pd breakup at approximately 300 MeV. Being far from the optimal
energy of about 150 MeV, this would result in degraded Bell states
(see Fig.~\ref{fig1bb}), with spin correlations taking values of
magnitude of about $0.7-0.8$.

The first possibility, however, can be realized in the pd breakup
in the FSI(pp) geometry using an incoming proton with an energy
of approximately $350$~MeV.

In the actual experiment, depicted in Fig.~\ref{exp_scheme}, an
unpolarized proton beam impinges on an unpolarized proton or deuteron
target, producing protons 2 (moving in direction $\Theta_2$) and 3
(in direction $\Theta_3$), both in an entangled Bell state, represented
by an orange connection. Proton 2 then scatters on the hydrogen target,
whose protons are in a pure state $| \psi \rangle_1$  with target
polarization $P_y$ as given in Eq.~(\ref{eq_0}).

After this scattering occurs, protons 4 and 5 are found in an
entangled Bell state, the type of which depends on the entangled state
of protons 2 and 3. Proton 3 ends up in a teleported state, which
depends both on the type of entangled state of protons 2 and 3 and
on the initial state of proton 1.

Regarding the practical feasibility of such an experiment, two
questions arise. First, what are the observable signals in this experiment
that would indicate that the teleportation process has indeed occurred?
Second, is the practical implementation of such an experiment feasible at all?

Concerning the experimental signals indicating that teleportation
has occurred, all possible observables are associated with
the polarizations of the outgoing protons. Proton 3, which ends up
in the teleported state  $| \psi \rangle_3$, exhibits a polarization
$\pm P_y$ that is strongly correlated with the polarization of the hydrogen
target. This polarization can be measured via left–right scattering on
a target with a known proton analyzing power.

In practice, the measurement requires selecting the appropriate events
using triple coincidences,   $N_L^{234}$ and  $N_R^{234}$. The coincidence
between protons 2 and 3 ensures that the entangled state of protons 2–3
has been formed, while the detection of proton 4 provides additional
evidence that proton 2 scattered on proton 1, thus triggering
the actual teleportation. As a control, the asymmetry based on
double coincidences  $N_{L,R}^{23}$  can be measured with the polarized
hydrogen target removed; in this case, no teleportation occurs, and the
asymmetry in proton 3 scattering should vanish.

Another independent signature of teleportation arises from
the concurrent transfer of entanglement. Teleportation not only transfers
the quantum state from proton 1 to proton 3, but also transfers
the entanglement from the initial proton pair 2–3 to the pair 1–2.
Consequently, protons 4 and 5 are expected to appear in a Bell-like state,
with vanishing polarizations, which leads to a null asymmetry in
their scattering on a target. Together, these observations provide
a clear experimental signature of the teleportation process.

Before addressing the second question, it is important to note
two components that are essential for a successful teleportation experiment.
The first is the formation, at the energy of the experiment, of high-quality entangled Bell-like states of two protons. The second is the proper structure
of the pp elastic scattering matrix M. This scattering process is actually
responsible for triggering teleportation, and for it to proceed smoothly,
the M matrix at the incoming proton energy must be dominated to a high degree
by a single contributing term.

\begin{figure}
\begin{center}
\begin{tabular}{c}
\resizebox{128mm}{!}{\includegraphics[angle=270]{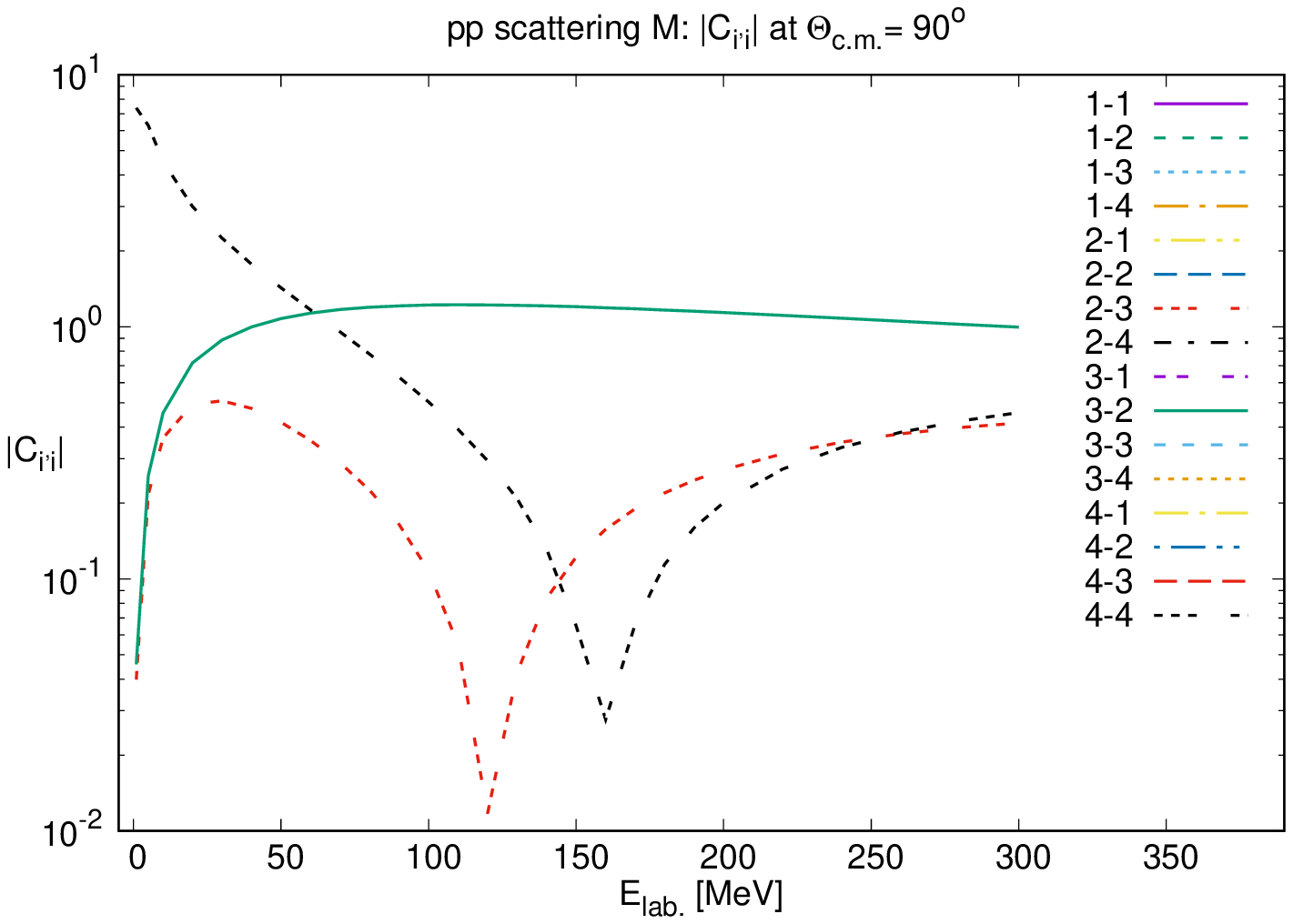}} \\
\resizebox{128mm}{!}{\includegraphics[angle=270]{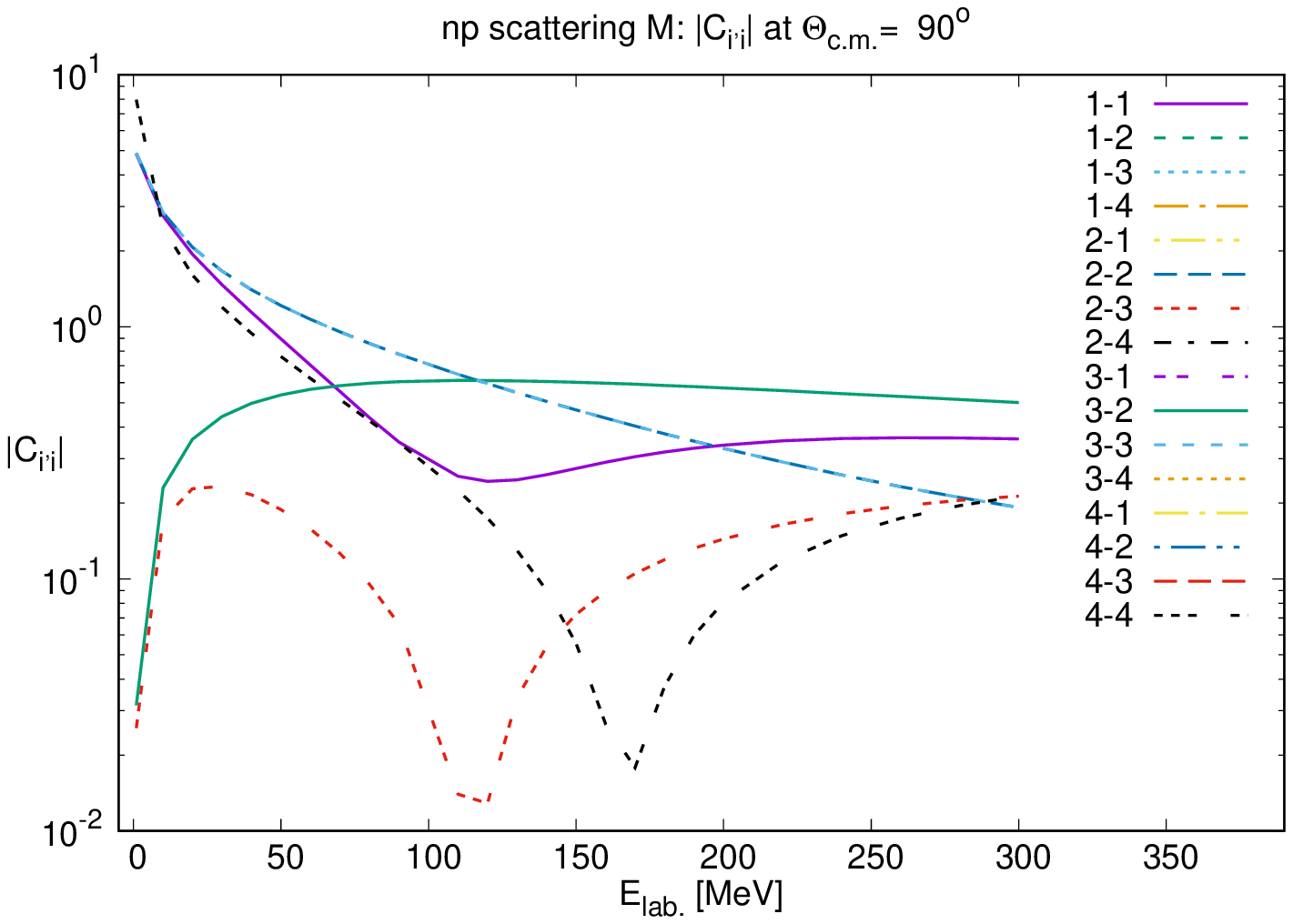}} \\
\end{tabular}%
\caption{
  (color online   Laboratory-energy dependence of the absolute values
  of the M-matrix expansion coefficients $|C_{i'i}(\Theta_{c.m.}=90^o)|$
  (Eq.~(\ref{eq_6}))  in unpolarized pp and np scattering.
  Note that at this angle $C_{22}=C_{33}$, and the corresponding curves overlap.
  The calculations were performed using the AV18 NN potential and a
  partial-wave set with $j_{max} = 5$.
  In the legend, the index $i'-i$  of the coefficients and the
  explanation of the lines are given.
}    
\label{fig1aa}
\end{center}
\end{figure}

\begin{figure}
\begin{center}
\begin{tabular}{c}
\resizebox{128mm}{!}{\includegraphics[angle=270]{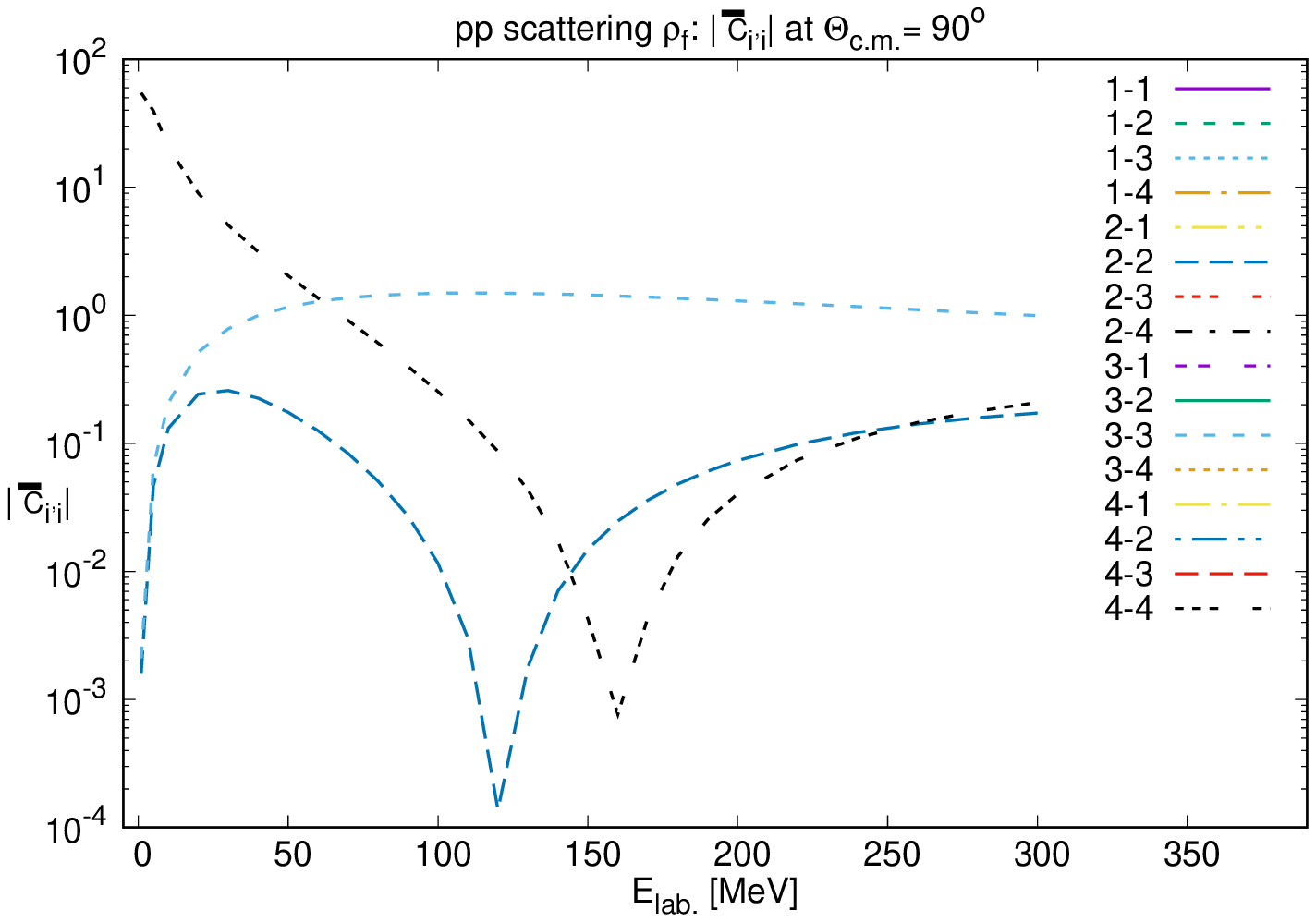}} \\
\resizebox{128mm}{!}{\includegraphics[angle=270]{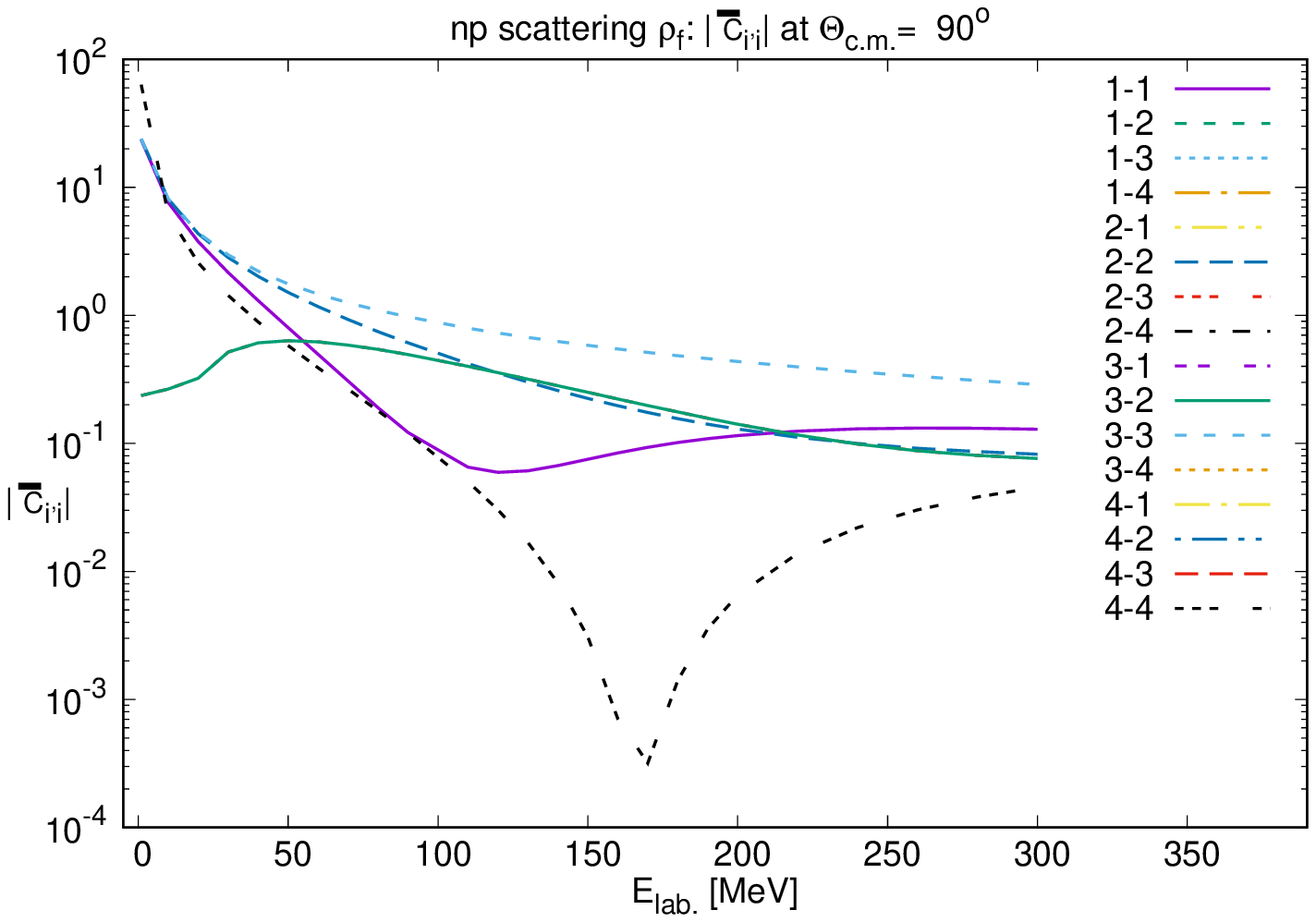}} \\
\end{tabular}%
\caption{
  (color online)  The same as in Fig.~\ref{fig1aa} but for the 
  spin density matrix $\rho_f$ and   the coefficients 
  $|\bar C_{i'i}(\Theta_{c.m.}=90^o)|$
  (Eq.~(\ref{eq_9})). Note that the absolute values of $|\bar C_{23}|$ and
  $|\bar C_{32}|$ are identical and the corresponding curves overlap.
 }
\label{fig1bb}
\end{center}
\end{figure}

\begin{figure}
\includegraphics[angle=270,scale=0.5]{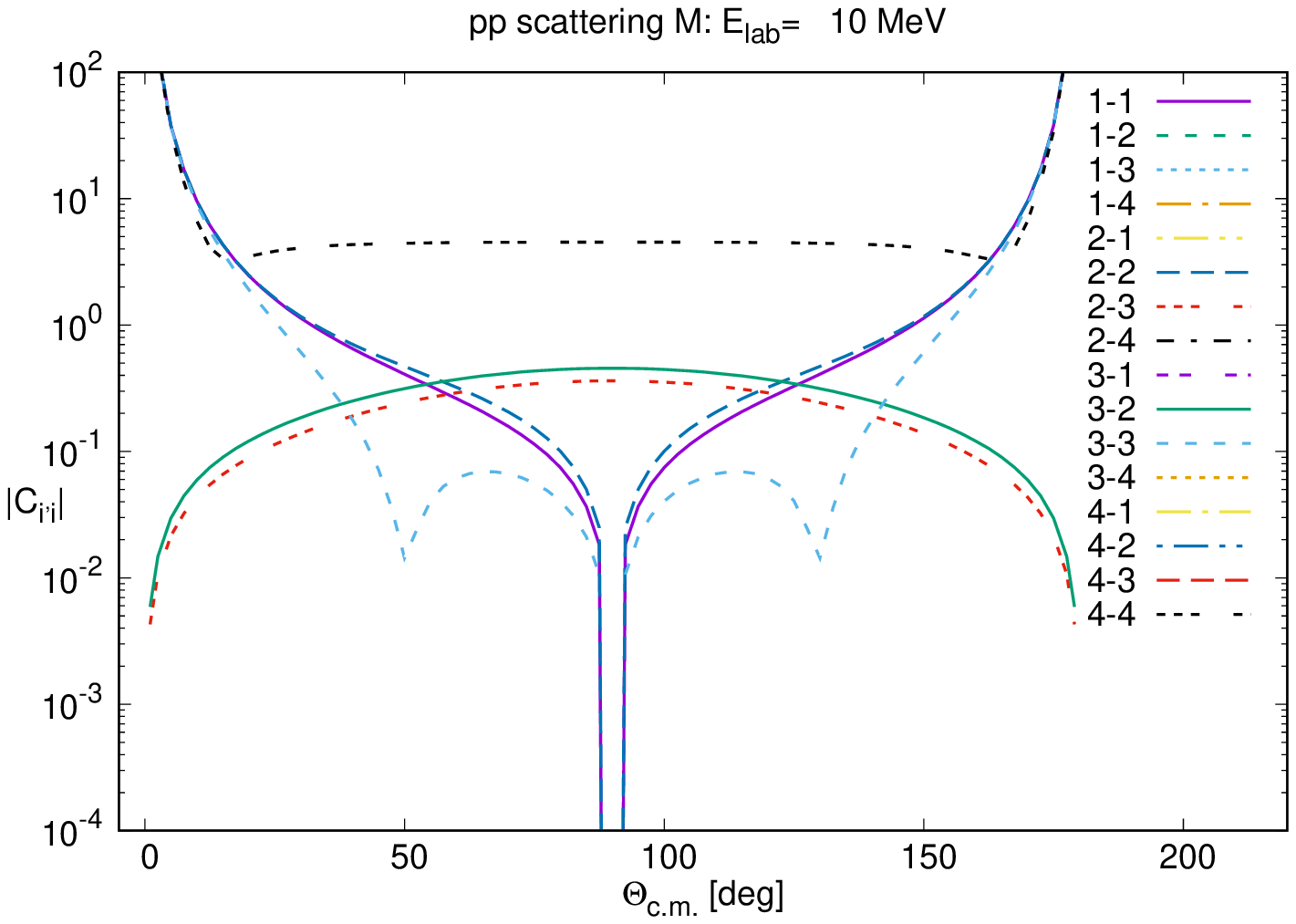}  
\caption{
  (color online) Angular distribution of the absolute values of the M-matrix
  expansion  coefficients $|C_{i^,i}|$ (Eq.~(\ref{eq_6})) in unpolarized pp
  scattering at $E_{lab}=10$~MeV.
  }
\label{fig1}
\end{figure}

\begin{figure}
\includegraphics[angle=270,scale=0.5]{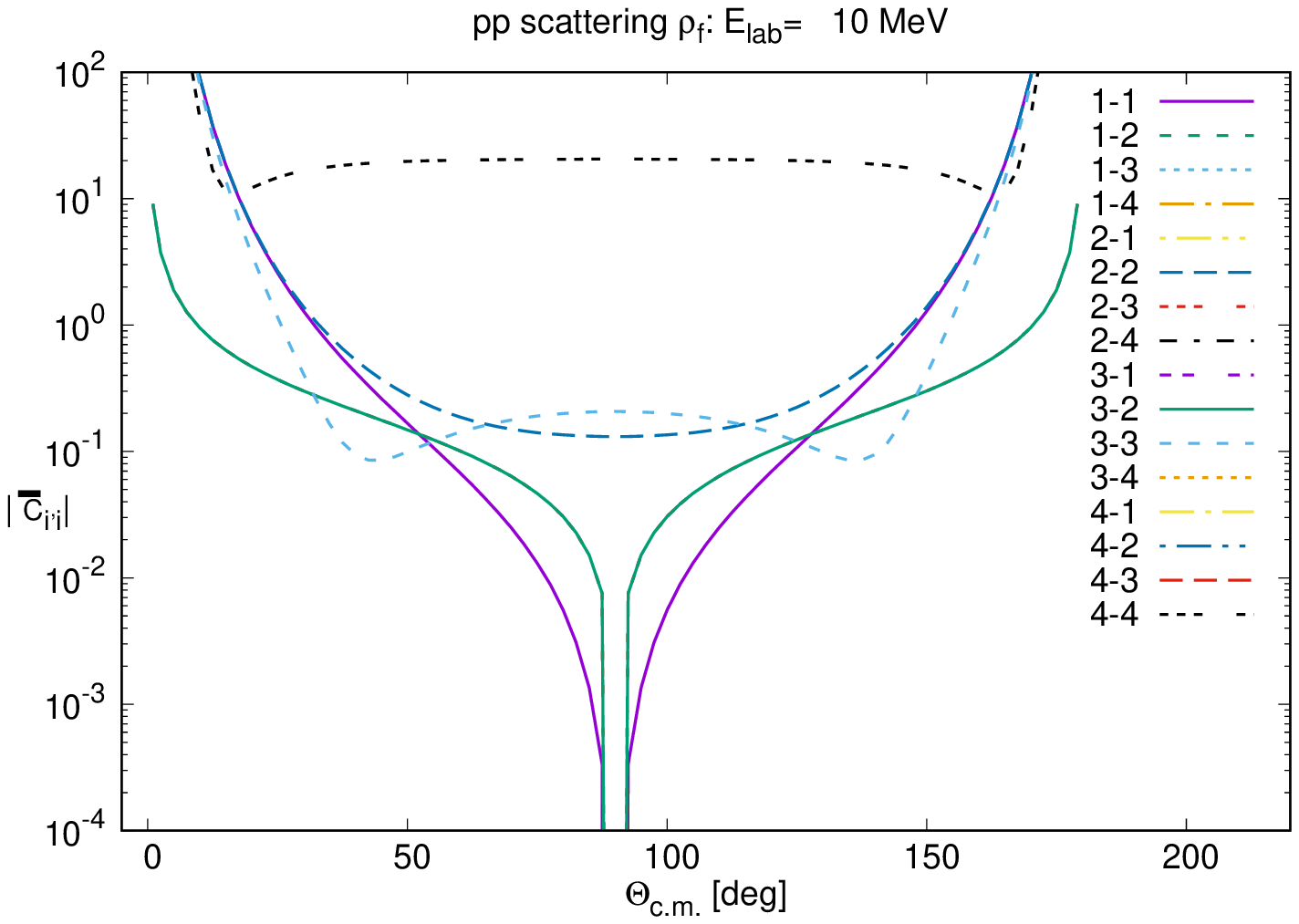}  
\caption{
  (color online) Angular distribution of the absolute values of the final spin 
  density matrix expansion
  coefficients $|\bar C_{i^,i}|$ (Eq.~(\ref{eq_9})) in unpolarized pp
  scattering at $E_{lab}=10$~MeV.
  }
\label{fig2}
\end{figure}

\begin{figure}
\includegraphics[angle=270,scale=0.5]{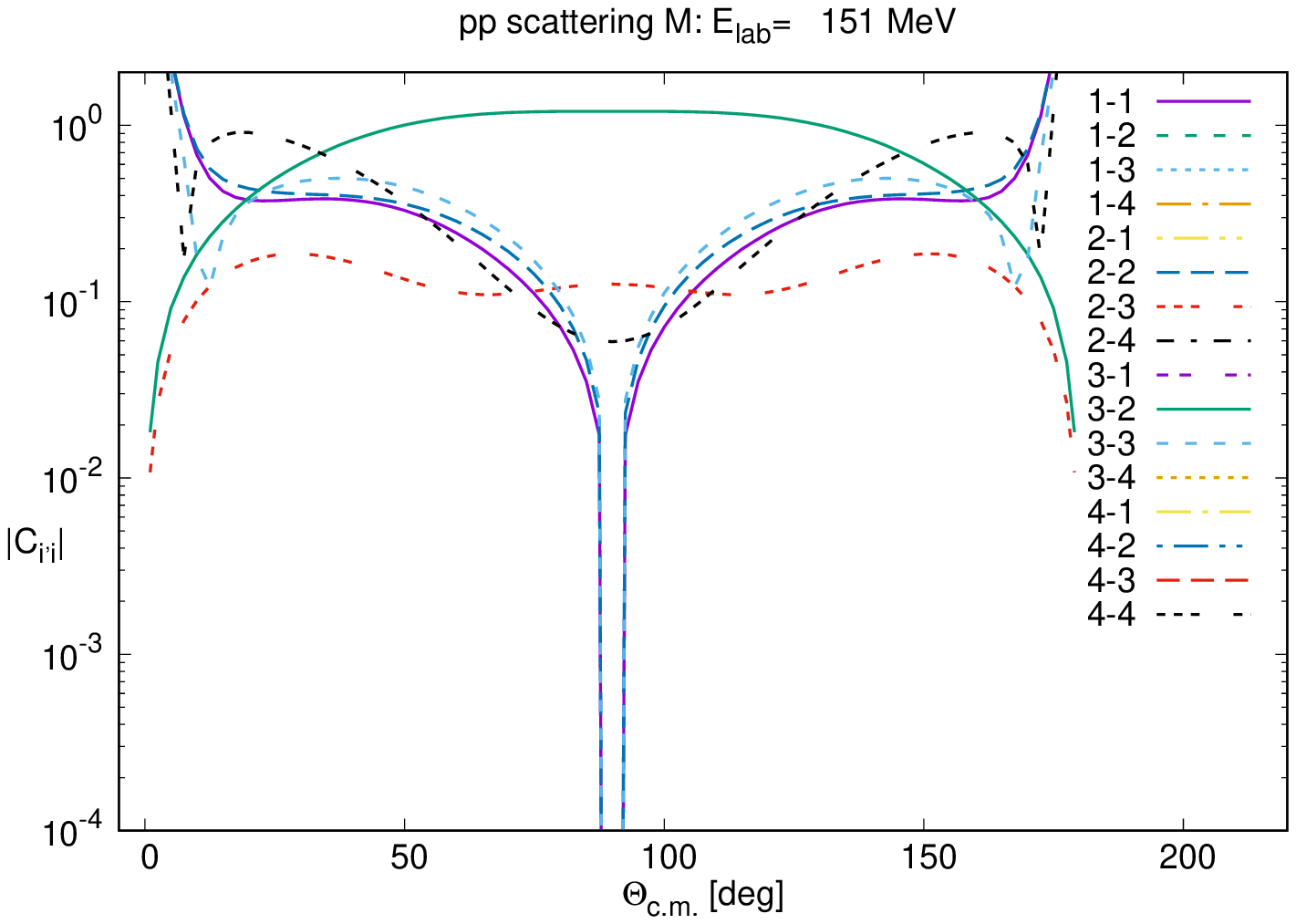}  
\caption{
  (color online) Same as in Fig.~\ref{fig1}, but for  $E_{lab}=151$~MeV.
  }
\label{fig3}
\end{figure}

\begin{figure}
\includegraphics[angle=270,scale=0.5]{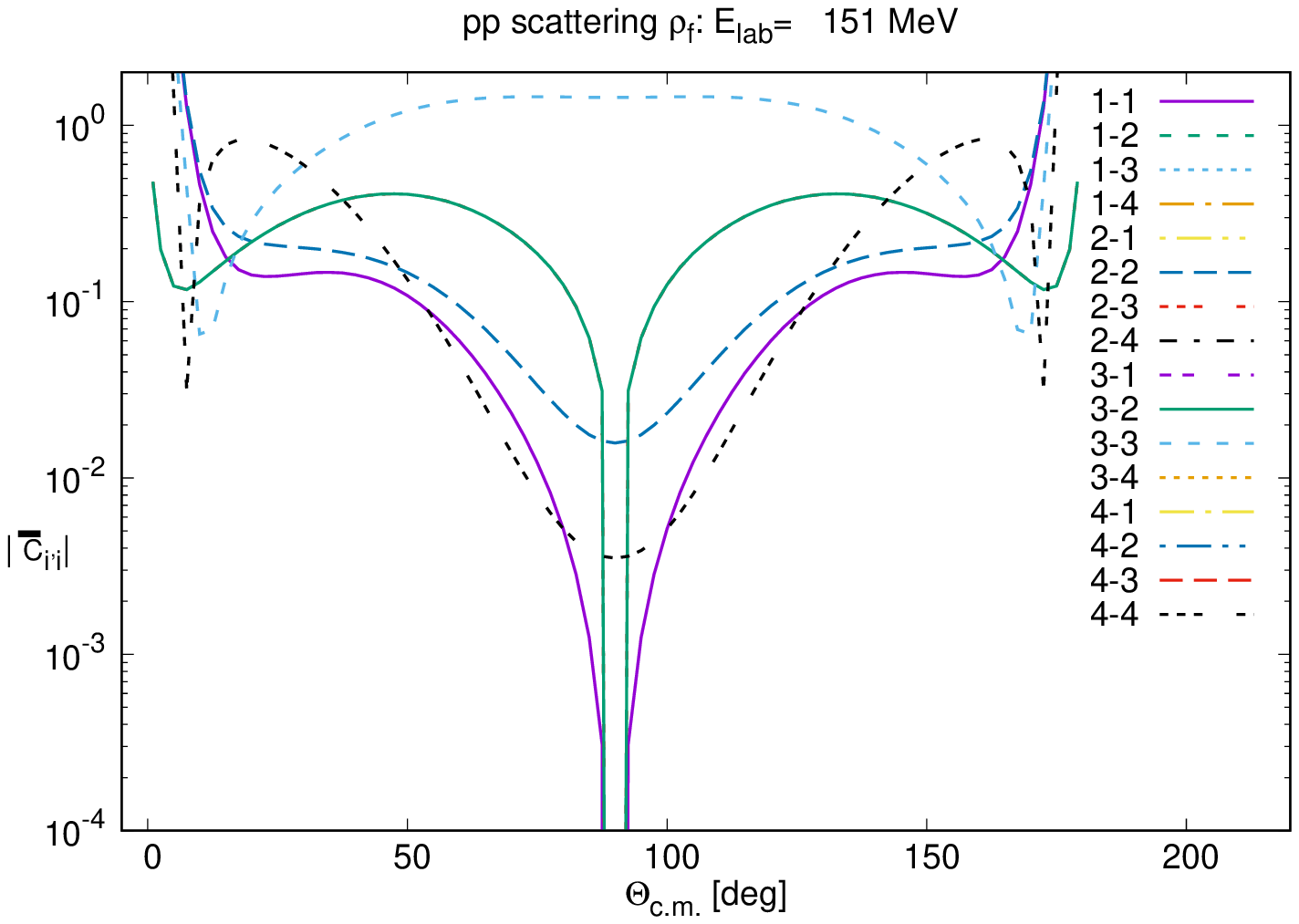}  
\caption{
  (color online)  Same as in Fig.~\ref{fig2}, but for $E_{lab}=151$~MeV.
  }
\label{fig4}
\end{figure}

\begin{figure}
\includegraphics[angle=270,scale=0.5]{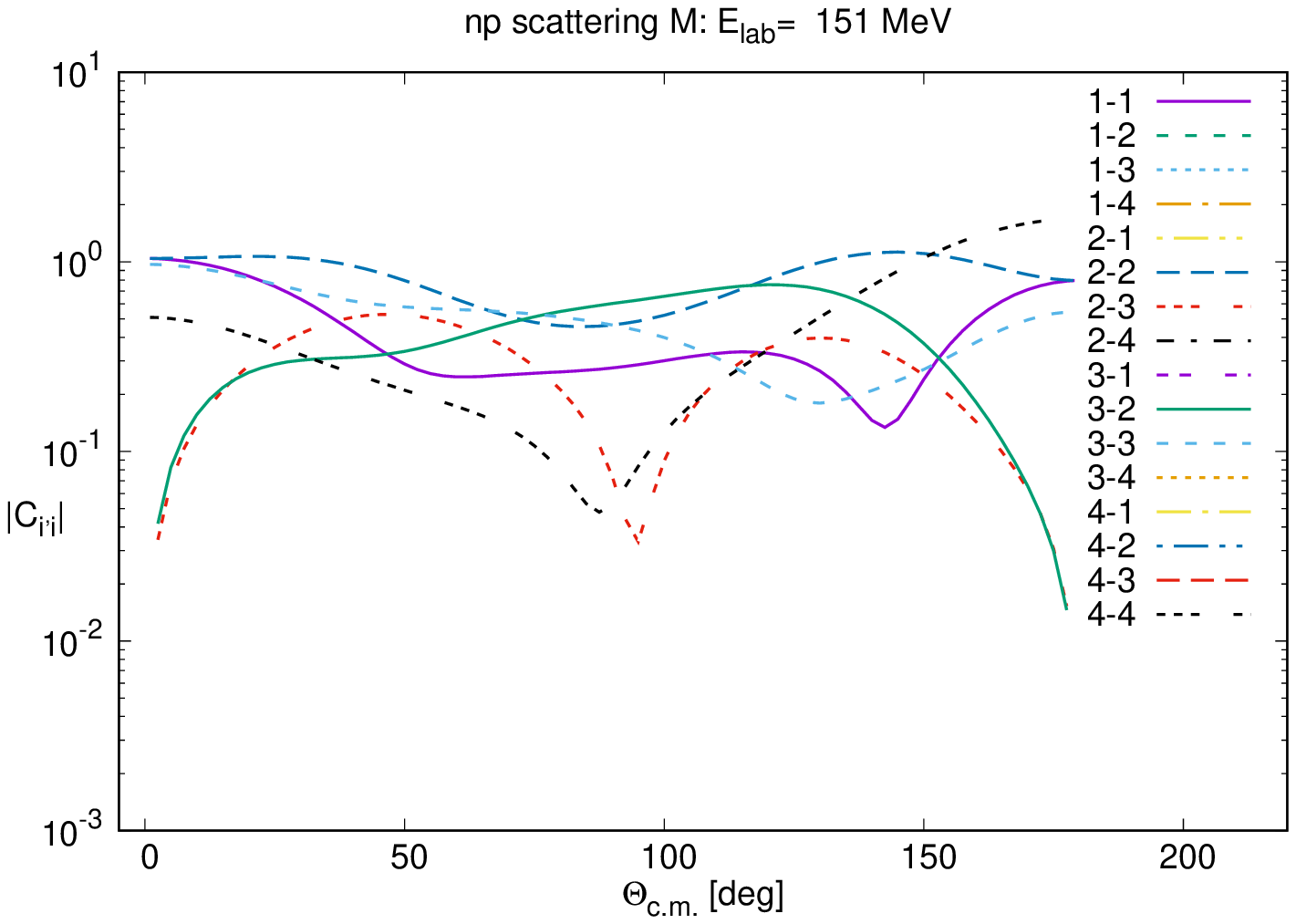}  
\caption{
  (color online) Angular distribution of the absolute values of the M-matrix
  expansion  coefficients $|C_{i^,i}|$ (Eq.~(\ref{eq_6})) in unpolarized np
  scattering at $E_{lab}=151$~MeV.
  }
\label{fig7}
\end{figure}

\begin{figure}
\includegraphics[angle=270,scale=0.5]{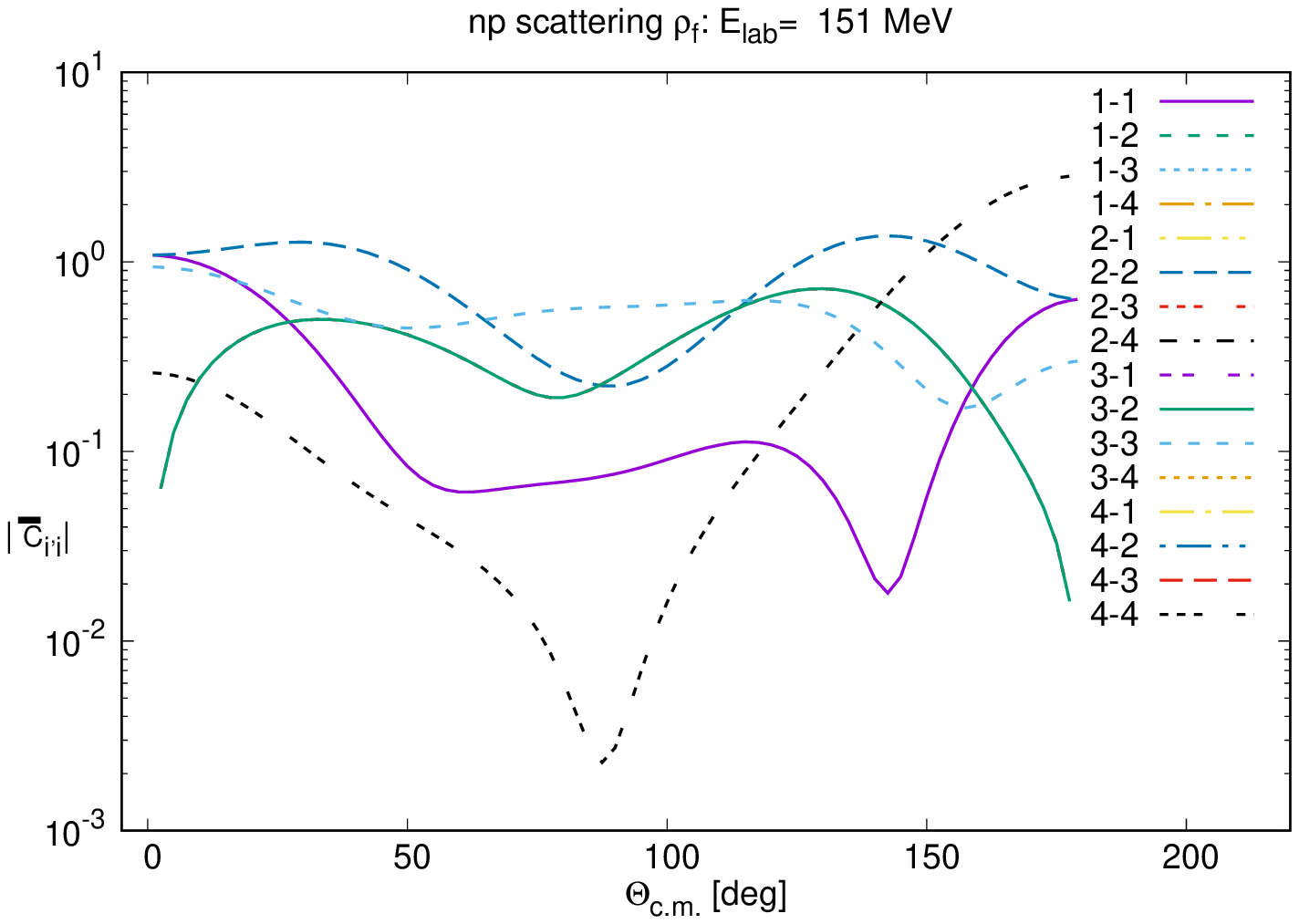}  
\caption{
  (color online) Angular distribution of the absolute values of the final spin 
  density matrix expansion
  coefficients $|\bar C_{i^,i}|$ (Eq.~(\ref{eq_9})) in unpolarized np
  scattering at $E_{lab}=151$~MeV.
  }
\label{fig8}
\end{figure}

\begin{figure}
\includegraphics[angle=270,scale=0.5]{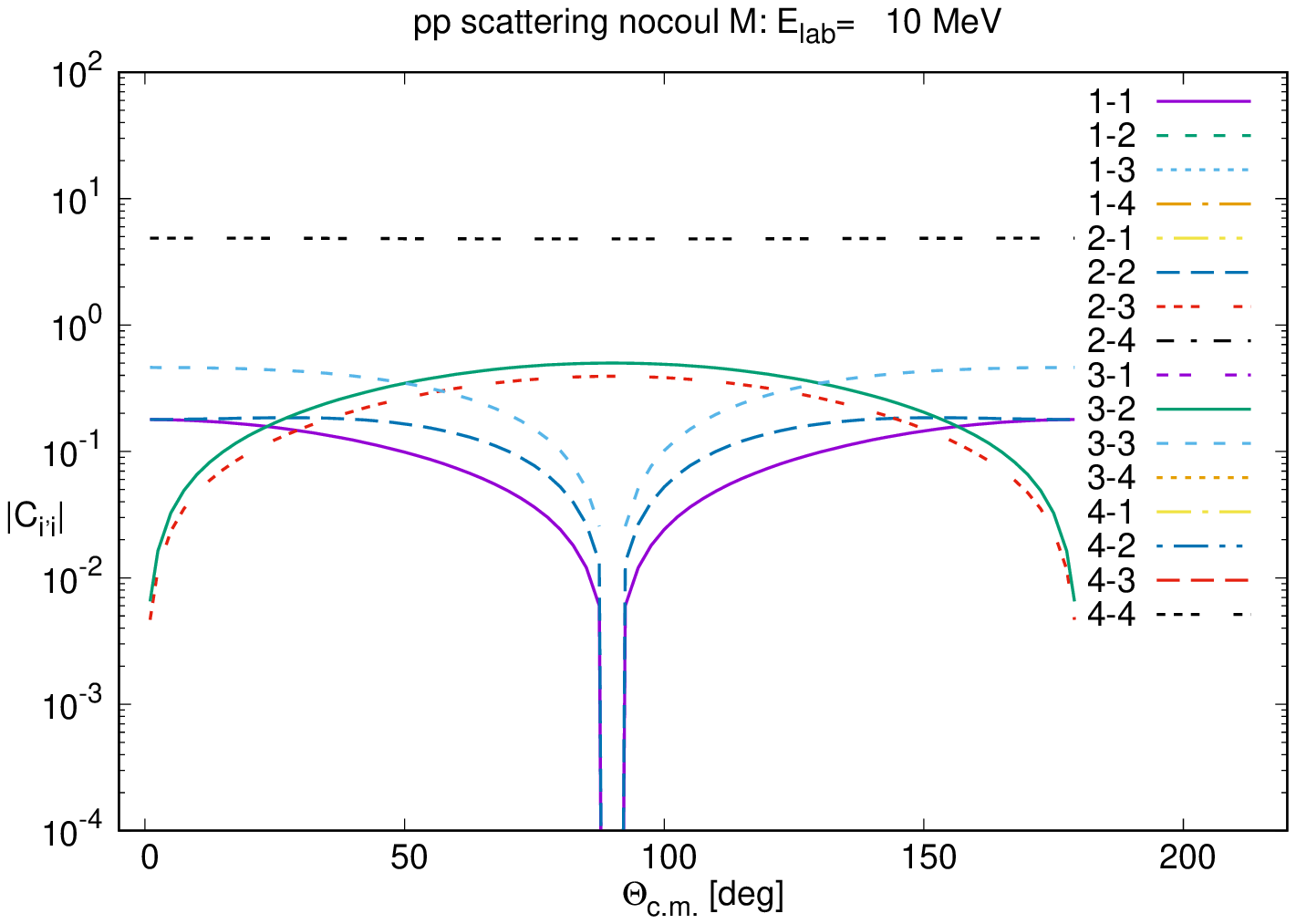}  
\caption{
  (color online) Angular distribution of the absolute values of the M-matrix
  expansion  coefficients $|C_{i^,i}|$ (Eq.~(\ref{eq_6})) in unpolarized pp
  scattering at $E_{lab}=10$~MeV with the pp Coulomb force switched off.
  }
\label{fig1a}
\end{figure}

\begin{figure}
\includegraphics[angle=270,scale=0.5]{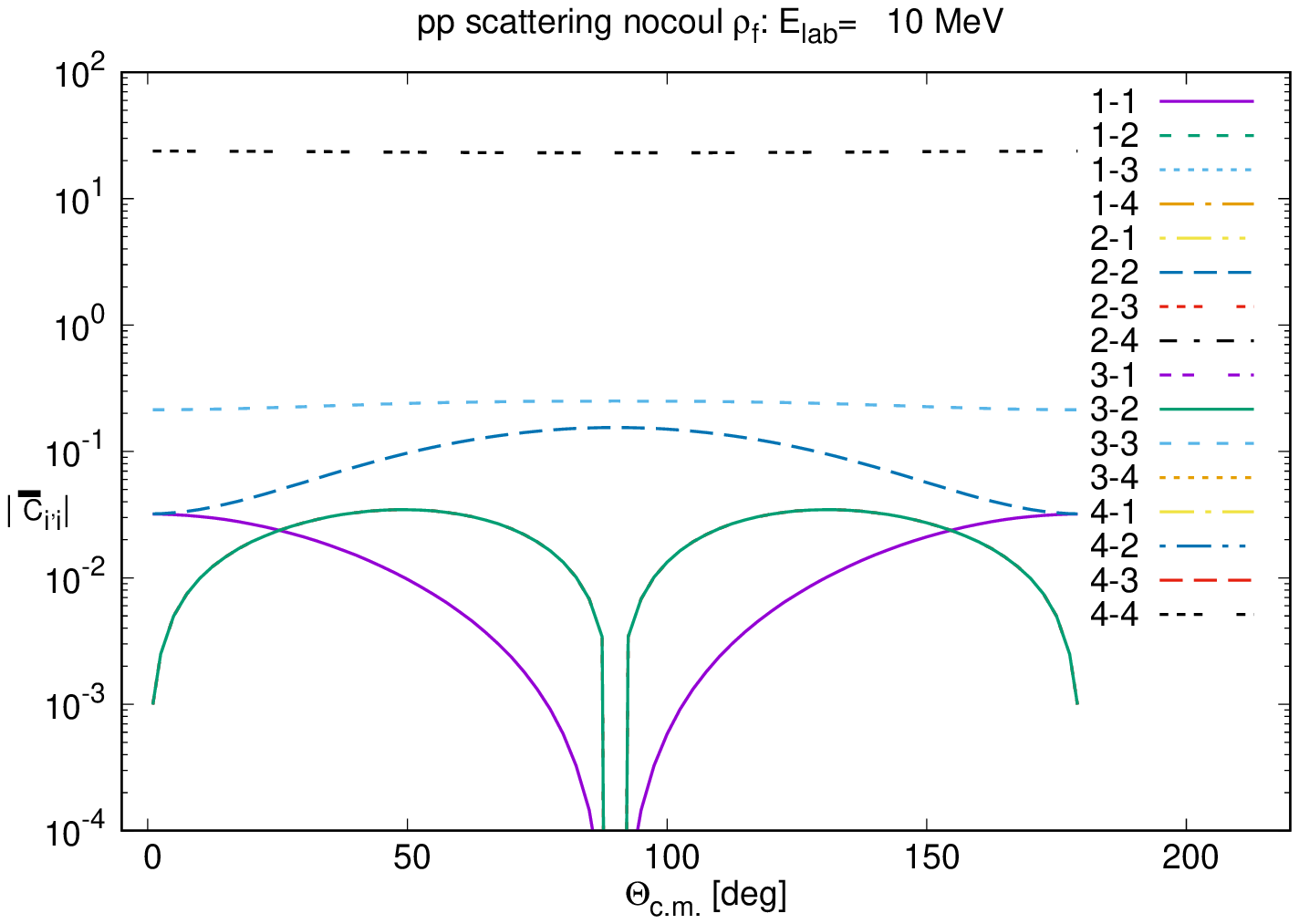}  
\caption{
  (color online) Angular distribution of the absolute values of the final spin 
  density matrix expansion
  coefficients $|\bar C_{i^,i}|$ (Eq.~(\ref{eq_9})) in unpolarized pp
  scattering at $E_{lab}=10$~MeV with the pp Coulomb force switched off.
  }
\label{fig2a}
\end{figure}

\begin{figure}
\includegraphics[angle=270,scale=0.5]{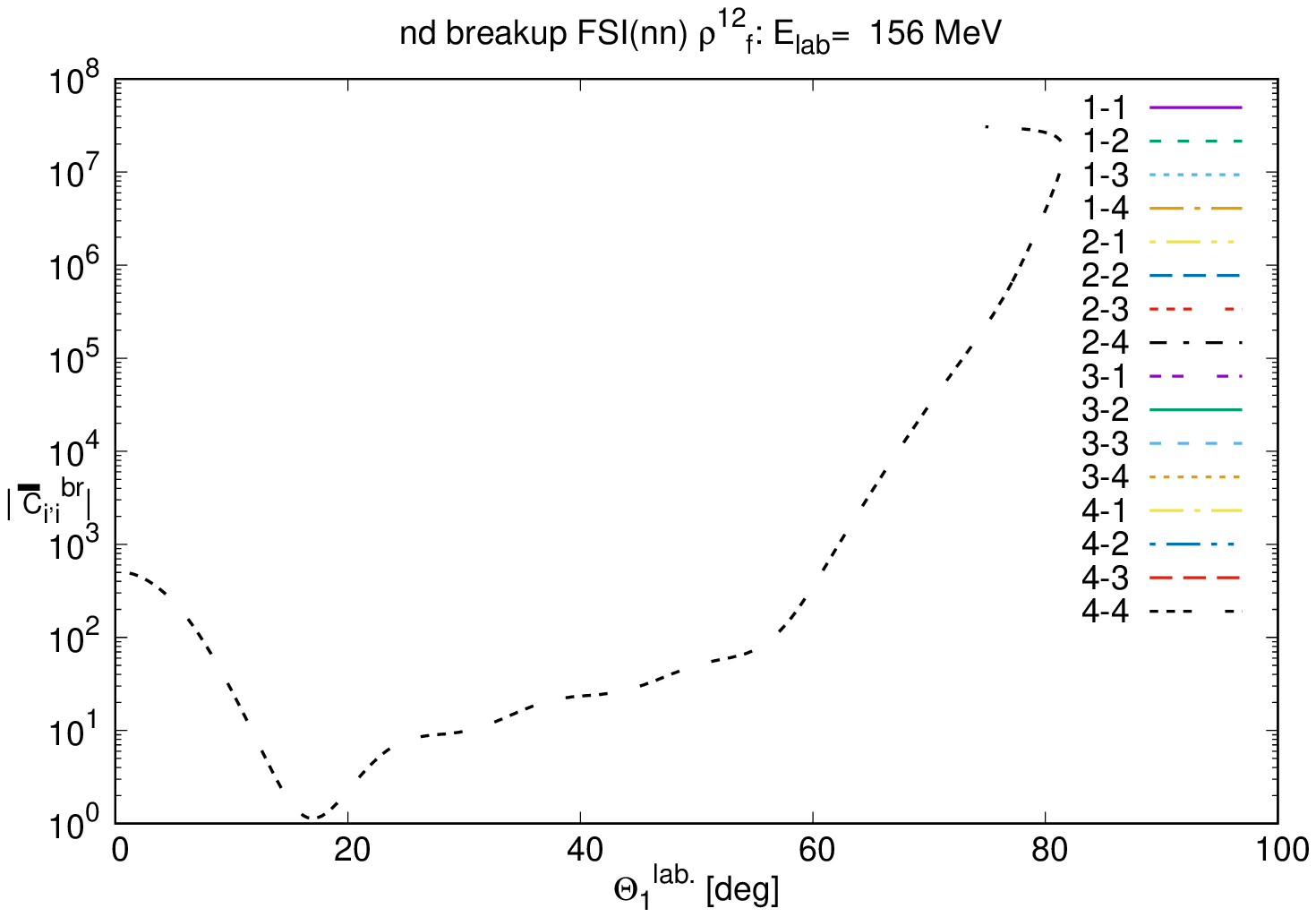}  
\caption{
  (color online) Absolute values of the expansion
  coefficients $|\bar C_{i^,i}^{br}|$ (Eq.~(\ref{eq_22})) for the final spin 
  density matrix $\rho_f^{12}$ in unpolarized nd
  breakup $d(n,nn)p$ at $E_{lab}=156$~MeV for a kinematically complete
  configuration 
  under the exact FSI(nn) condition. Shown as a function of the laboratory angle
  of the final-state interacting neutron, $\Theta_1^{lab}$. 
  }
\label{fig9}
\end{figure}

\begin{figure}
\includegraphics[angle=270,scale=0.5]{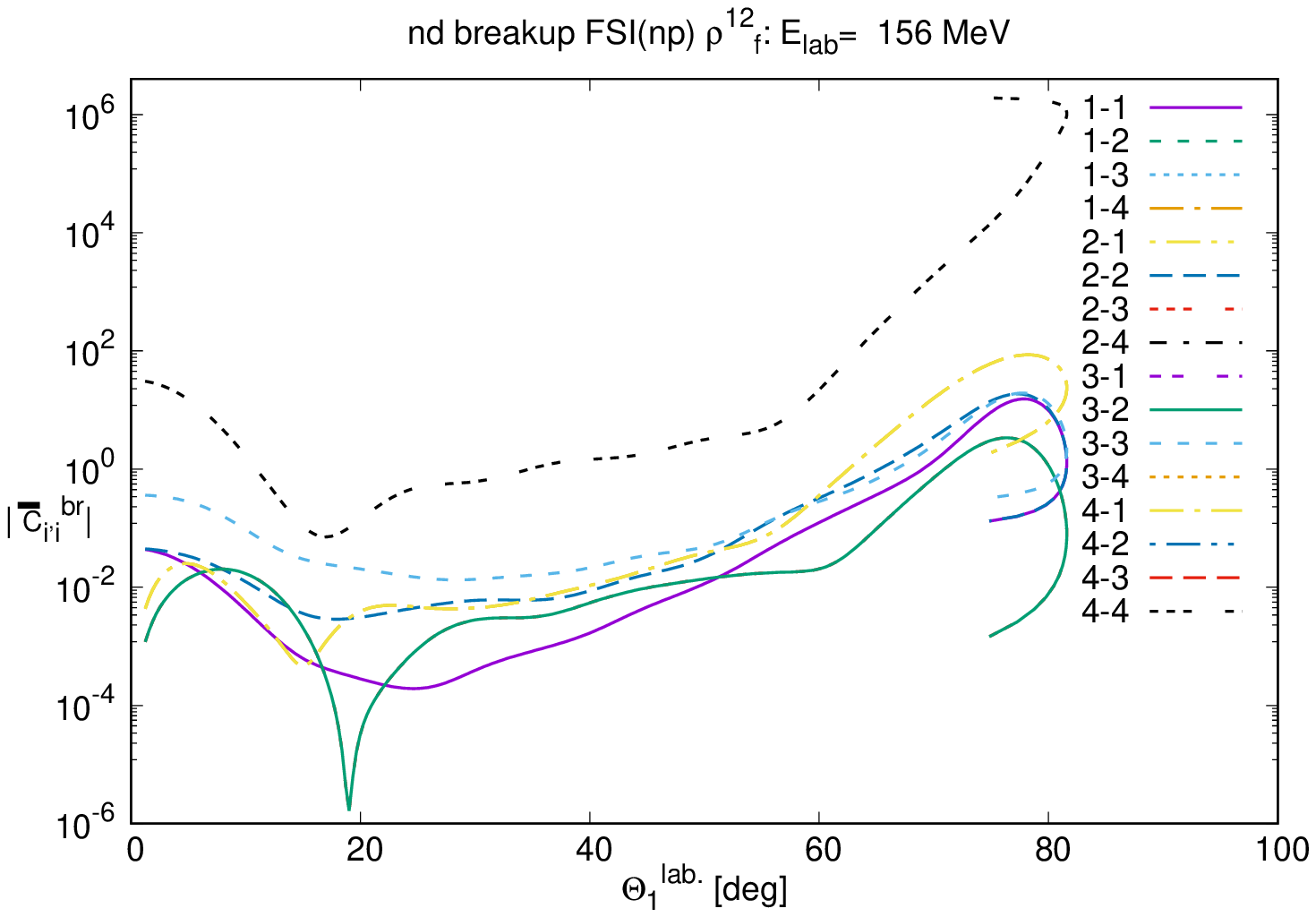}  
\caption{
  (color online) Same as in Fig.~\ref{fig9},  but for FSI(np)
  in the unpolarized nd  breakup $d(n,np)n$ reaction.
  }
\label{fig11}
\end{figure}

\begin{figure}
\includegraphics[angle=270,scale=0.5]{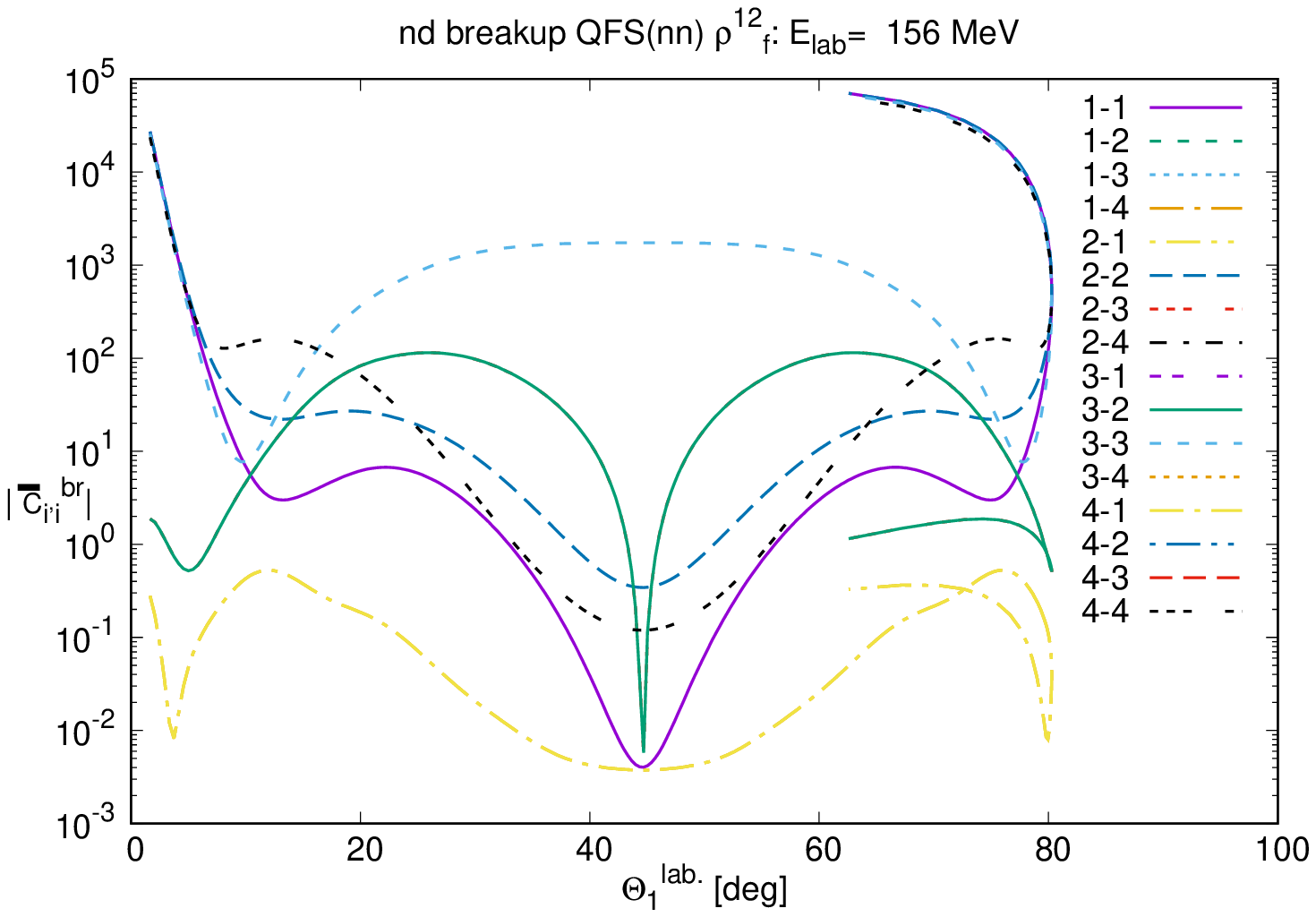}  
\caption{
  (color online) Same as in Fig.~\ref{fig9},  but for QFS(nn)
  in the unpolarized nd  breakup $d(n,nn)p$ reaction.
  }
\label{fig13}
\end{figure}

\begin{figure}
\includegraphics[angle=270,scale=0.5]{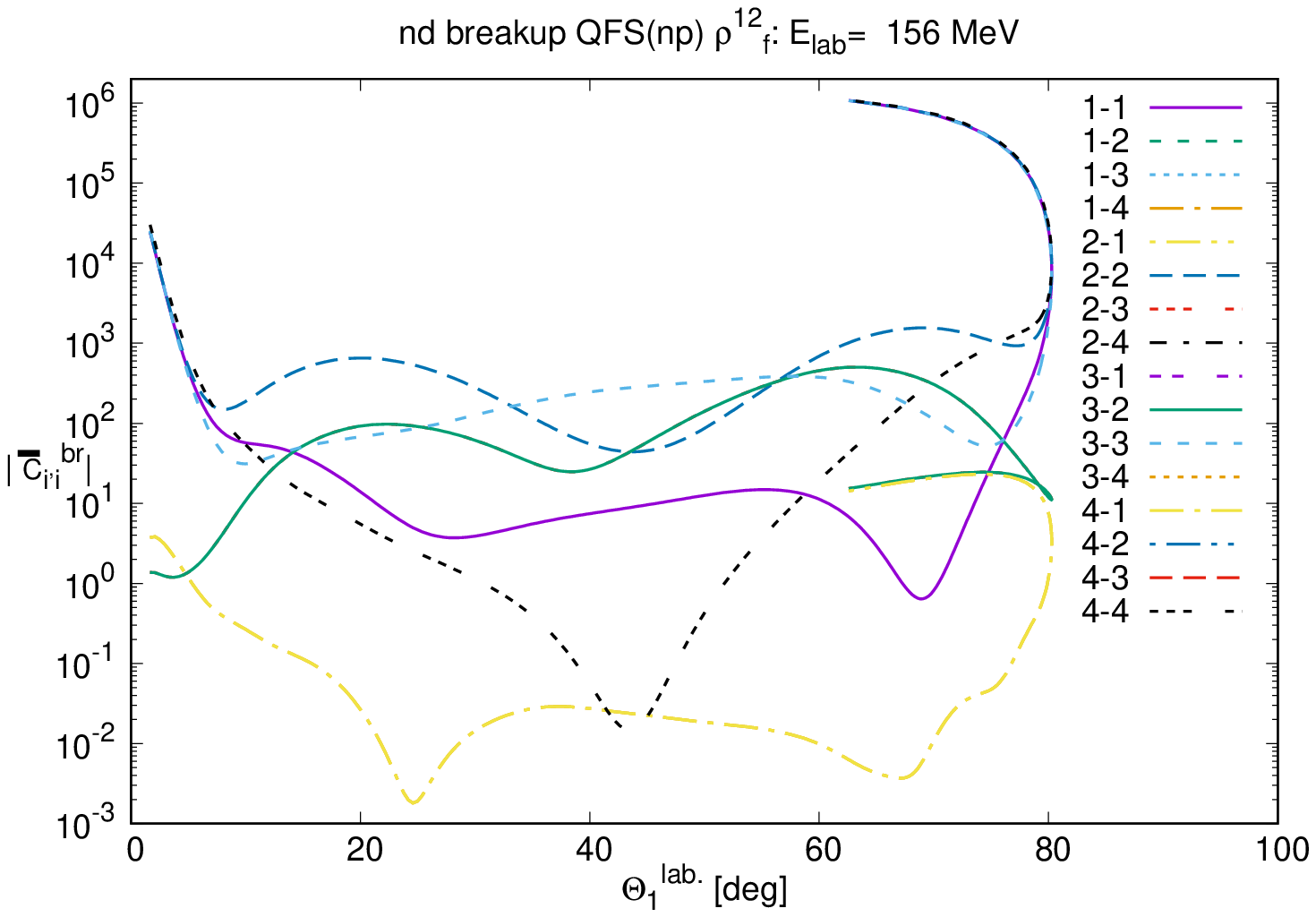}  
\caption{
  (color online) Same as in Fig.~\ref{fig9},  but for QFS(np)
  in the unpolarized nd  breakup $d(n,np)n$ reaction.
  }
\label{fig15}
\end{figure}

\begin{figure}
\includegraphics[angle=270,scale=0.5]{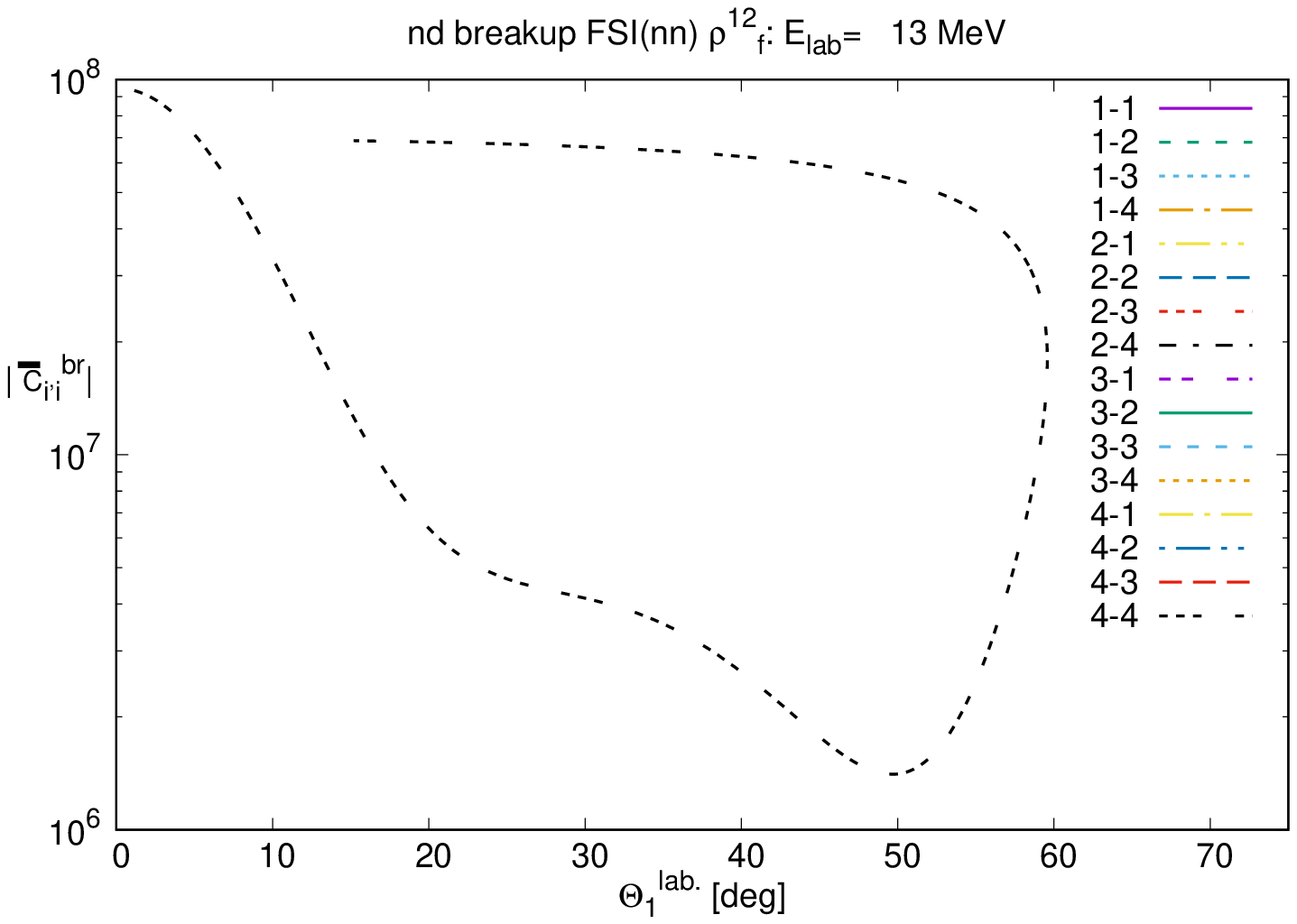}  
\caption{
  (color online) Absolute values of the expansion
  coefficients $|\bar C_{i^,i}^{br}|$ (Eq.~(\ref{eq_22})) for the final spin 
  density matrix $\rho_f^{12}$ in unpolarized nd
  breakup $d(n,nn)p$ at $E_{lab}=13$~MeV for a kinematically complete
  configuration 
  under the exact FSI(nn) condition. Shown as a function of the laboratory angle
  of the final-state interacting neutron, $\Theta_1^{lab}$. 
  }
\label{fig17}
\end{figure}

\begin{figure}
\includegraphics[angle=270,scale=0.5]{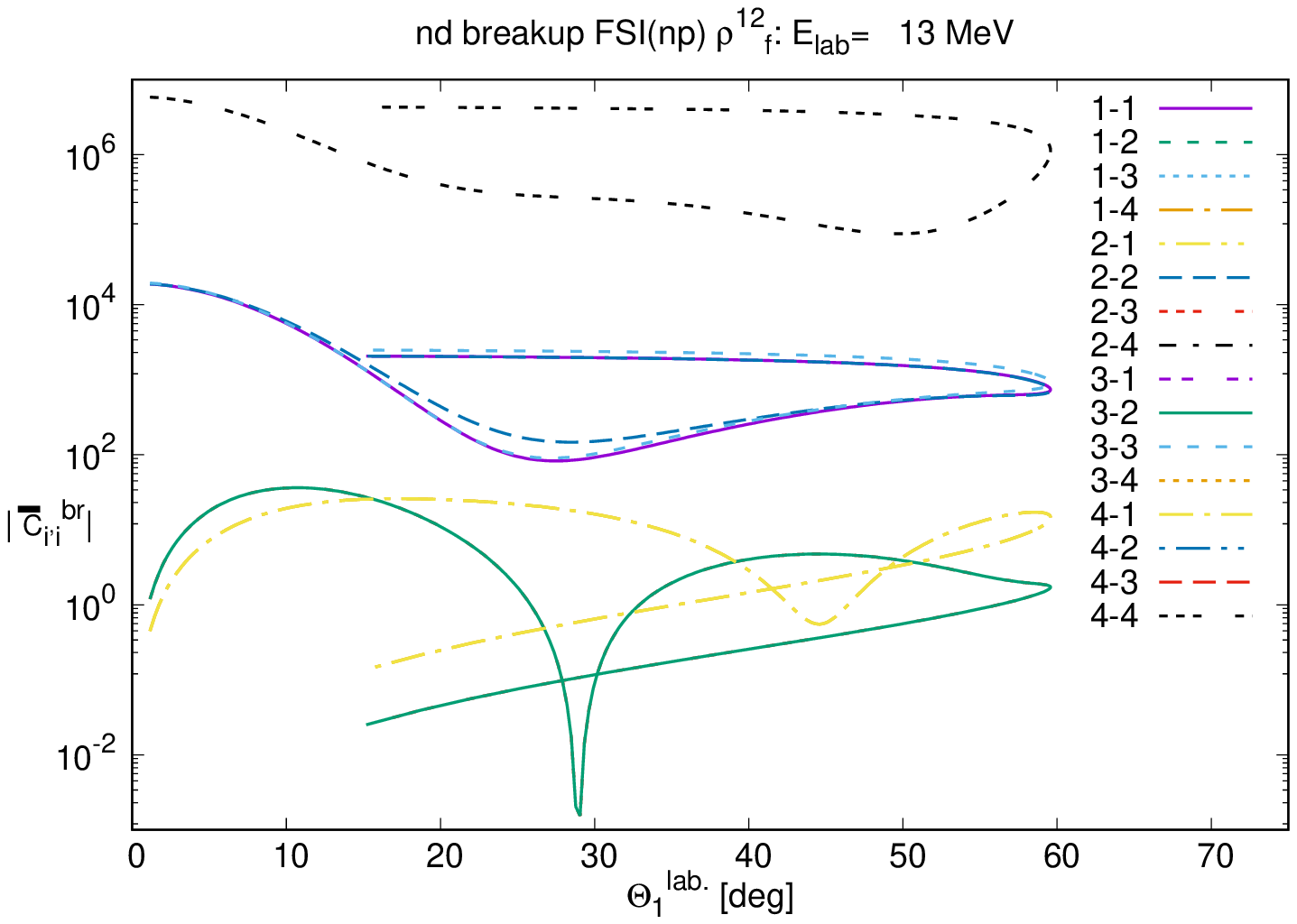}  
\caption{
  (color online) Same as in Fig.~\ref{fig17}  but for FSI(np)
  in the unpolarized nd  breakup $d(n,np)n$ reaction at $E_{lab}=13$~MeV.
  }
\label{fig19}
\end{figure}

\begin{figure}
\includegraphics[angle=270,scale=0.5]{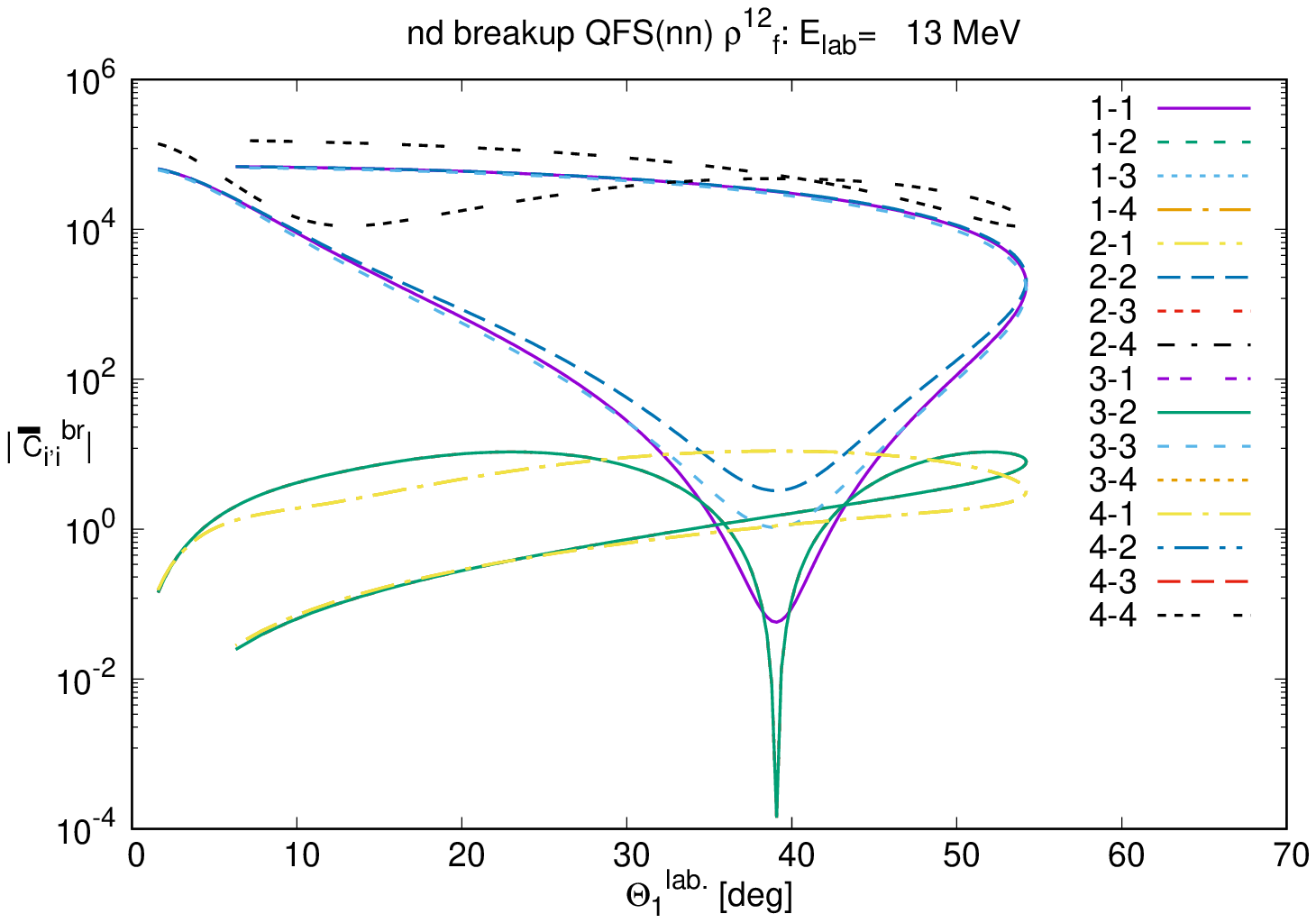}  
\caption{
  (color online) Same as in Fig.~\ref{fig17}  but for QFS(nn)
  in the unpolarized nd  breakup $d(n,nn)p$ reaction at $E_{lab}=13$~MeV.
  }
\label{fig21}
\end{figure}

\begin{figure}
\includegraphics[angle=270,scale=0.5]{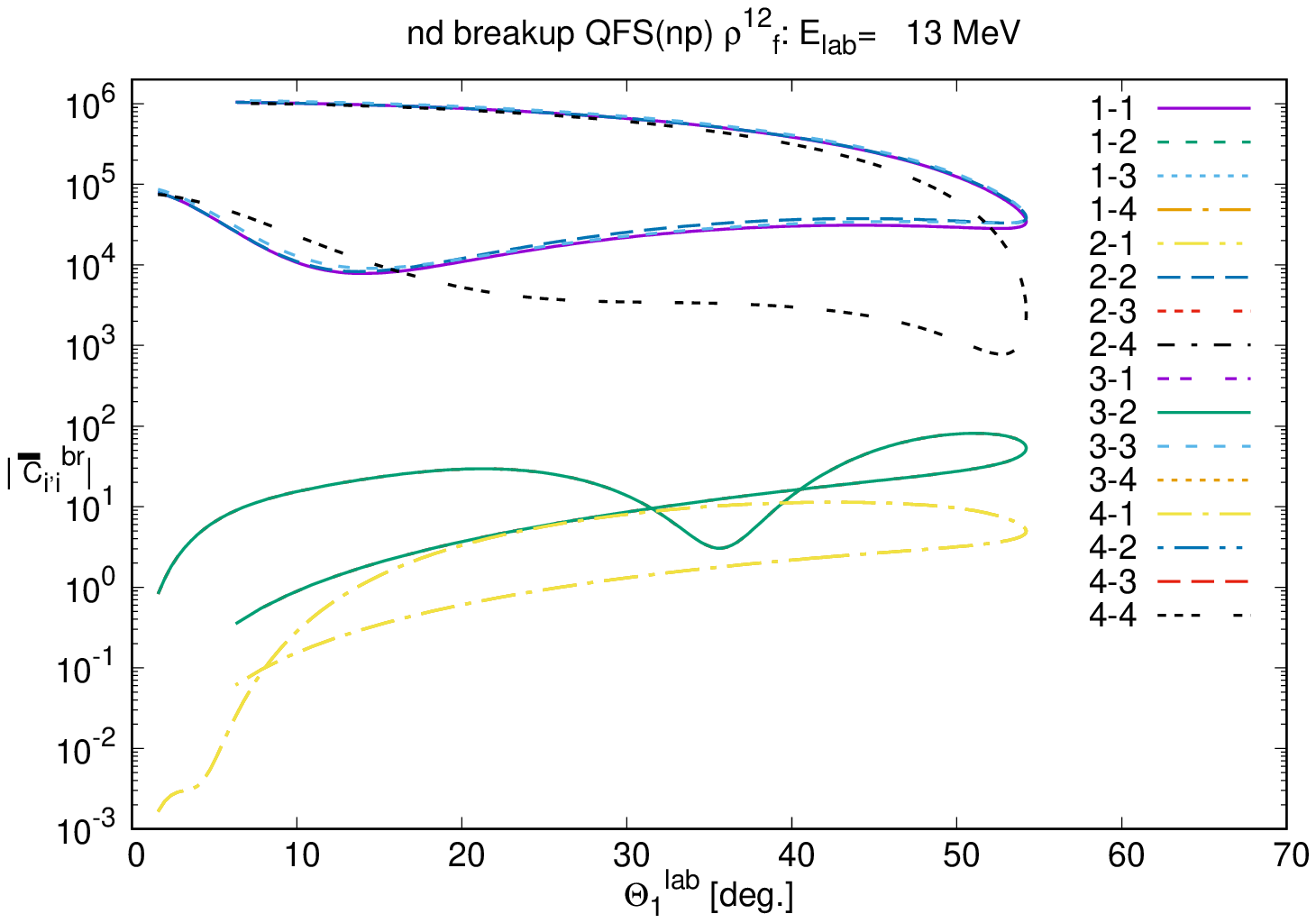}  
\caption{
  (color online) Same as in Fig.~\ref{fig17}  but for QFS(np)
  in the unpolarized nd  breakup $d(n,np)n$ reaction at $E_{lab}=13$~MeV.
  }
\label{fig23}
\end{figure}

\begin{figure}
\includegraphics[angle=270,scale=0.5]{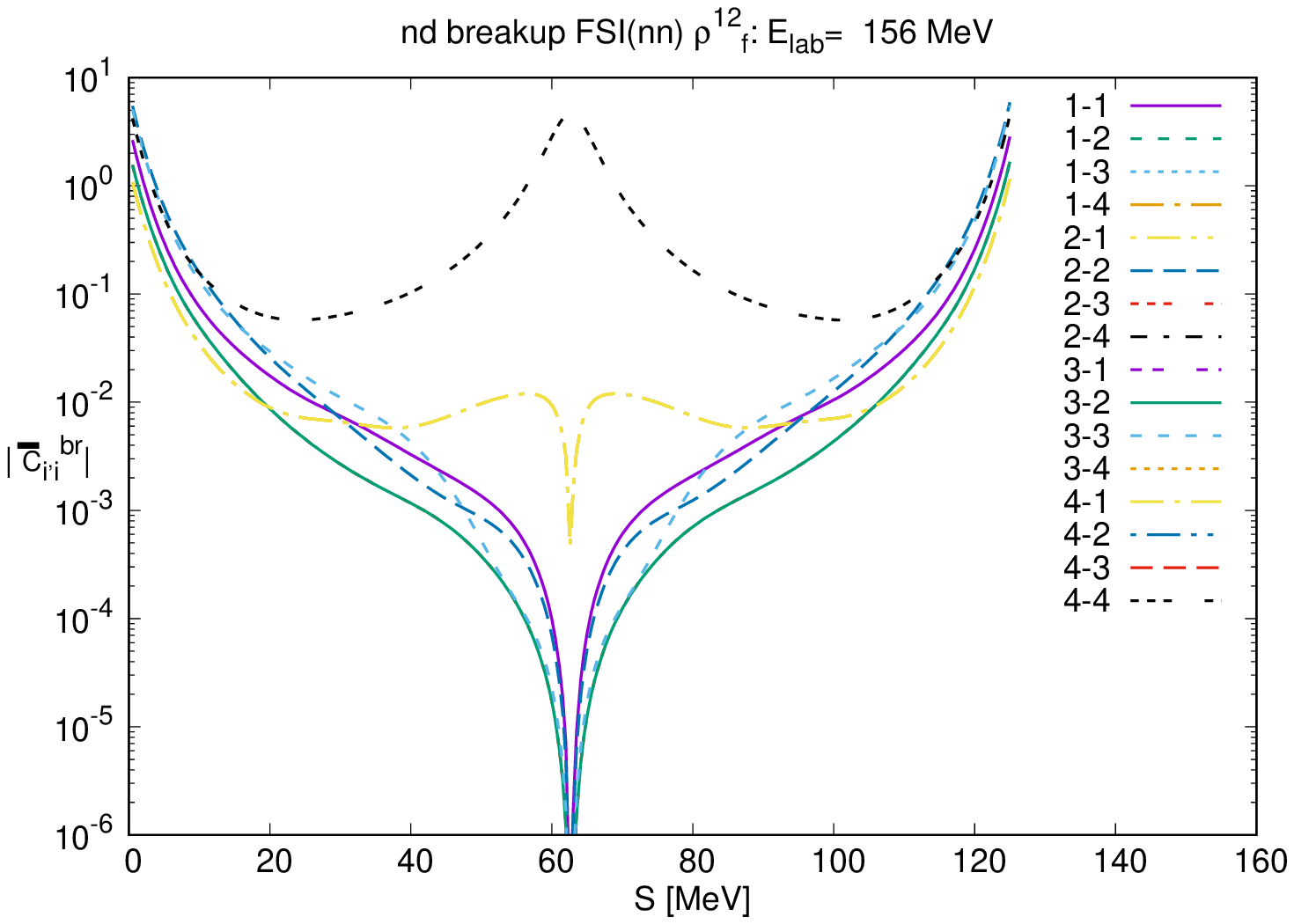}  
\caption{
  (color online) Absolute values of the expansion
  coefficients $|\bar C_{i^,i}^{br}|$ (Eq.~(\ref{eq_22})) for the final spin 
  density matrix $\rho_f^{12}$ in unpolarized nd
  breakup $d(n,nn)p$ at $E_{lab}=156$~MeV for a complete geometry with 
  $\Theta_1^{lab}=\Theta_2^{lab}=40.5^o$ and $\Phi_{12}=0^o$. 
  Shown as a function of the arc length along the S-curve.
  The exact FSI(nn) condition occurs at $S=62$~MeV.
  }
\label{fig25}
\end{figure}

\begin{figure}
\includegraphics[angle=270,scale=0.5]{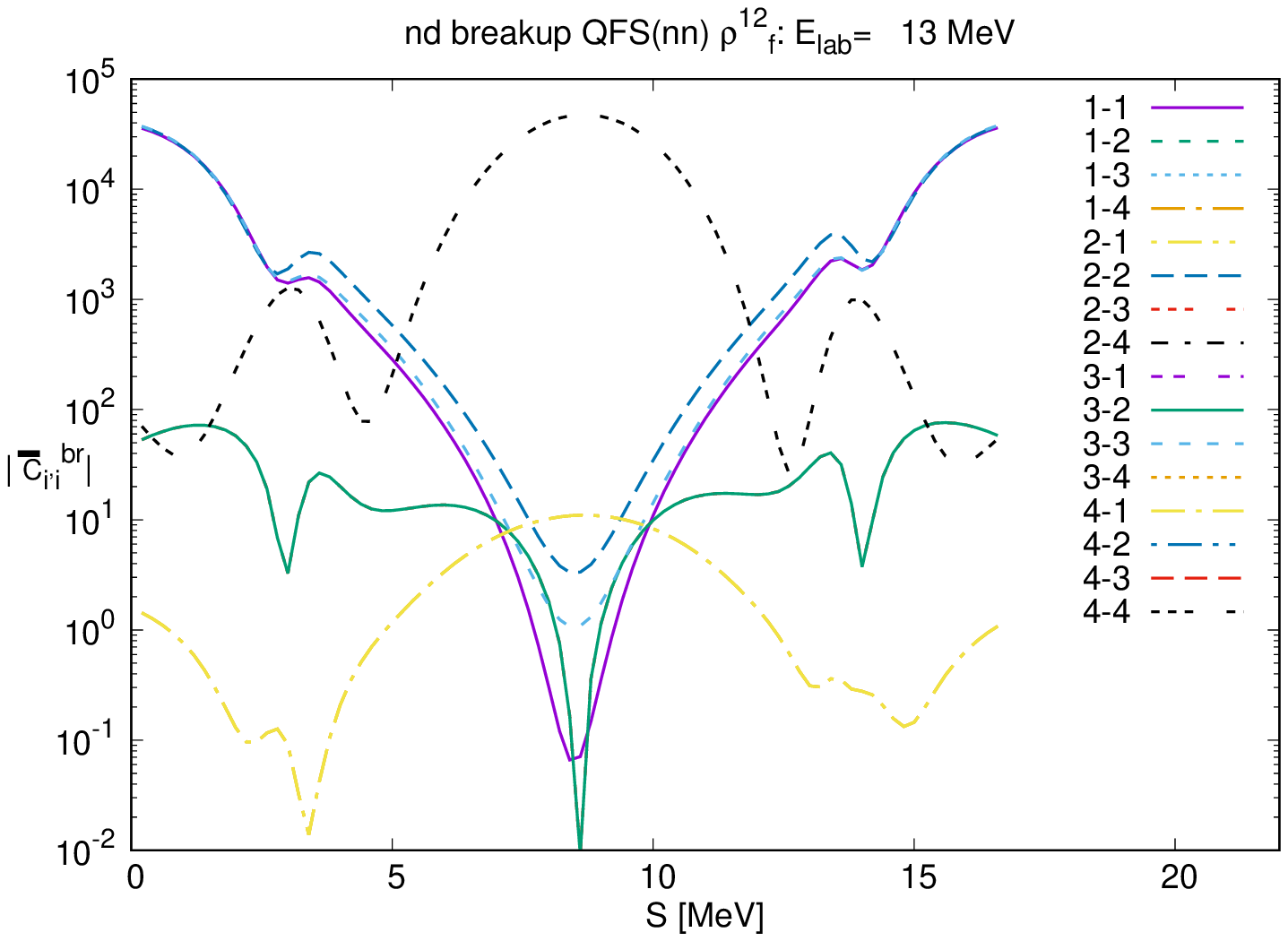}
\caption{
  (color online) Same as in Fig.~\ref{fig25},  but for 
  $E_{lab}=13$~MeV and a complete geometry
  with $\Theta_1^{lab}=40.23^o$, $\Theta_2^{lab}=37.85^o$ and $\Phi_{12}=180^o$.
  The exact QFS(nn) condition occurs at $S=8.5$~MeV. 
  }
\label{fig31}
\end{figure}

\begin{figure}
\includegraphics[scale=0.8]{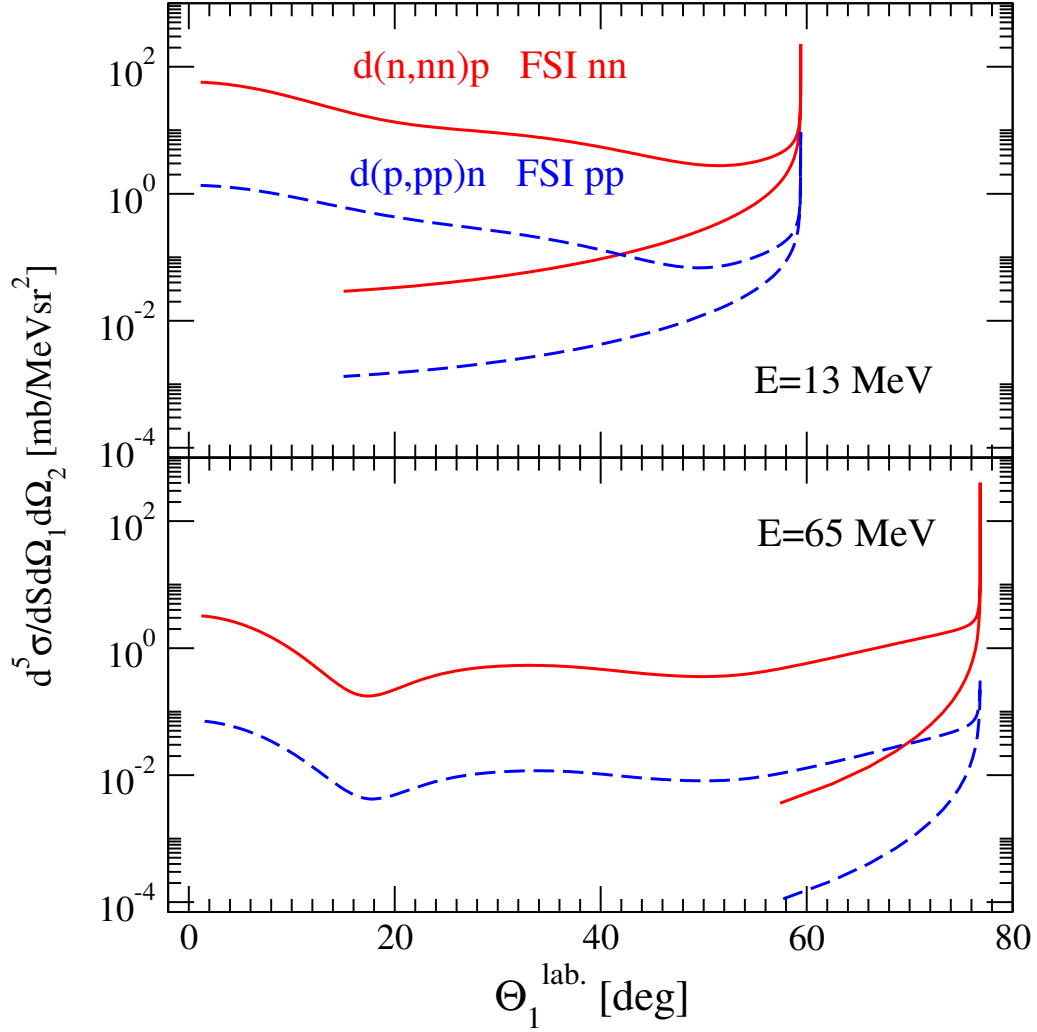}  
\caption{
  (color online) The cross section $\frac {d^5\sigma} {dSd\Omega_1 d\Omega_2}$
    in exclusive breakup reactions $d(n,nn)p$ (red line) and
    $d(p,pp)n$ (blue dashed line) at $E_{lab}=13$~MeV and $65$~MeV, 
    under the exact FSI(12) condition,
    shown as a function of the laboratory angle
    $\Theta_1^{lab}$.
    Calculations were performed using the AV18 NN potential, with and
    without the inclusion of the Coulomb force \cite{wit_coul}.
  }
\label{fig33}
\end{figure}

\begin{figure}
\includegraphics[scale=0.8]{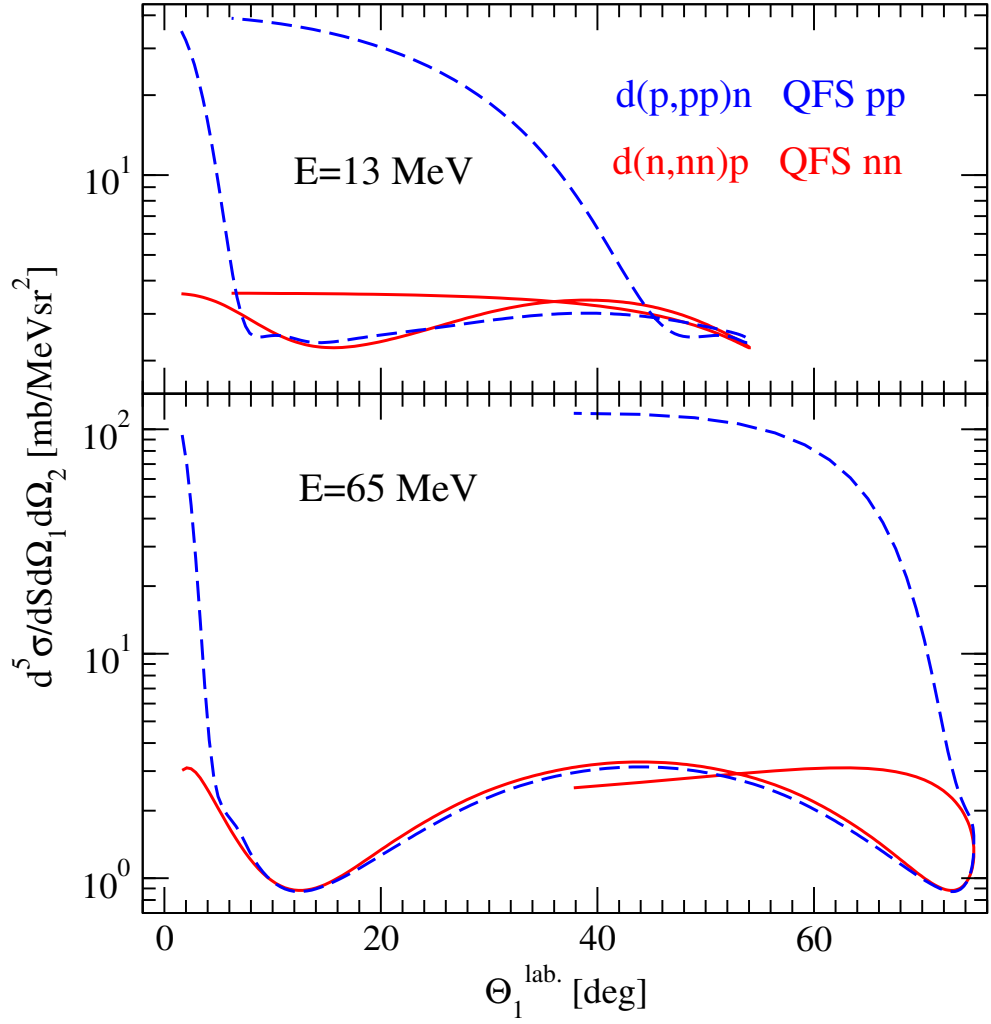}  
\caption{
  (color online) Same as in Fig.~\ref{fig33}, but for the QFS(12) condition.
  }
\label{fig34}
\end{figure}

\begin{figure}
\includegraphics[scale=0.8]{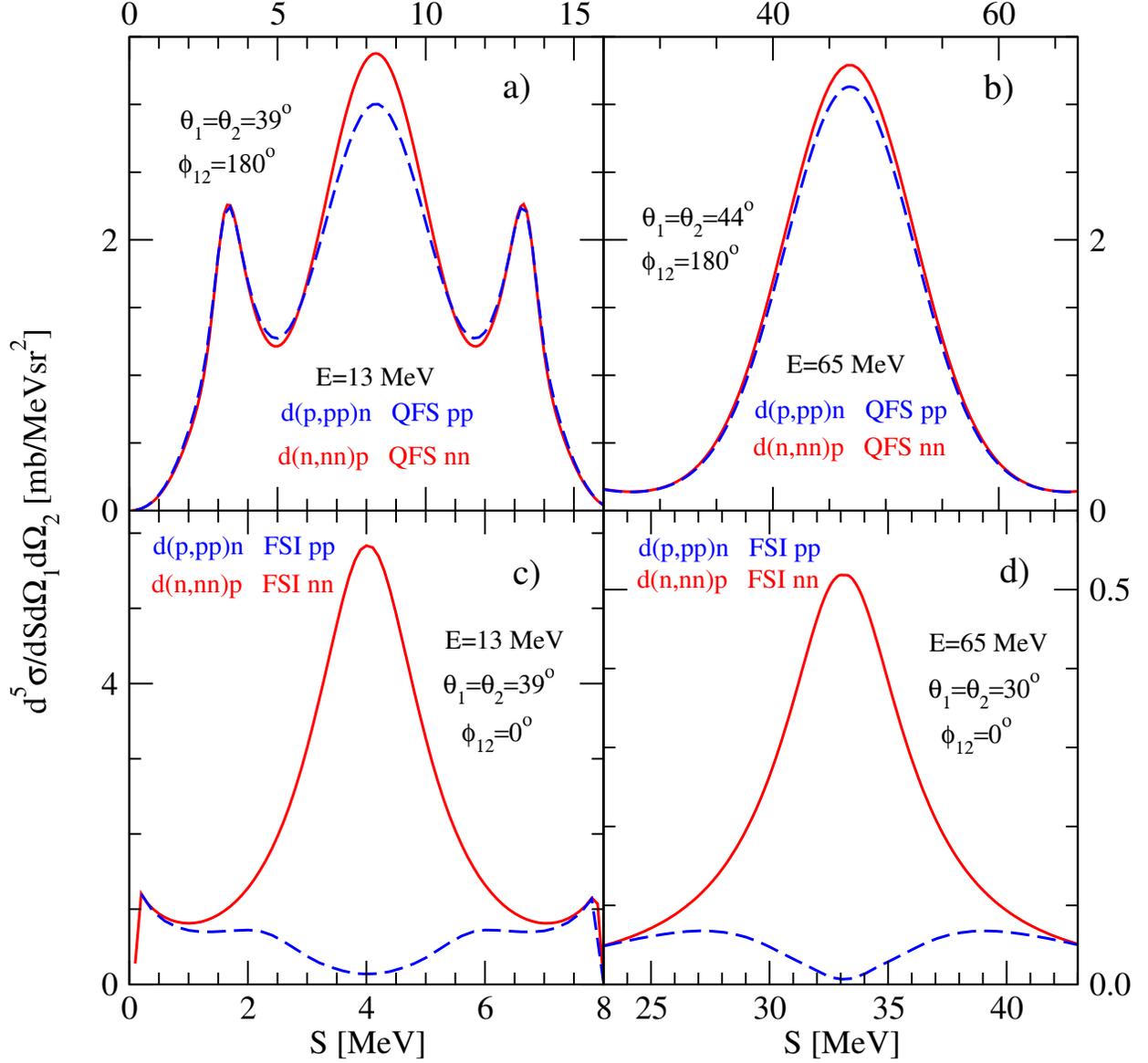}  
\caption{
  (color online) The cross section $\frac {d^5\sigma} {dSd\Omega_1 d\Omega_2}$
    in exclusive breakup reactions $d(n,nn)p$ (red line) and
    $d(p,pp)n$ (blue dashed line) at the QFS(12) condition [(a) $E_{lab}=13$~MeV,
      (b) $E_{lab}=65$~MeV], and the FSI(12) condition [(c) $E_{lab}=13$~MeV,
      (d) $E_{lab}=65$~MeV],
    shown as a function of the arc-length S.
    Calculations were performed using the AV18 NN potential, with and
    without the inclusion of the Coulomb force \cite{wit_coul}.
  }
\label{fig35}
\end{figure}

\begin{figure}
\begin{center}
\begin{tabular}{c}
\resizebox{125mm}{!}{\includegraphics[angle=270]{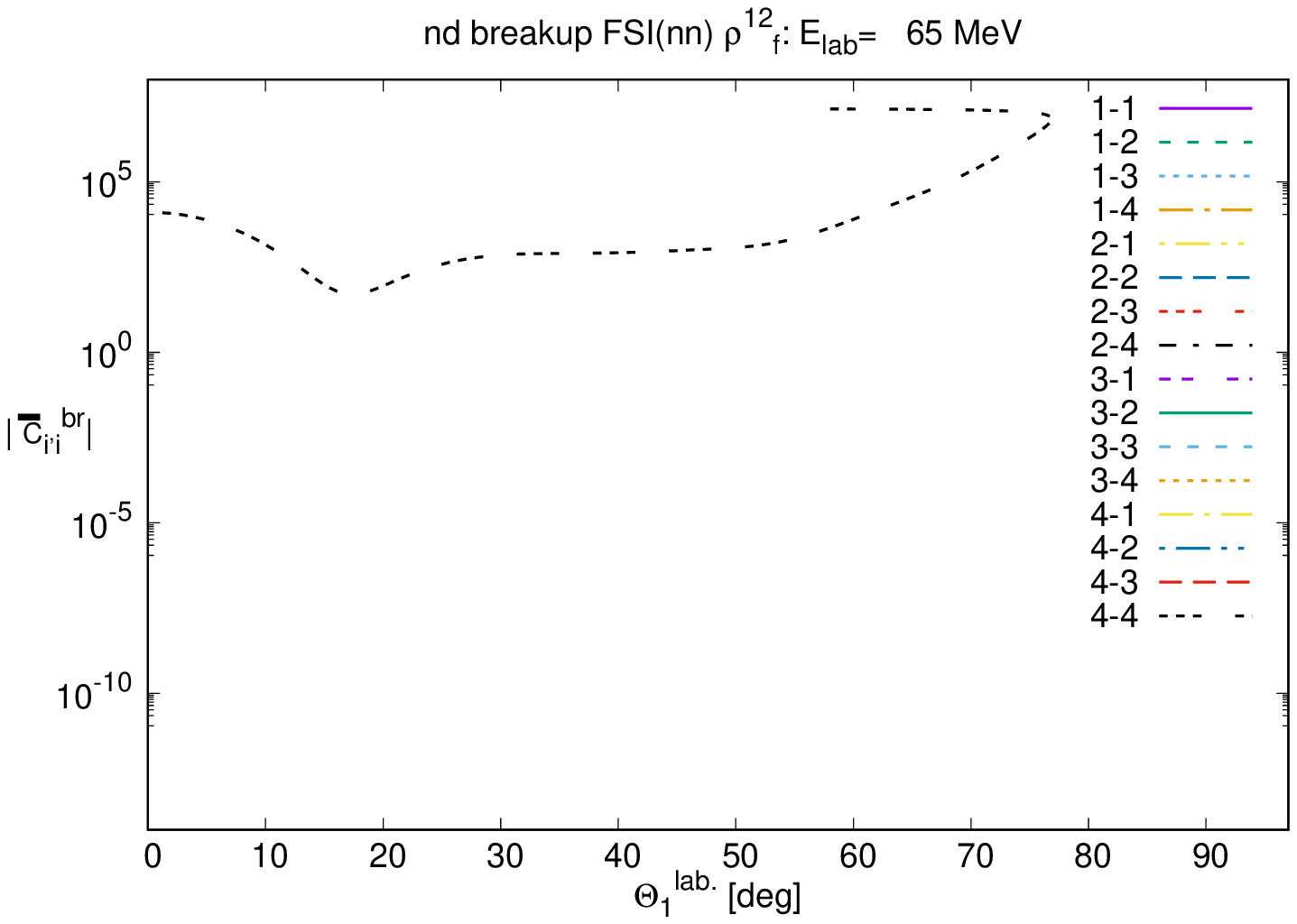}} \\
\resizebox{125mm}{!}{\includegraphics[angle=270]{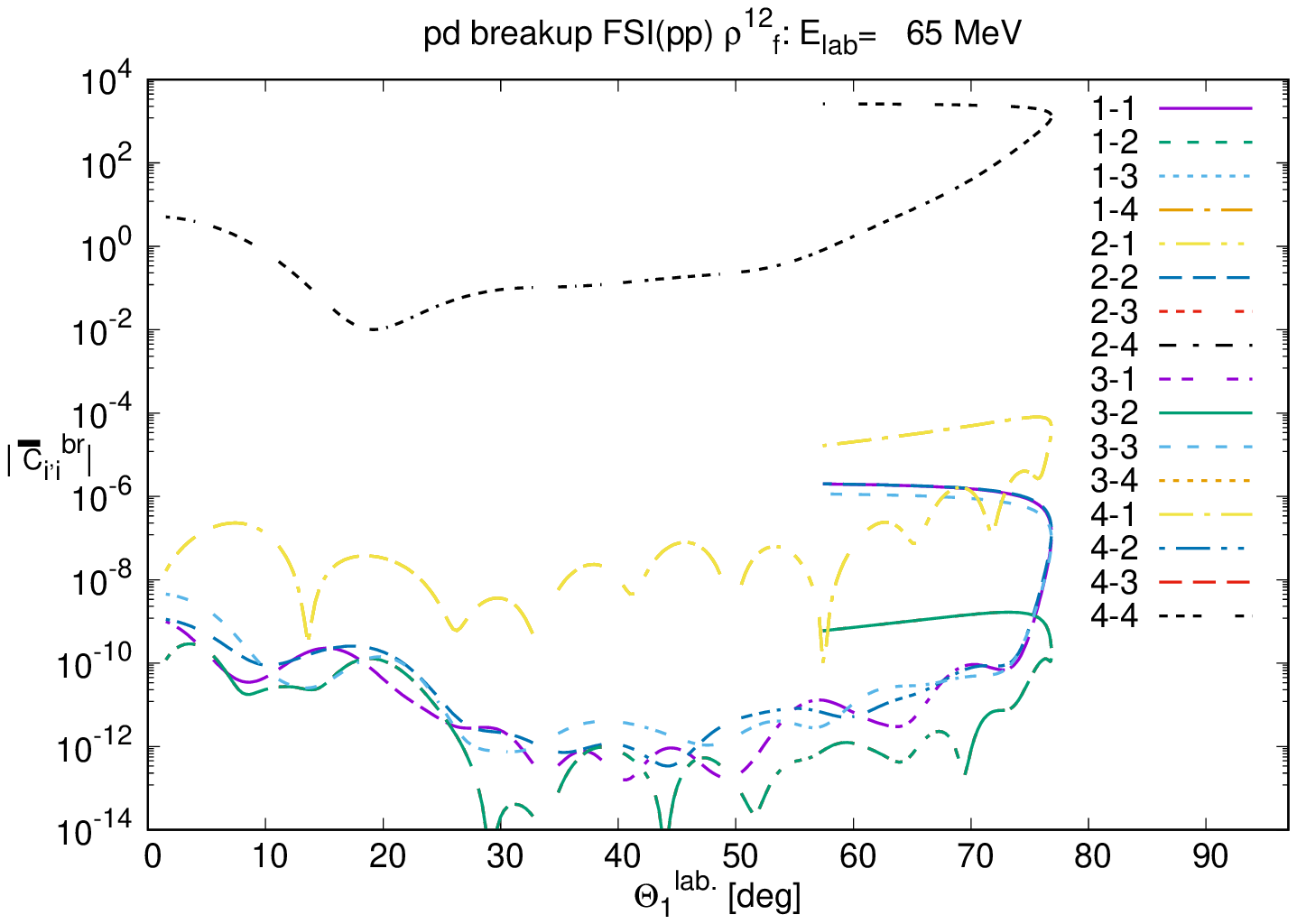}} \\
\end{tabular}%
\caption{
(color online) Absolute values of the expansion
  coefficients $|\bar C_{i^,i}^{br}|$ (Eq.~(\ref{eq_22})) for the final spin 
  density matrix $\rho_f^{12}$ in unpolarized nd and pd 
  breakup  $d(N,N_1N_2)N_3$ at $E_{lab}=65$~MeV for a kinematically
  complete geometry
  under the exact FSI($N_1N_2$) condition. Shown as a function of the
  laboratory angle of the final-state interacting first nucleon
  $\Theta_1^{lab}$.
   Calculations were performed using the AV18 NN potential, with and
    without the inclusion of the Coulomb force \cite{wit_coul}.  
 }
\label{fig36}
\end{center}
\end{figure}

\begin{figure}
\begin{center}
\begin{tabular}{c}
\resizebox{122mm}{!}{\includegraphics[angle=270]{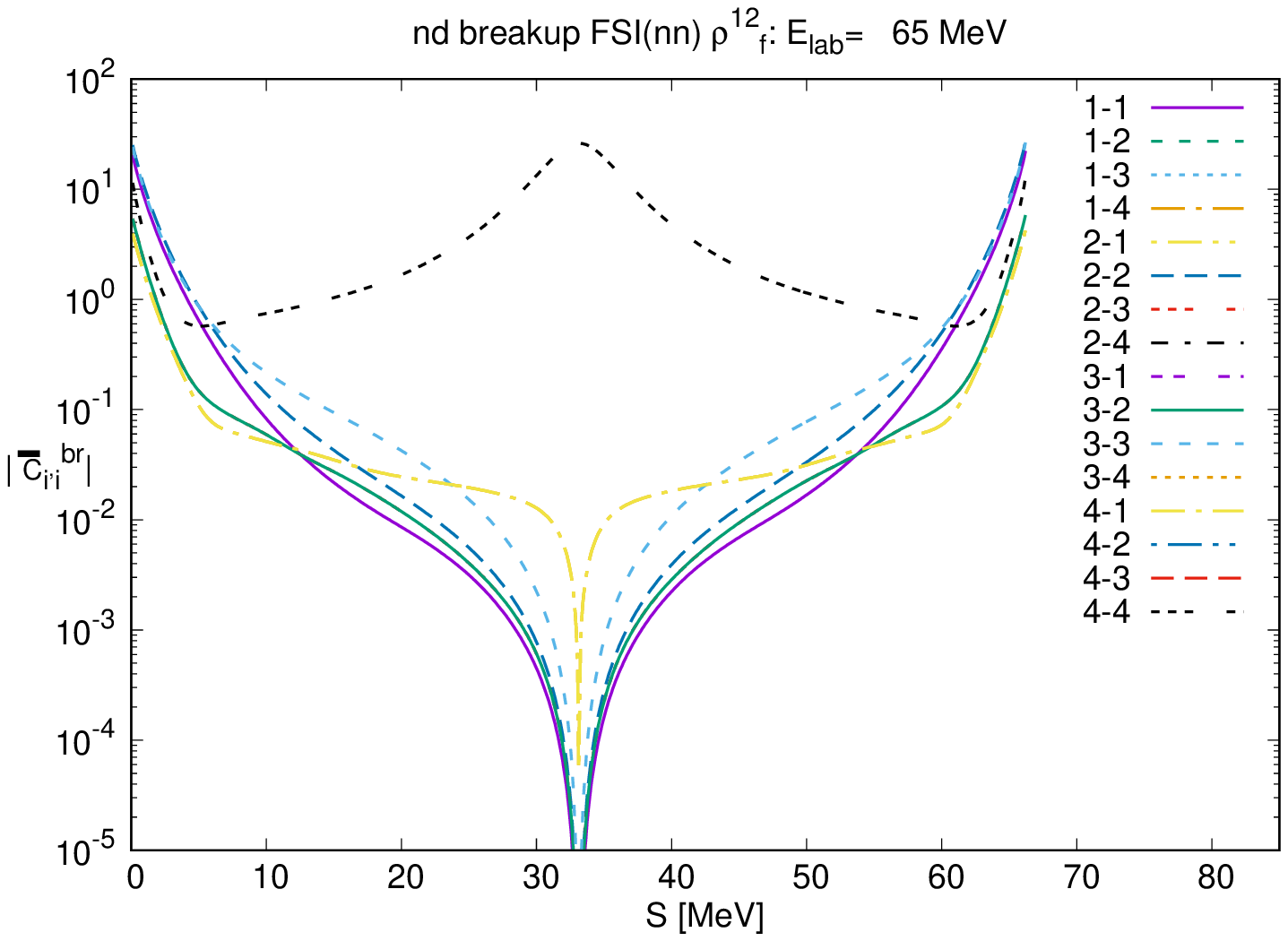}} \\
\resizebox{122mm}{!}{\includegraphics[angle=270]{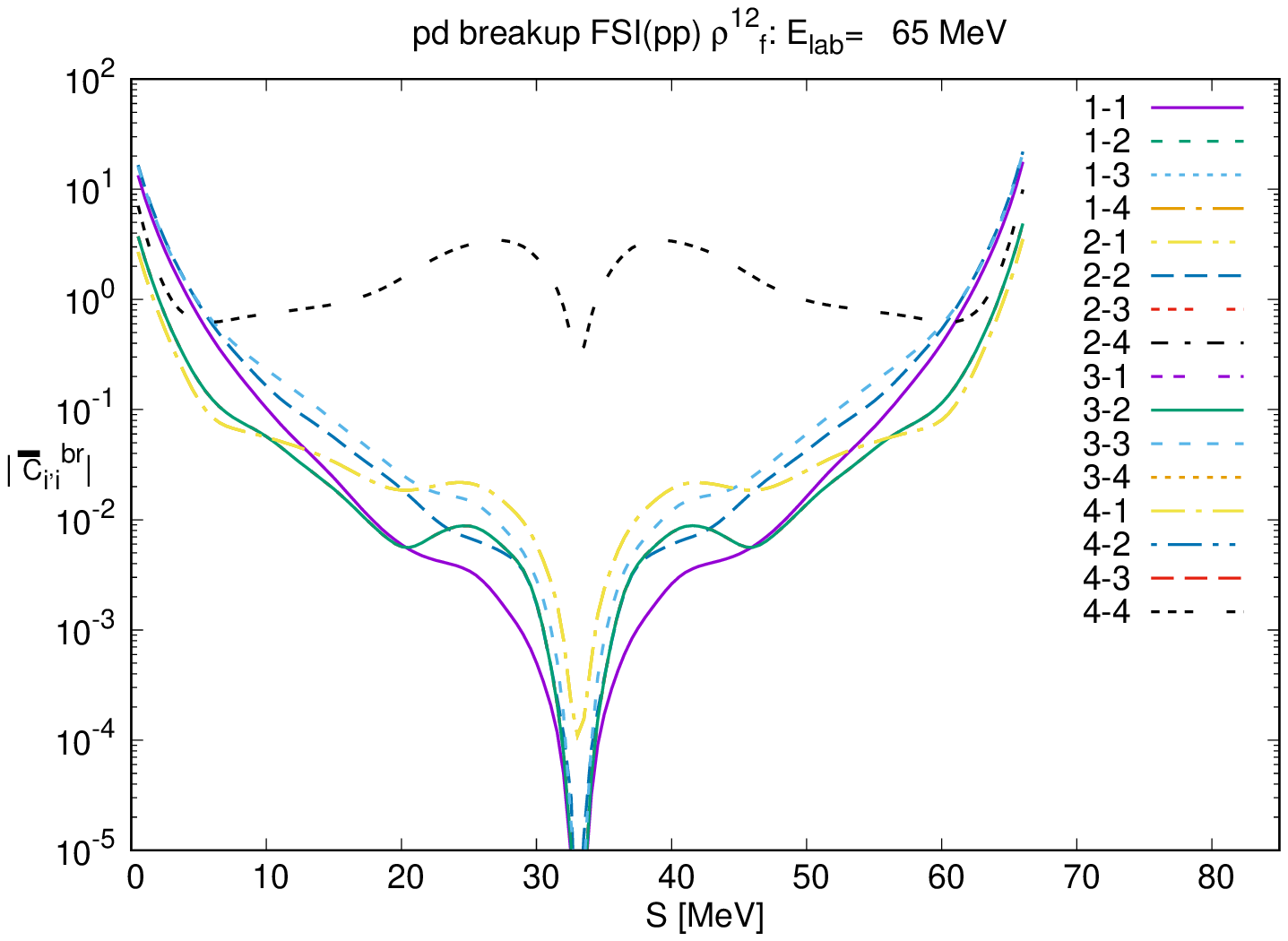}} \\
\end{tabular}%
\caption{
(color online) Absolute values of the expansion
  coefficients $|\bar C_{i^,i}^{br}|$ (Eq.~(\ref{eq_22})) for the final spin 
  density matrix $\rho_f^{12}$ in unpolarized nd and pd 
breakup  $d(N,N_1N_2)N_3$ at $E_{lab}=65$~MeV for a kinematically complete 
configuration with $\Theta_1^{lab}=\Theta_2^{lab}=30^o$ and $\Phi_{12}=0^o$. 
 Shown as a function of the
 arc length S, with the FSI($N_1N_2$) condition fulfilled at $S=37$~MeV.
Calculations were performed using the AV18 NN potential, with and
    without the inclusion of the Coulomb force \cite{wit_coul}. 
 }
\label{fig38}
\end{center}
\end{figure}

\begin{figure}
\begin{center}
\begin{tabular}{c}
\resizebox{125mm}{!}{\includegraphics[angle=270]{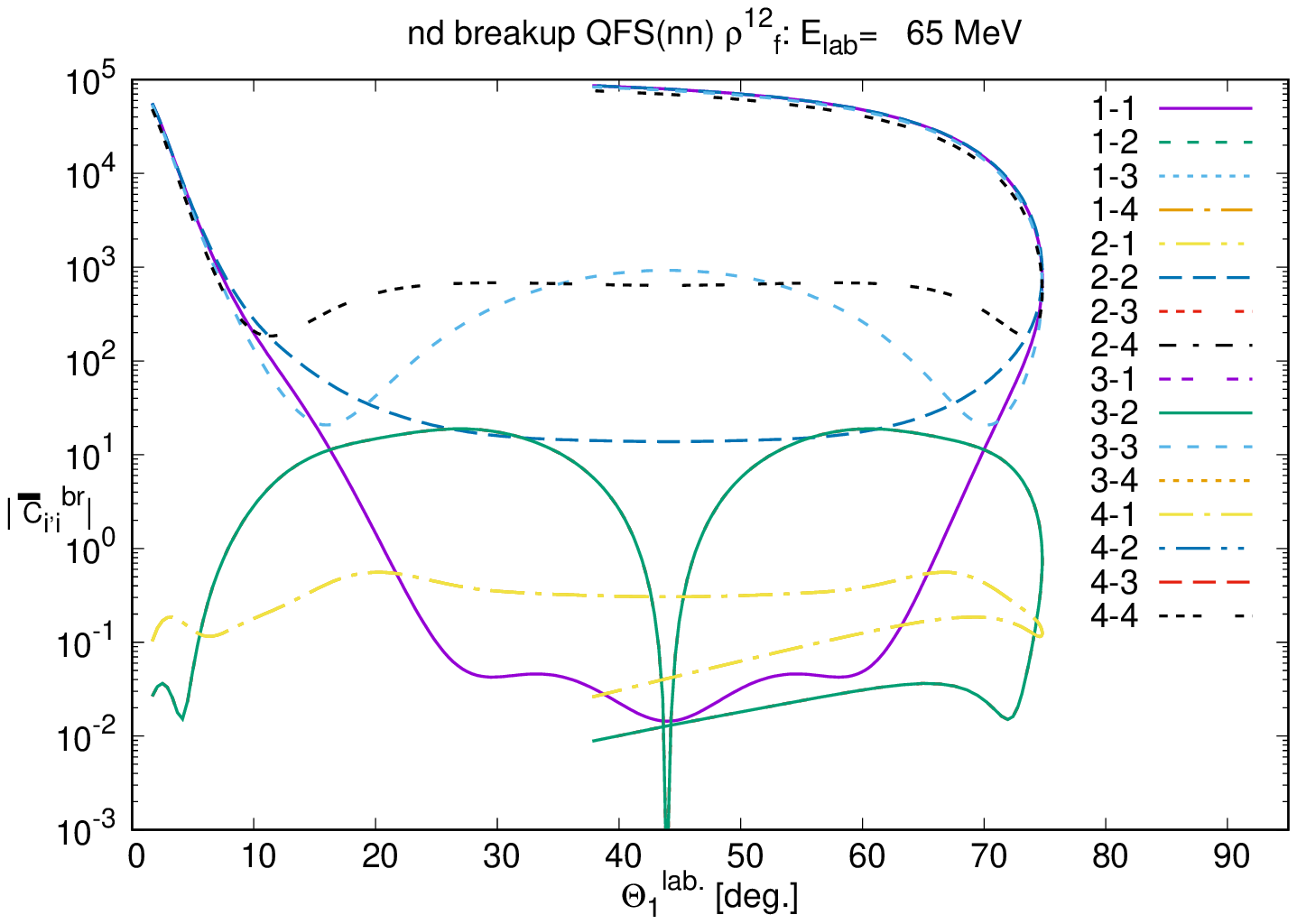}} \\
\resizebox{125mm}{!}{\includegraphics[angle=270]{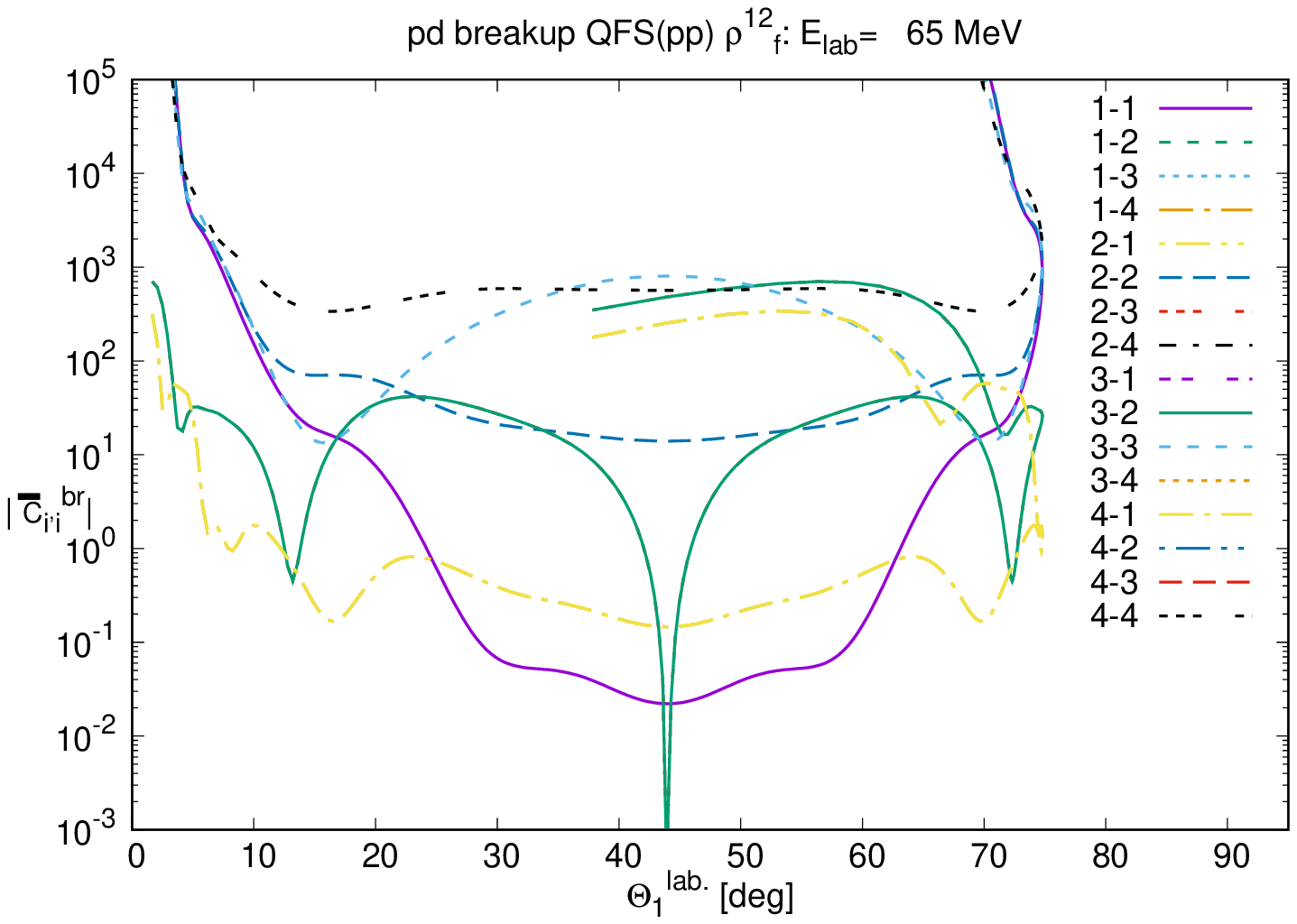}} \\
\end{tabular}%
\caption{
(color online) Absolute values of the expansion
  coefficients $|\bar C_{i^,i}^{br}|$ (Eq.~(\ref{eq_22})) for the final spin 
  density matrix $\rho_f^{12}$ in unpolarized nd and pd 
  breakup  $d(N,N_1N_2)N_3$ at $E_{lab}=65$~MeV, for a kinematically
  complete geometry
  under exact the QFS($N_1N_2$) condition. They are shown as a function of the
  laboratory angle of the QFS-interacting first nucleon
  $\Theta_1^{lab}$.
Calculations were performed using the AV18 NN potential, with and
    without the inclusion of the Coulomb force \cite{wit_coul}.  
 }
\label{fig40}
\end{center}
\end{figure}

\begin{figure}
\begin{center}
\begin{tabular}{c}
\resizebox{122mm}{!}{\includegraphics[angle=270]{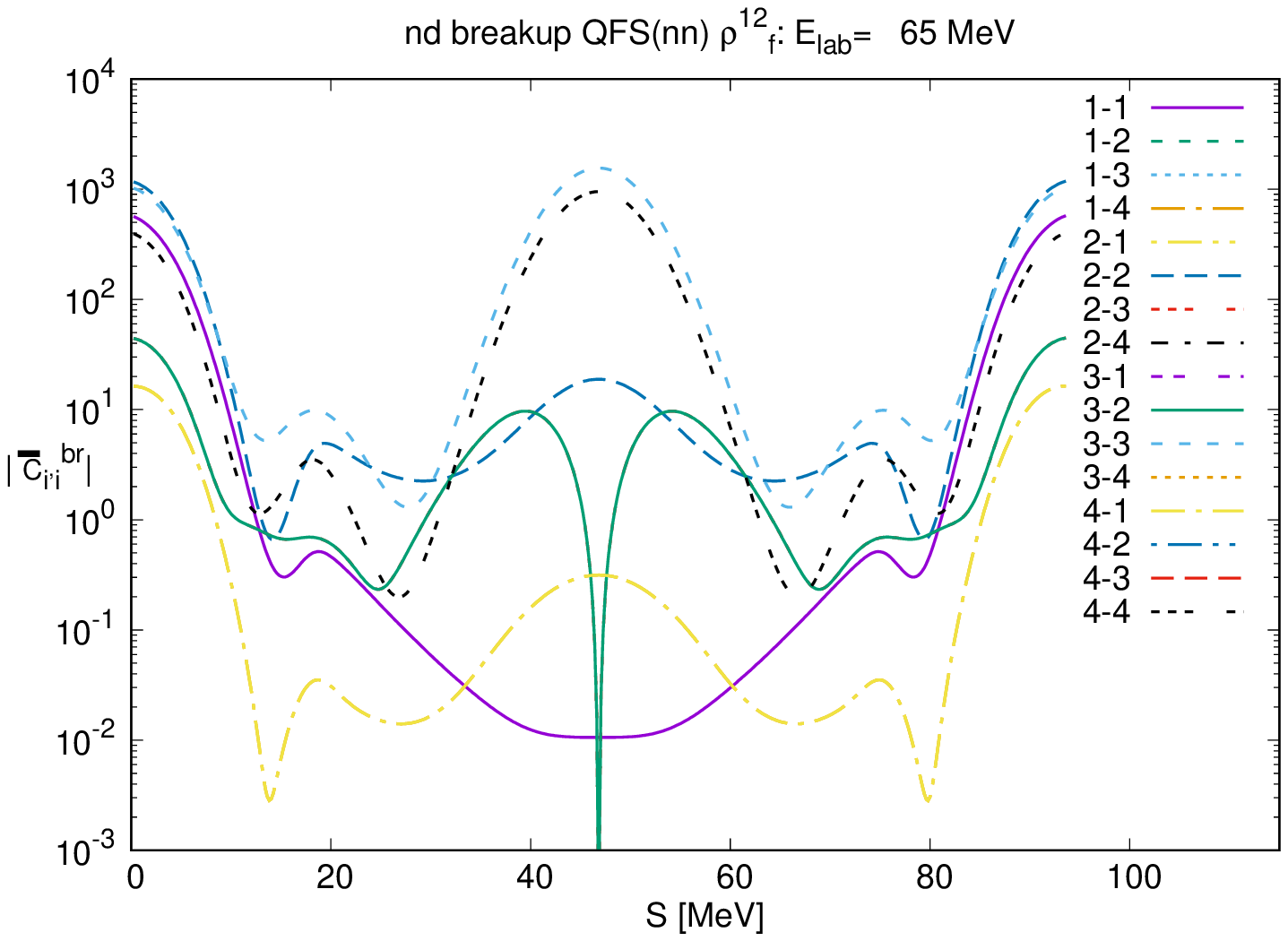}} \\
\resizebox{122mm}{!}{\includegraphics[angle=270]{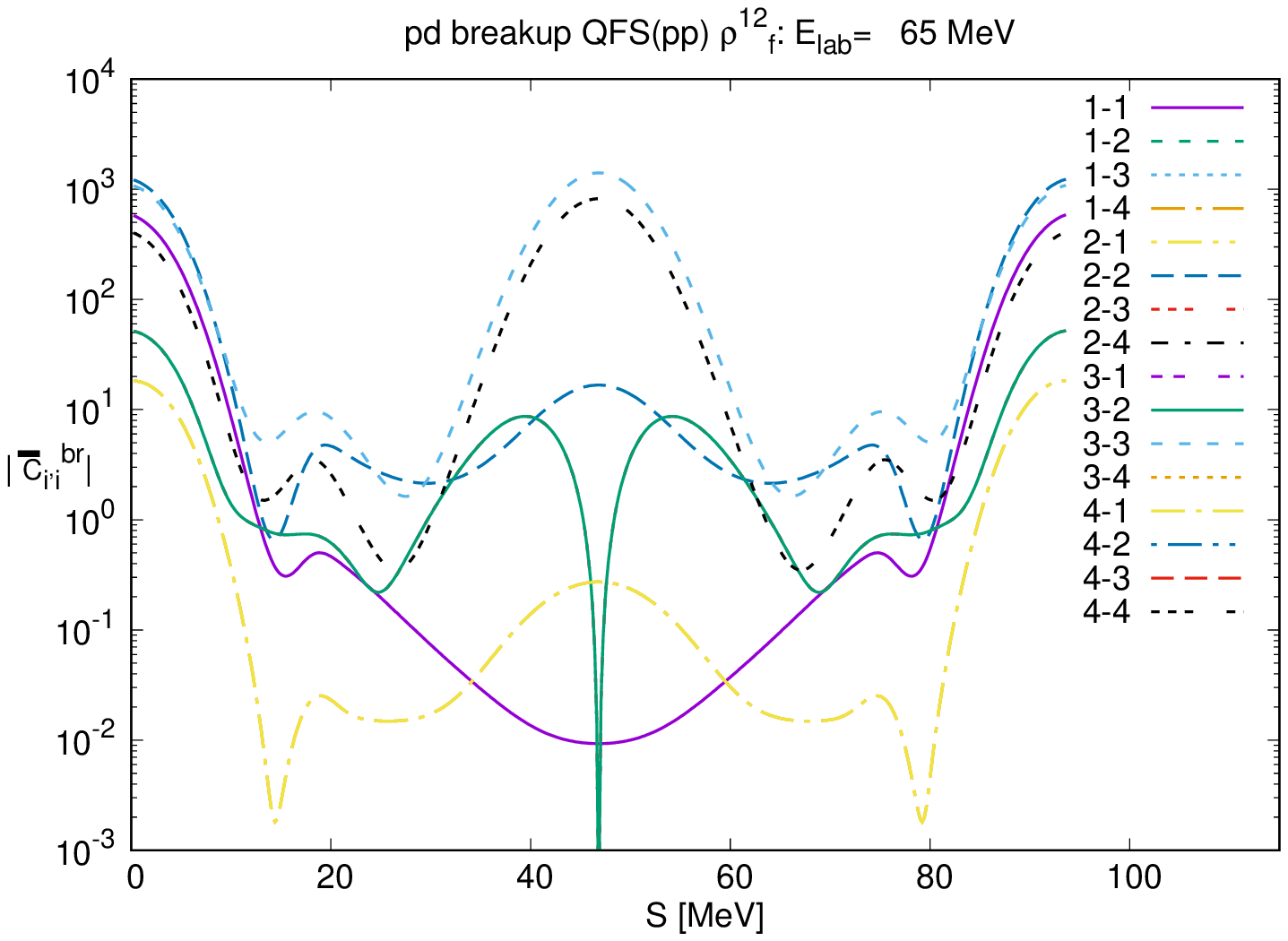}} \\
\end{tabular}%
\caption{
(color online) Absolute values of the expansion
  coefficients $|\bar C_{i^,i}^{br}|$ (Eq.~(\ref{eq_22})) for the final spin 
  density matrix $\rho_f^{12}$ in unpolarized nd and pd 
breakup  $d(N,N_1N_2)N_3$ at $E_{lab}=65$~MeV for a kinematically complete 
configuration with $\Theta_1^{lab}=\Theta_2^{lab}=44^o$ and $\Phi_{12}=180^o$. 
 They are shown as a function of the
arc length S, with the  QFS($N_1N_2$) condition fulfilled at $S=45$~MeV.
Calculations were performed using the AV18 NN potential, with and
    without the inclusion of the Coulomb force \cite{wit_coul}.
 }
\label{fig42}
\end{center}
\end{figure}

\begin{figure}
\includegraphics[scale=0.7]{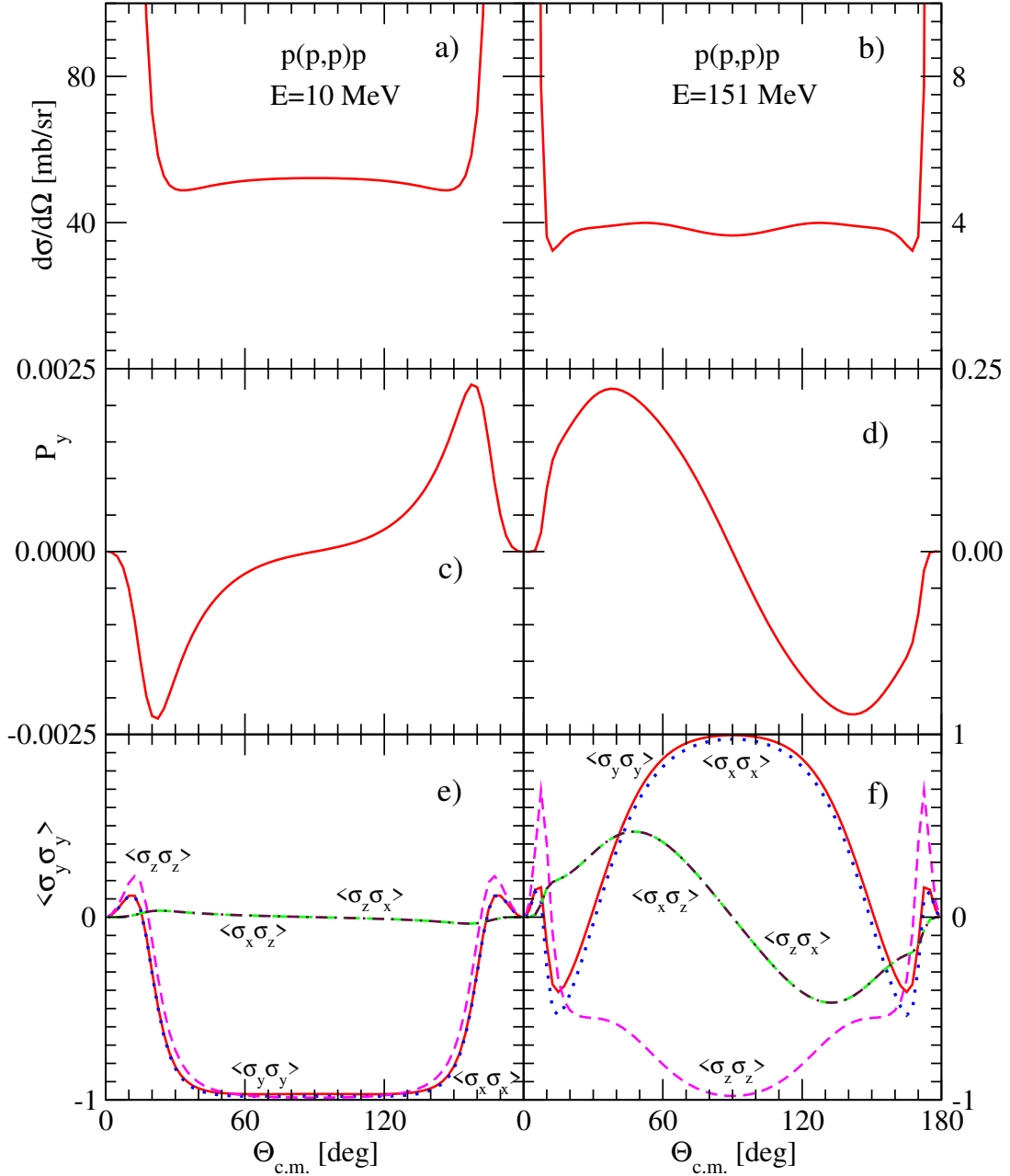}  
\caption{
  (color online) The cross section (a, b), induced polarization $P_y$ of the
  outgoing protons (c, d), and their induced spin correlations (shown with
  different line styles)
  $<\sigma_x^1 \sigma_x^2>$ (blue dotted),
  $<\sigma_y^1 \sigma_y^2>$ (red solid),
  $<\sigma_z^1 \sigma_z^2>$ (magenta short dashed),
  $<\sigma_x^1 \sigma_z^2>$ (green long dashed), and
  $<\sigma_z^1 \sigma_x^2>$ (maroon dash-dotted) (e, f), in unpolarized
  elastic $p(p,p)p$ scattering at
  $E_{lab}=10$~MeV (left) and $151$~MeV (right).
  Calculations were performed using the AV18 NN potential, including  
   the pp Coulomb force, with partial waves up to $j_{max}=5$.  
  }
\label{fig44}
\end{figure}

\begin{figure}
\begin{center}
\begin{tabular}{c}
\resizebox{122mm}{!}{\includegraphics[angle=270]{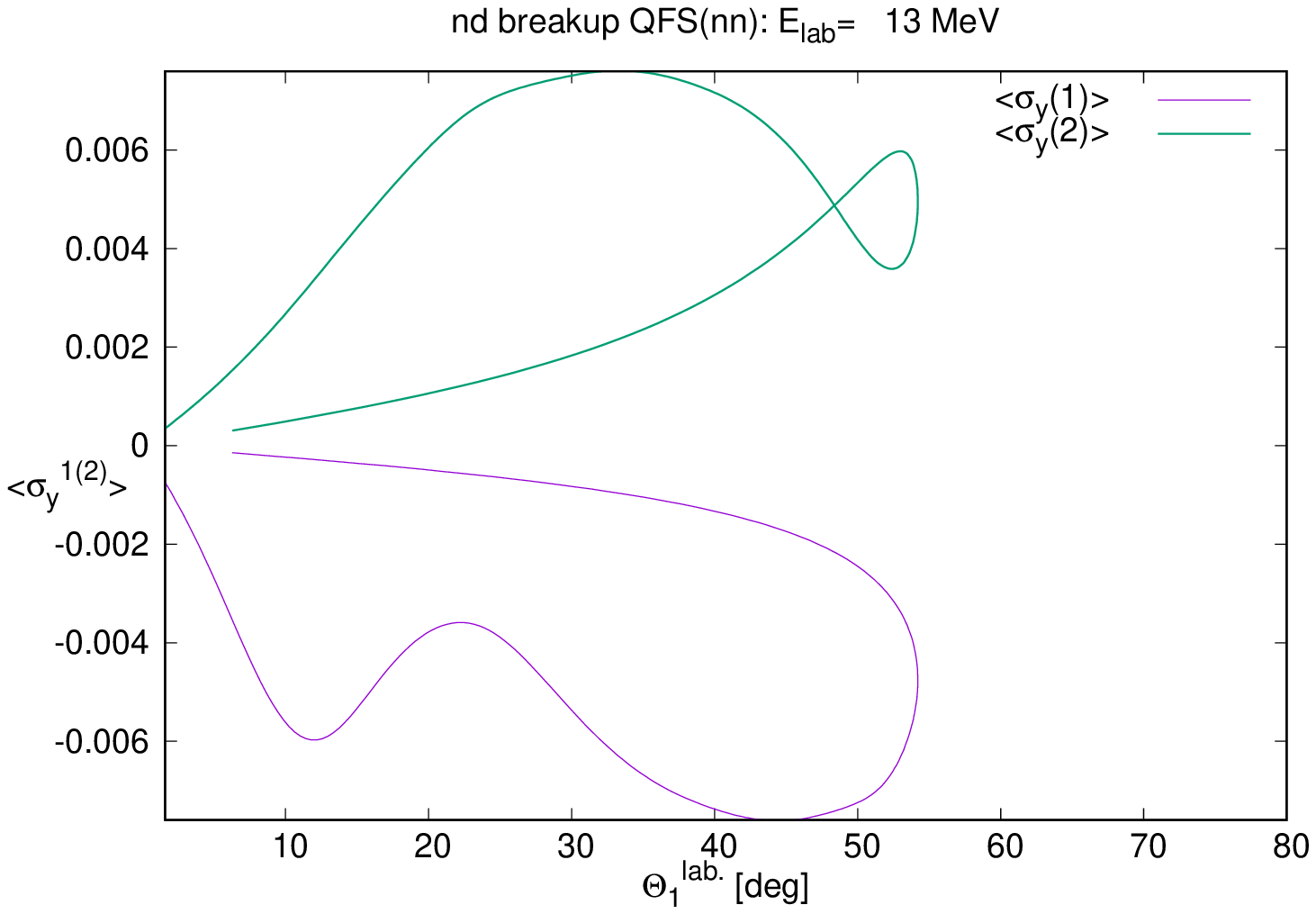}} \\
\resizebox{122mm}{!}{\includegraphics[angle=270]{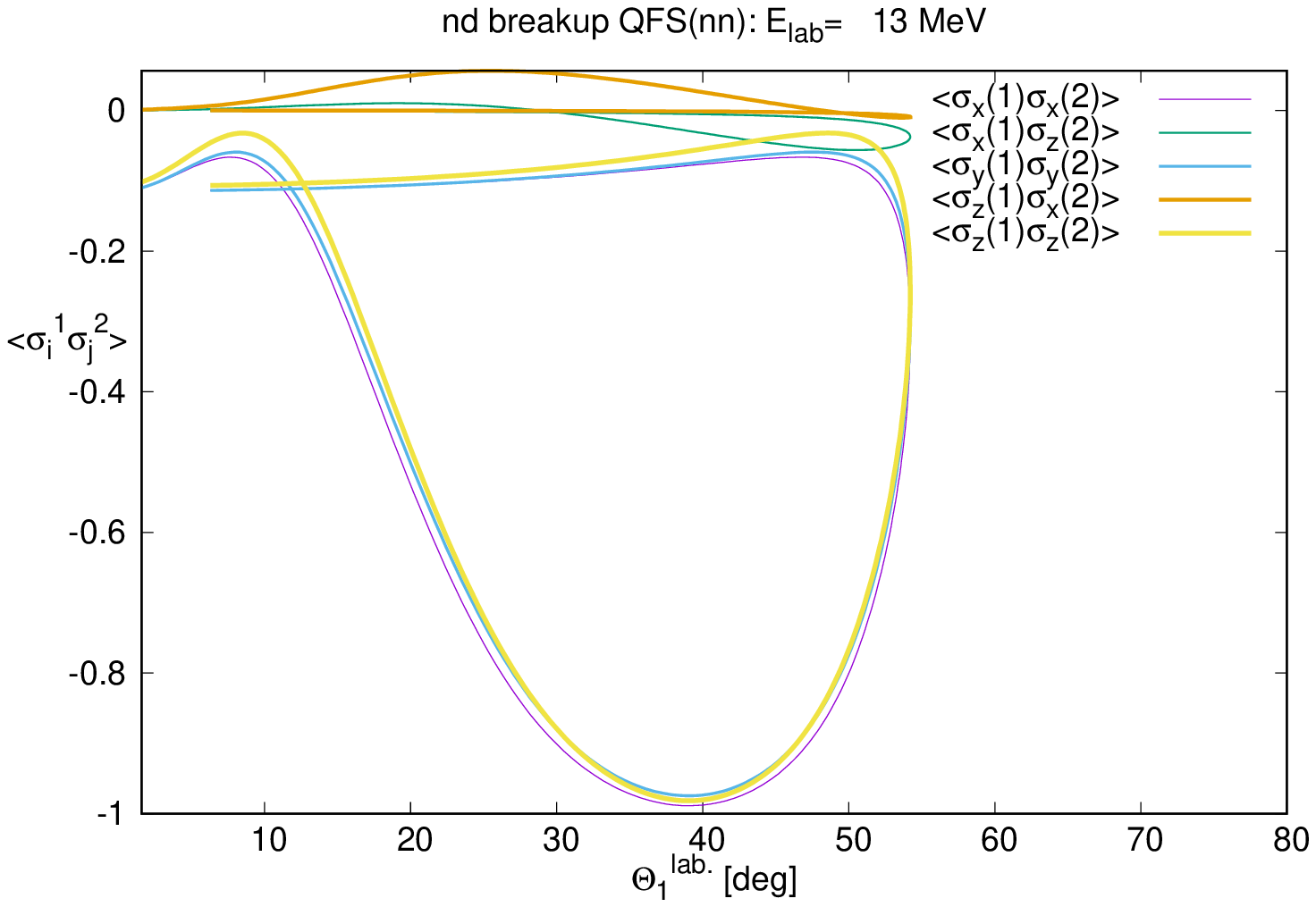}} \\
\end{tabular}%
\caption{
  (color online) Induced polarizations $P_y^{1(2)}=< \sigma_y^{1(2)} >
  $ and induced spin correlations
  $< \sigma_i^1 \sigma_j^2> $ of the outgoing nucleons  in unpolarized nd  
  breakup  $d(n,N_1N_2)N_3$ at $E_{lab}=13$~MeV for a kinematically complete
  geometry
  under the exact QFS(nn) condition. They are shown as a function of the
  laboratory angle of the first nucleon,  $\Theta_1^{lab}$.
      Calculations were performed using the CD Bonn NN potential with 
      partial waves up to $j_{max}=5$.
 }
\label{fig45}
\end{center}
\end{figure}

\begin{figure}
\begin{center}
\begin{tabular}{c}
\resizebox{122mm}{!}{\includegraphics[angle=270]{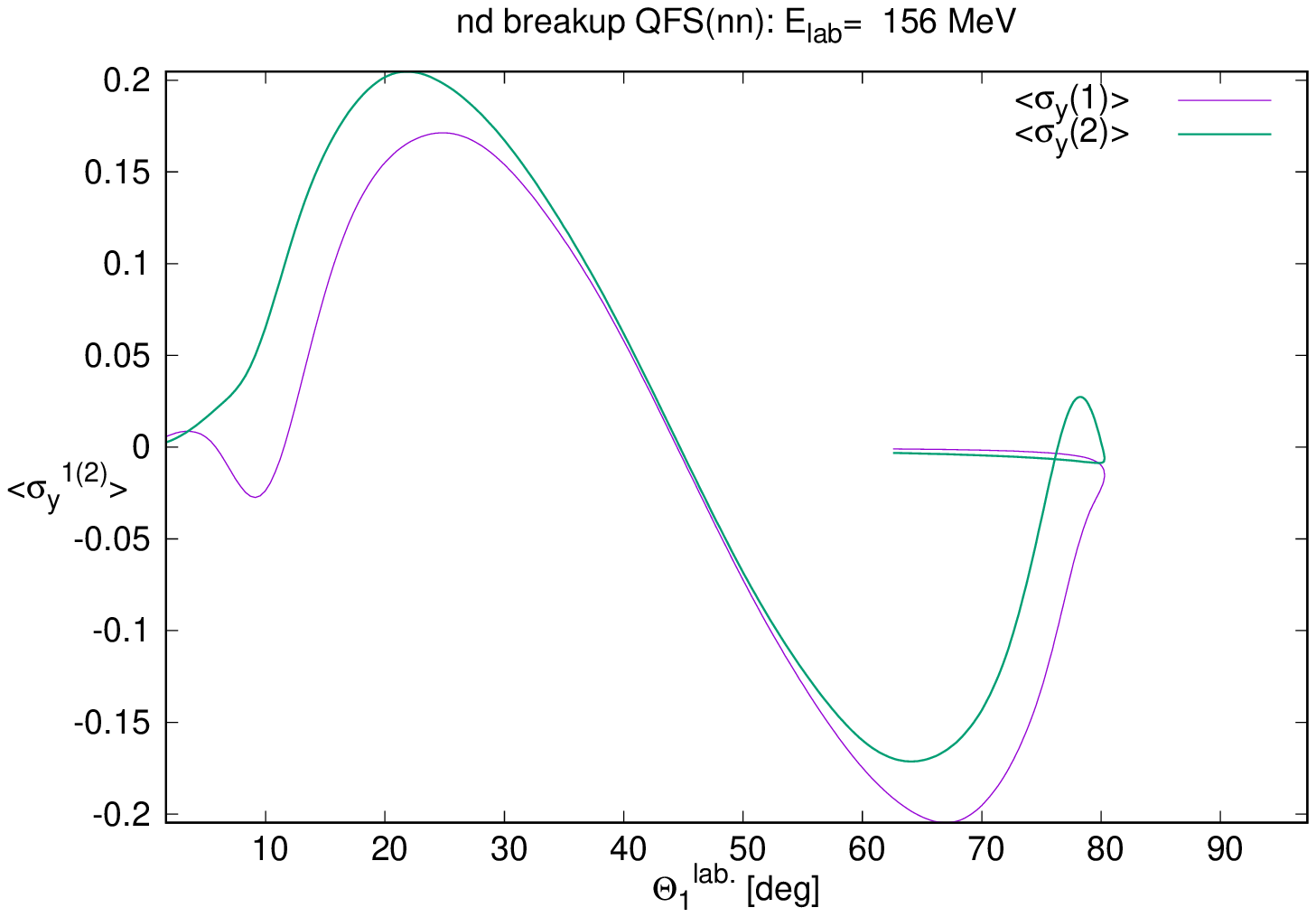}} \\
\resizebox{122mm}{!}{\includegraphics[angle=270]{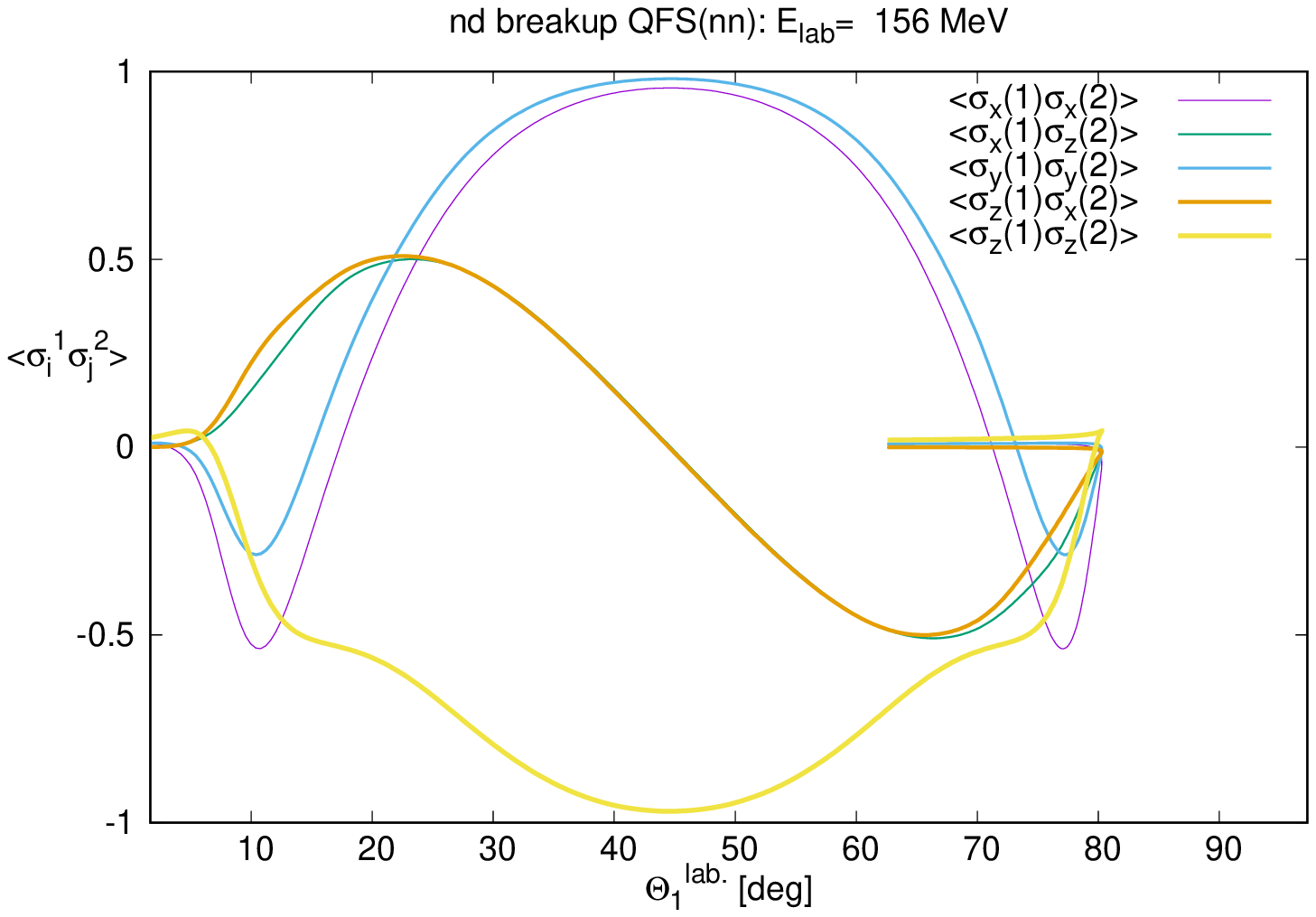}} \\
\end{tabular}%
\caption{
  (color online) Same as in Fig.~\ref{fig45}, but for QFS(nn)
  at $E_{lab}=156$~MeV.
 }
\label{fig46}
\end{center}
\end{figure}

\begin{figure}
\begin{center}
\begin{tabular}{c}
\resizebox{122mm}{!}{\includegraphics[angle=270]{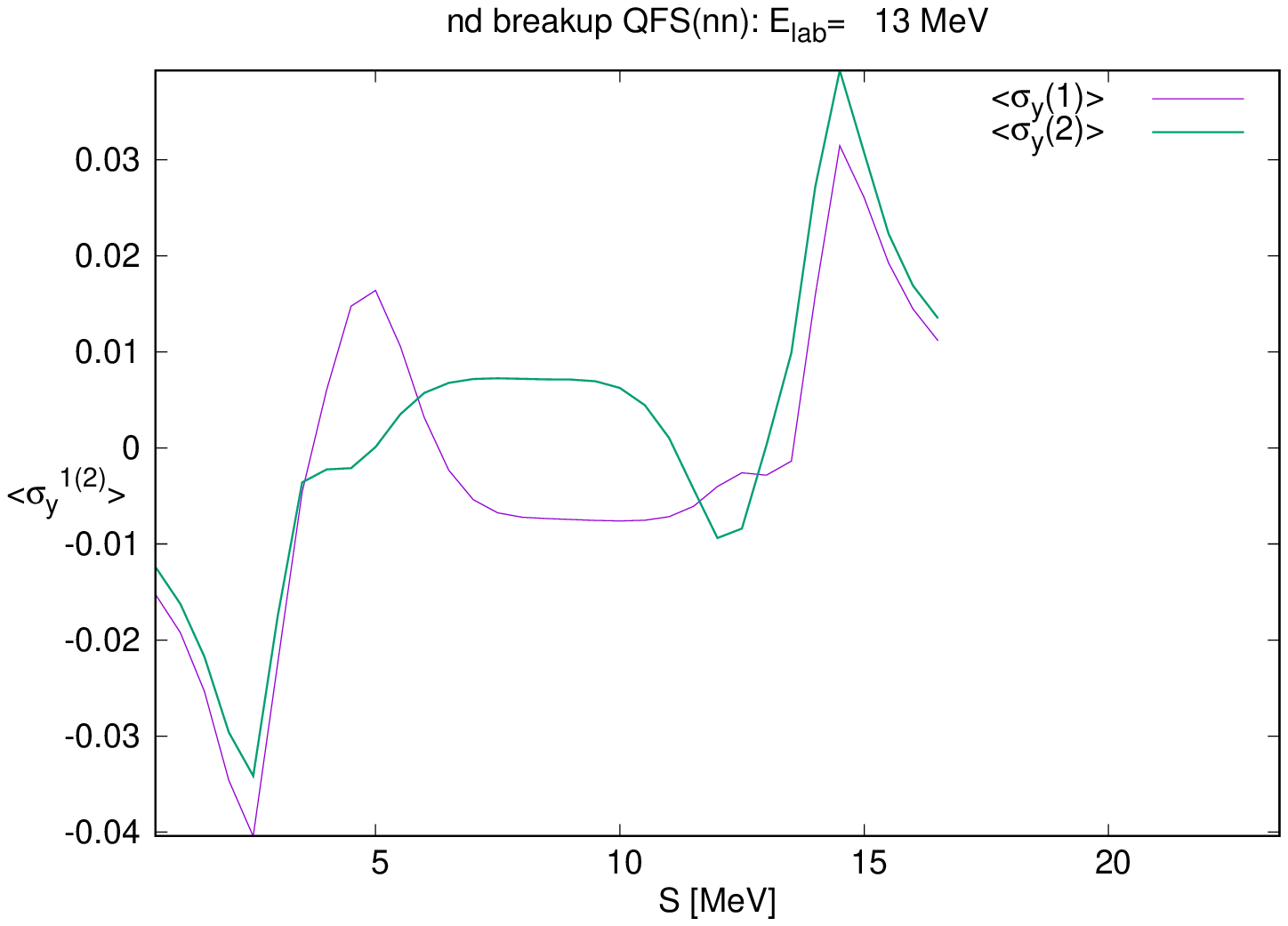}} \\
\resizebox{122mm}{!}{\includegraphics[angle=270]{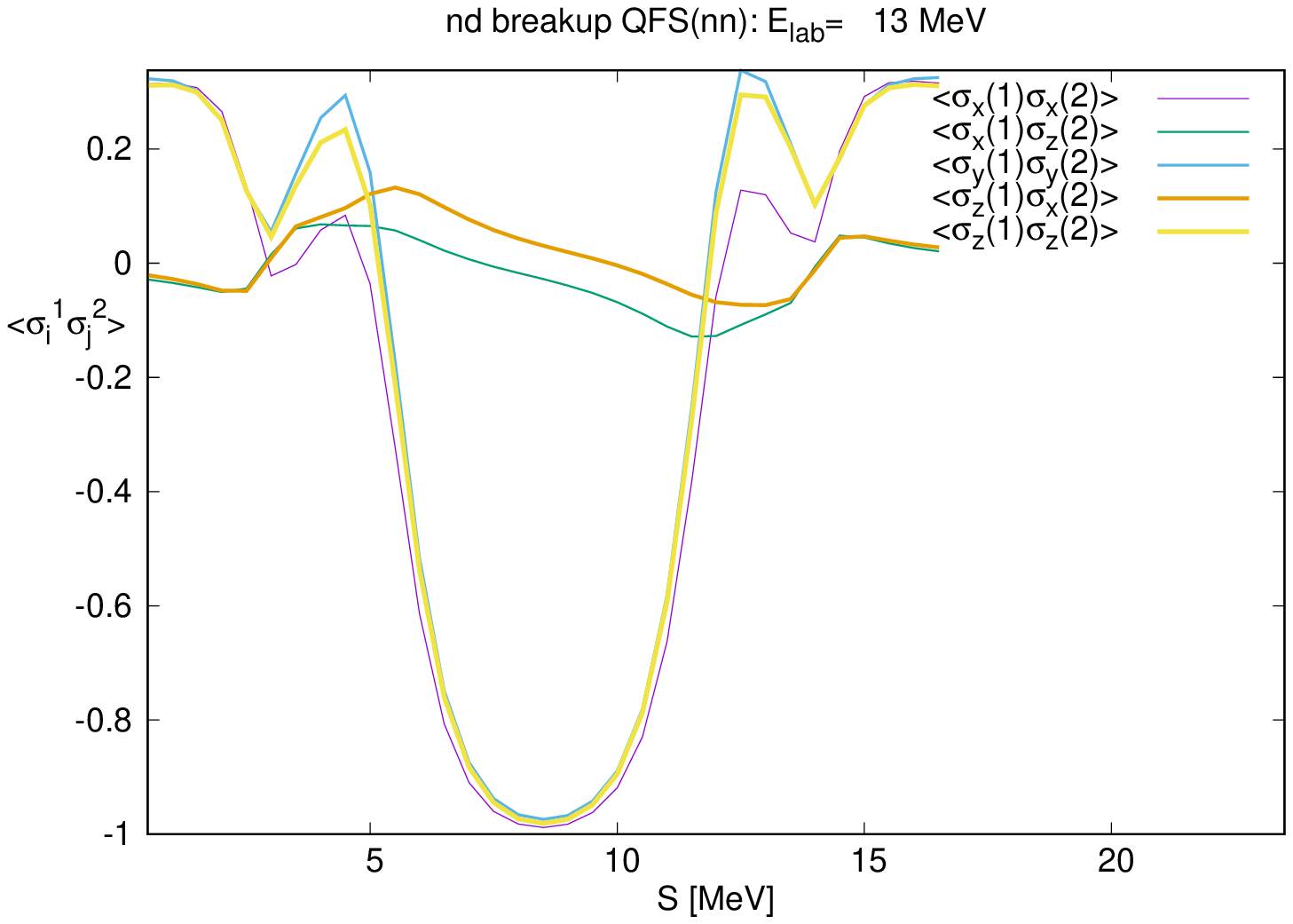}} \\
\end{tabular}%
\caption{
  (color online) Induced polarizations $P_y^{1(2)}=< \sigma_y^{1(2)} >
  $ and induced spin correlations
  $< \sigma_i^1 \sigma_j^2> $ of the outgoing nucleons  in unpolarized nd  
  breakup  $d(n,N_1N_2)N_3$ at $E_{lab}=13$~MeV for a kinematically
  complete geometry
with $\Theta_1^{lab}=40.23^o$, $\Theta_2^{lab}=37.85^o$, and $\Phi_{12}=180^o$.
 They are shown as a function of the arc length S of the S-curve, with 
  the exact QFS(nn) condition occuring at $S=8.5$~MeV.
      Calculations were performed using the CD Bonn NN potential with 
      partial waves up to $j_{max}=5$.
 }
\label{fig51}
\end{center}
\end{figure}

\begin{figure}
\begin{center}
\begin{tabular}{c}
\resizebox{122mm}{!}{\includegraphics[angle=270]{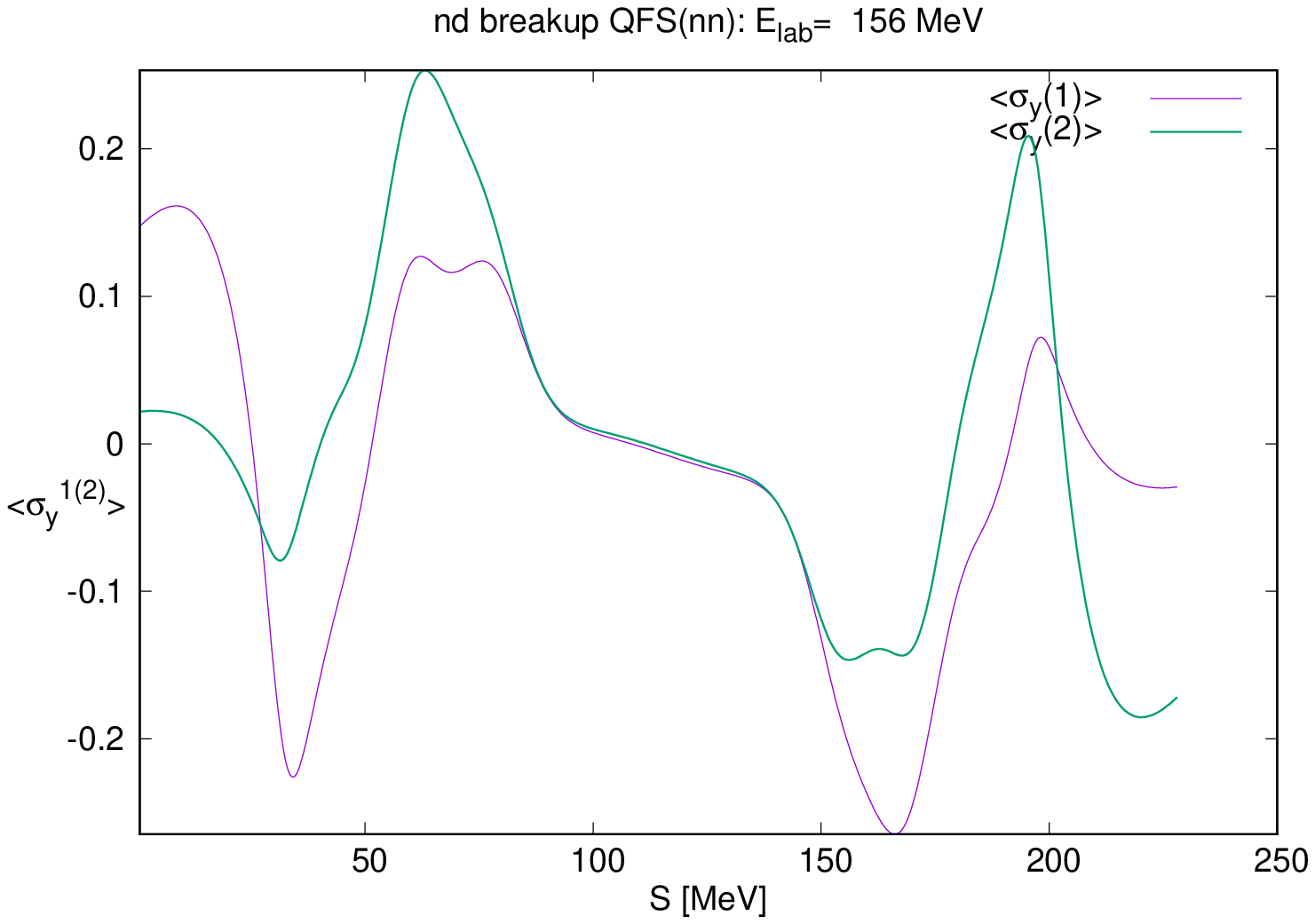}} \\
\resizebox{122mm}{!}{\includegraphics[angle=270]{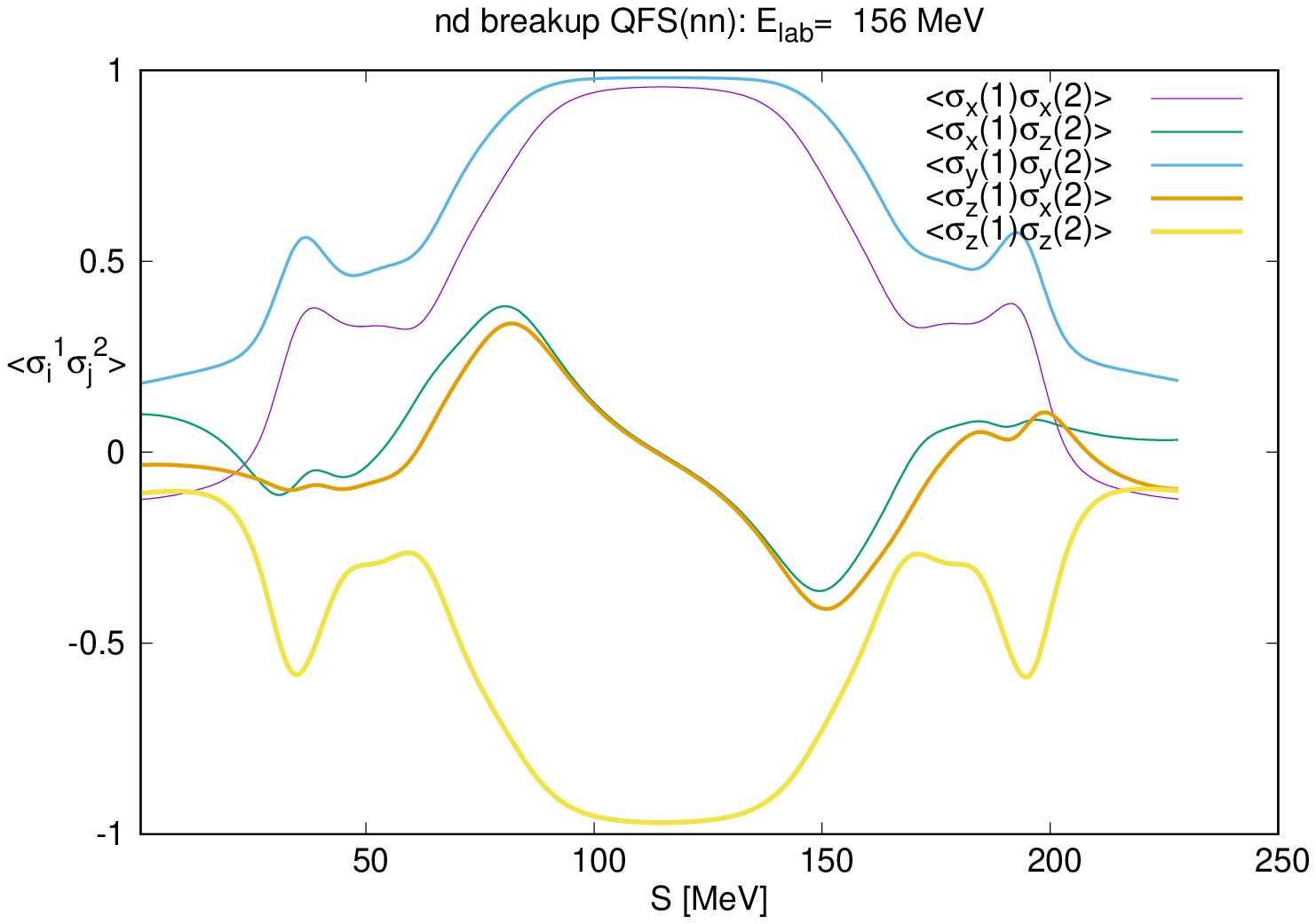}} \\
\end{tabular}%
\caption{
  (color online) Same as in Fig.~\ref{fig51},  but at $E_{lab}=156$~MeV and
  for a kinematically complete geometry
  with $\Theta_1^{lab}=45.03^o$, $\Theta_2^{lab}=44.14^o$, and $\Phi_{12}=180^o$.
  The exact QFS(nn) condition occurs at $S=115$~MeV.
 }
\label{fig52}
\end{center}
\end{figure}

\begin{figure}
\begin{center}
\begin{tabular}{c}
\resizebox{122mm}{!}{\includegraphics[angle=270]{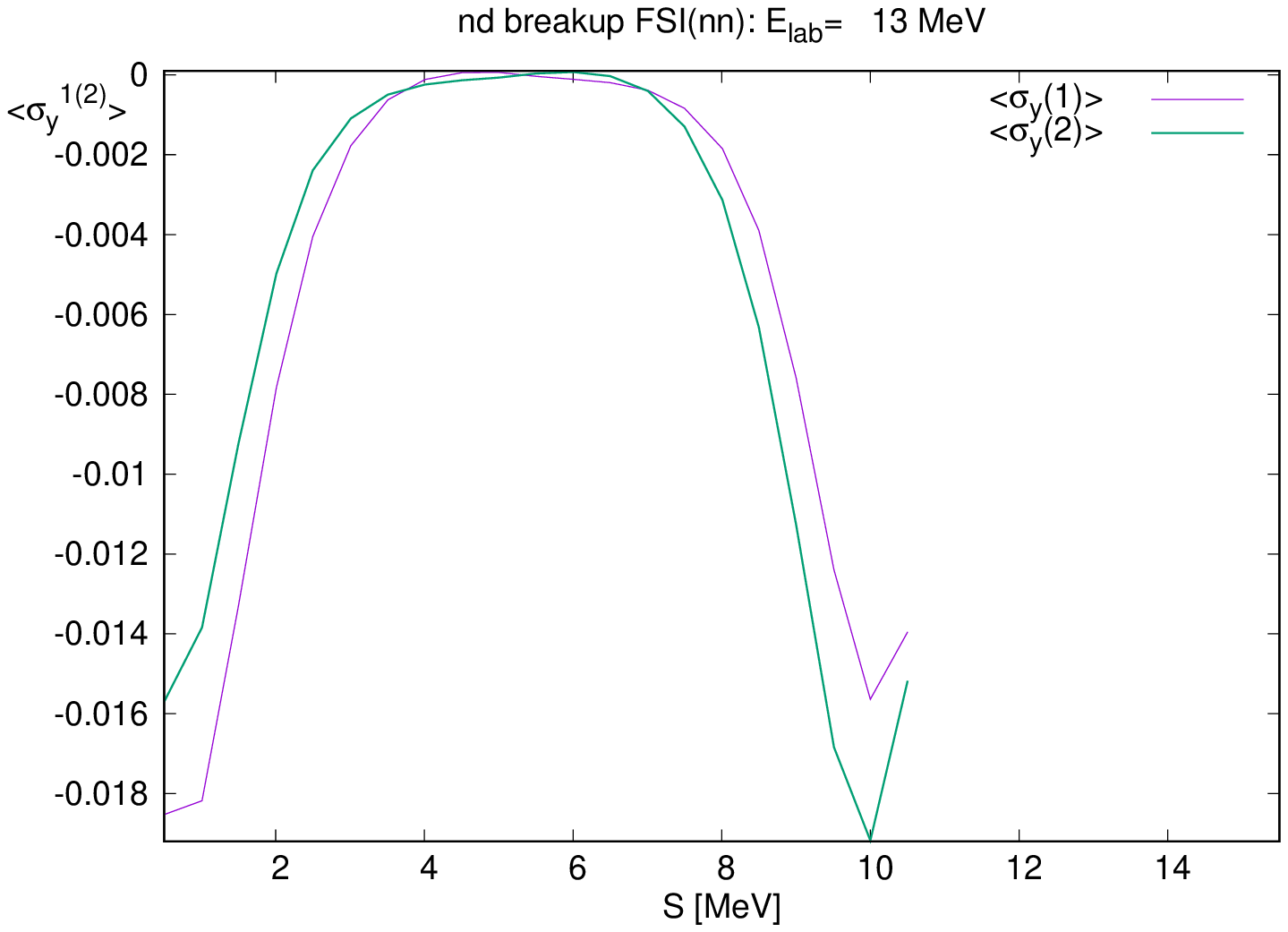}} \\
\resizebox{122mm}{!}{\includegraphics[angle=270]{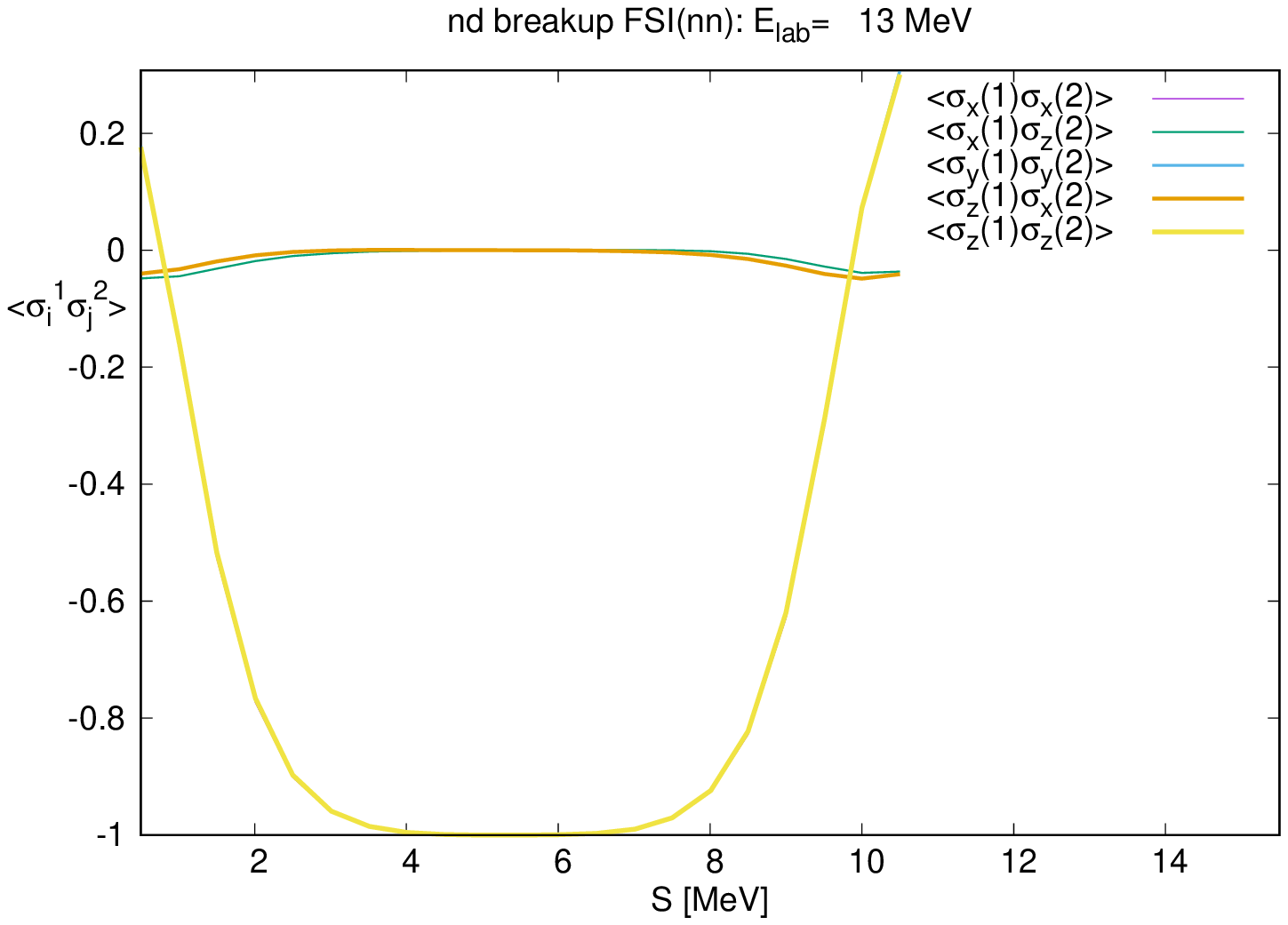}} \\
\end{tabular}%
\caption{
  (color online) Same as in Fig.~\ref{fig51},  but at $E_{lab}=13$~MeV and 
  for a kinematically complete geometry
  with $\Theta_1^{lab}=\Theta_2^{lab}=30.18^o$ and $\Phi_{12}=0^o$.
  The exact FSI(nn) condition occurs at $S=5.5$~MeV.
 }
\label{fig53}
\end{center}
\end{figure}

\begin{figure}
\begin{center}
\begin{tabular}{c}
\resizebox{122mm}{!}{\includegraphics[angle=270]{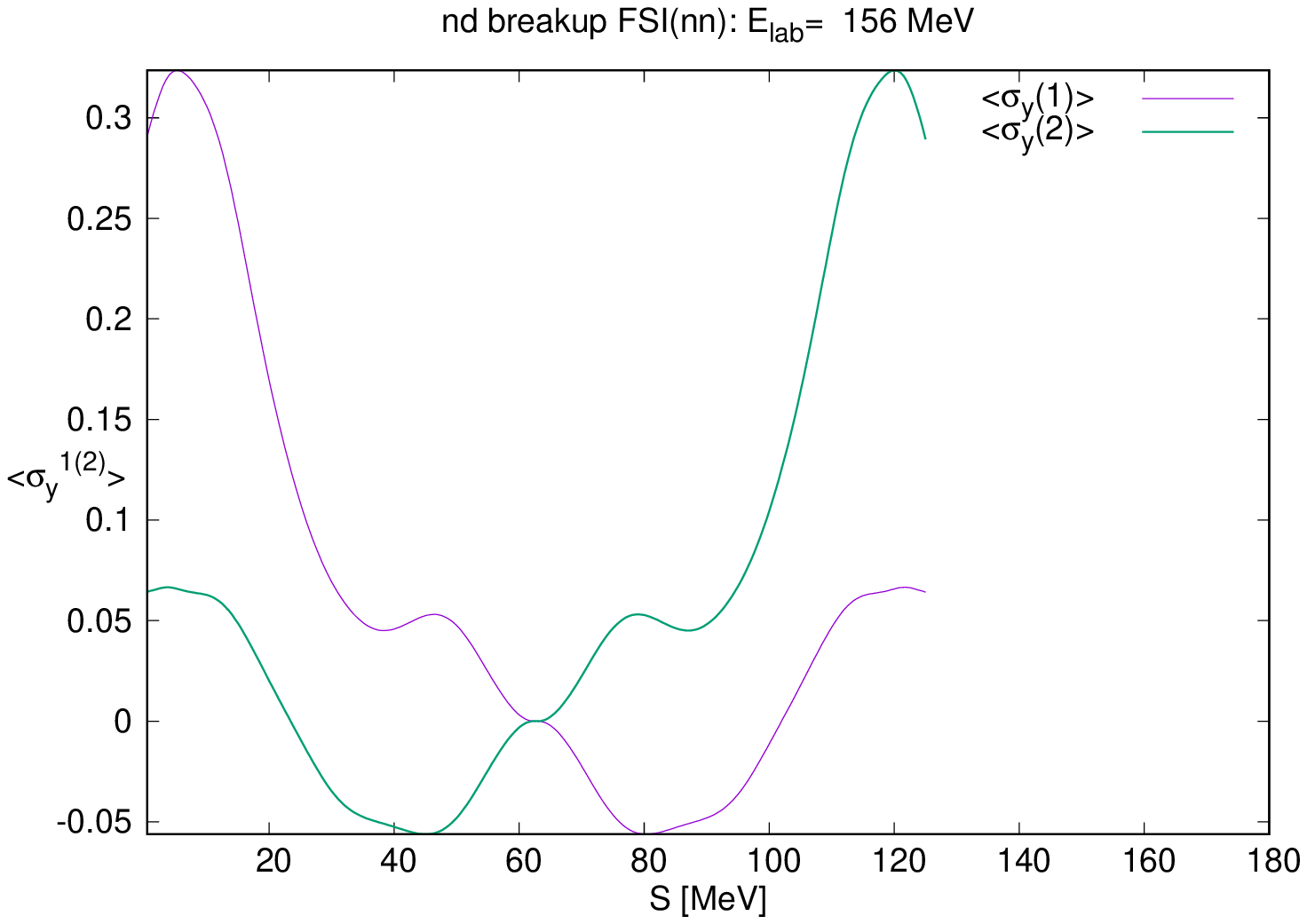}} \\
\resizebox{122mm}{!}{\includegraphics[angle=270]{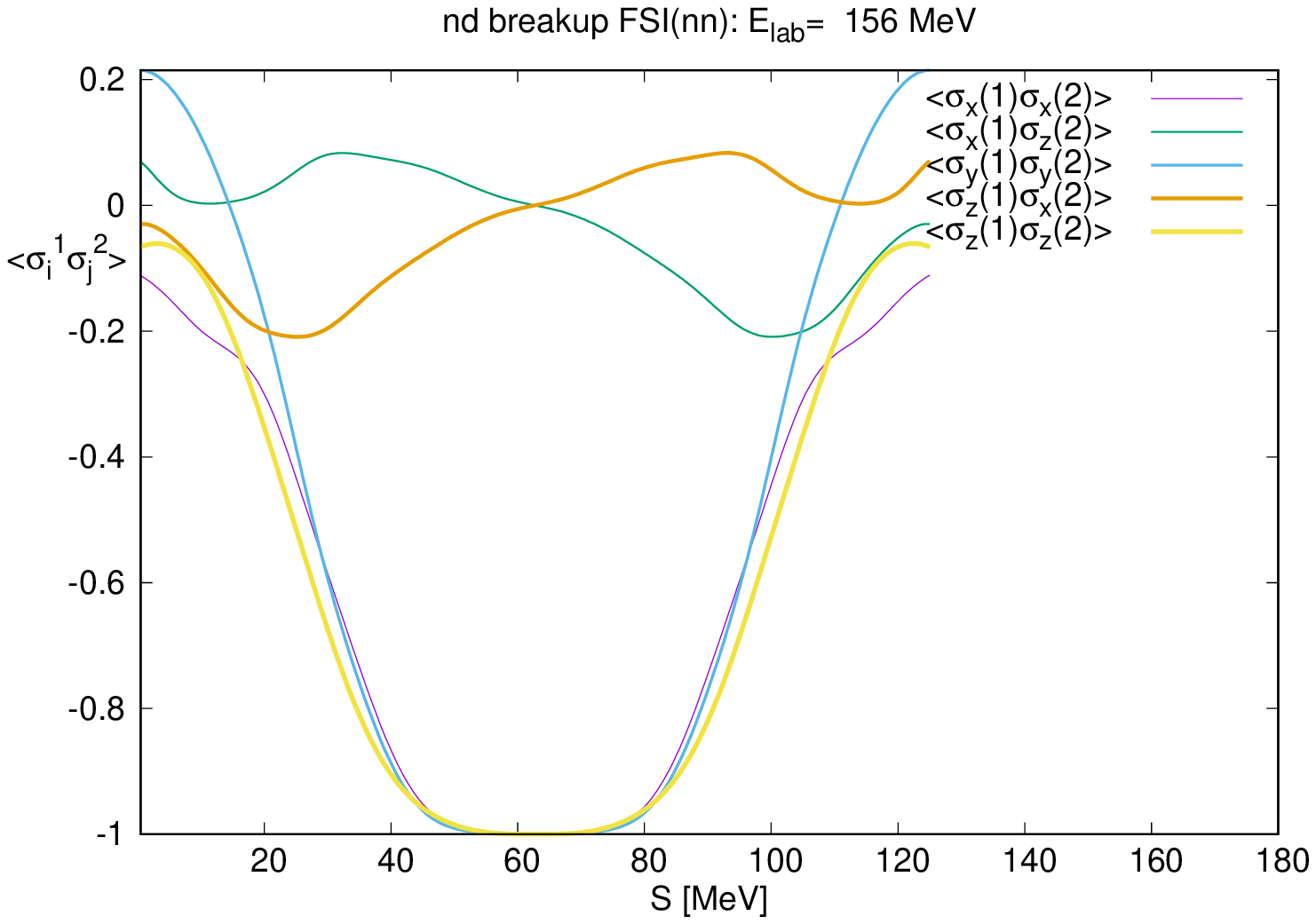}} \\
\end{tabular}%
\caption{
  (color online) Same as in Fig.~\ref{fig51},  but at $E_{lab}=156$~MeV and 
  for a kinematically complete geometry
  with $\Theta_1^{lab}=\Theta_2^{lab}=40.5^o$ and $\Phi_{12}=0^o$.
  The exact FSI(nn) condition occurs at $S=62$~MeV.
 }
\label{fig54}
\end{center}
\end{figure}

This requirement excludes the use, at $E_{lab}=151$~MeV,  of
protons from entangled pp pairs formed in pp elastic scattering or
in quasi-free scattering (QFS) geometries of the pd  breakup.
As seen in Fig.~\ref{fig1aa}, only at low energies and around
$E_{lab} \approx 150$~MeV is the M matrix dominated by a single term.
Since in all entangled Bell states formed in unpolarized pp scattering
or pd  breakup, the energies of the entangled protons are approximately
half of the incoming proton energy, it follows that a teleportation
experiment at higher energies must be performed with incoming protons
at $E_{lab} \approx 300$~MeV.

However, at such high energies, the quality of the produced pp
entangled states deteriorates, as reflected in smaller values of their
spin correlation coefficients, which drop to approximately $0.7-0.8$ compared
to the ideal value of $1$ for pure Bell states. This behavior is observed
both for pp elastic scattering and for the pd breakup reaction in QFS
geometries.

Thus, at higher energies, the only viable option is to produce highly
entangled Bell states of two protons, each with energy
$E_{lab} \approx 150$~MeV, using the pd breakup reaction under the FSI(pp)
condition. This requires a pd breakup experiment with a proton beam of
approximately  $350 $~MeV.  Although the reduction of the FSI cross section
due to the pp Coulomb barrier is admittedly a negative factor, experiments
performed in \cite{sakai2006} suggest that such a teleportation test
is likely feasible.

At low energies, specifically at $E_{lab} \approx 10$~MeV, the reduction of the
energies of the entangled protons by approximately a factor of
two relative to the incoming proton energy—both in pp scattering
and in pd breakup—is actually advantageous. In this regime, the
dominance of a single term in the  matrix, as well as the quality
of the produced entangled states, is further enhanced by the lower
energies of the outgoing protons (see Figs.~\ref{fig1aa} and \ref{fig1bb}).
Consequently,
entangled pp pairs produced in both elastic pp scattering and
exclusive pd breakup under QFS(pp) conditions can be utilized in
a teleportation experiment at low energies.

Before proceeding with actual measurements, a Monte Carlo simulation
under realistic experimental conditions should be performed to provide
estimates of the expected counting rates for triple and double coincidences.

\section{Summary and conclusions}
\label{sumary}

Encouraged  by the results of \cite{shen_2025}, we investigated the possibility
of producing strongly entangled pairs of nucleons in unpolarized proton–proton
elastic scattering and in the unpolarized exclusive pd breakup reaction.
Our results for pp scattering support those of \cite{shen_2025}.
We indeed find that pp scattering can produce strongly entangled proton pairs
in two relatively narrow ranges of the incoming proton laboratory energy,
centered around $E_{\mathrm{lab}} = 10$~MeV and $150$~MeV.
 This possibility is a consequence of the identity of
the two protons, which implies that at the center-of-mass angle
$\Theta_{c.m.} = 90^o$ the transition matrix for the pp system is fully
specified by only three components in the Bell-state basis.

At these two specific energies, one component strongly dominates, such
that around $\Theta_{c.m.} = 90^o$  the transition matrix is well
approximated by $M \approx | \psi^- \rangle \langle \psi^- |$ at low energies
around  $10$~MeV, and by  $M \approx | \psi^+ \rangle \langle \Phi^- |$
at higher energies around $150$~MeV.
As a consequence, pp scattering can produce strongly entangled, Bell-like
states $|\psi^-\rangle$ around $10$~MeV and $|\psi^+\rangle$ around $150$~MeV,
with each entangled proton carrying approximately half of the total energy.

In contrast, no such states can be formed in unpolarized elastic np scattering.
In this case, the transition matrix M is specified by six nonvanishing
components, at least two of which have comparable magnitudes at any energy.

The unpolarized exclusive pd breakup reaction also provides strongly
entangled pp pairs formed in specific, kinematically complete geometries.
An exceptional case is the FSI(pp) geometry, in which the two outgoing
protons are always—independent of the energy of the incoming proton and
the production angle of the final-state interacting protons—produced in
the exact Bell state $| \psi^- \rangle$.

The QFS(pp) geometry, which in many respects—especially at higher
energies—resembles free pp scattering, also allows the production of
strongly entangled pp pairs, but only in restricted regions of the
proton-beam energy: in the Bell state $| \psi^- \rangle$ at low
energies around $10$~MeV and in $| \psi^- \rangle$ at around $150$~MeV.

We have shown that, by using these strongly entangled proton pairs, it
is possible to design an experiment that would allow one to test the
feasibility of quantum-state teleportation in a three-proton system.
Such an experiment should be performed either at low ($\approx 20$~MeV)
or at higher ($\approx 350$~MeV) beam energies.

At low energies, unpolarized pp scattering or unpolarized exclusive
pd breakup in the QFS(pp) geometry, with an incoming proton energy of
$\approx 20$~MeV, can be used to produce entangled pp pairs with proton
energies of $\approx 10$~MeV. At higher energies, strongly entangled
pp pairs with proton energies of $\approx 150$~MeV can be formed only
in unpolarized exclusive pd breakup under FSI(pp) conditions, with an
incoming proton laboratory energy of $\approx 350$~MeV, yielding
entangled pp pairs with proton energies of $\approx 150$~MeV.

In both cases, the resulting Bell state is $| \psi^- \rangle$.
Evidence for the occurrence of teleportation is then provided by measuring
the polarization of one proton from the entangled pair after the second
proton has undergone scattering from a target with known polarization.

\appendix

\section{Teleportation of a mixed state}
\label{a1}

Let us assume that the state of proton 1 is not pure but instead a
statistical mixture of states, described by a spin density matrix
with polarization vector $\vec P = (0,P_y,0)$. Thus, the initial spin
density matrix of the three-proton system is:
\begin{eqnarray}
  \rho_{\text{in}} =    \rho_{\text{23}} \otimes \rho_{\text{1}} =
 | \psi_i \rangle_{23}   ~_{23}\langle \psi_i |  \otimes 
  \frac{1}{2} (I_1 + P_y \sigma_y^1)
   ~,
 \label{eq_ap1}
\end{eqnarray}
with the initial state of the entangled protons 2 and 3 given by
the Bell state $| \psi_i \rangle_{23}$.

The final density matrix of the system after proton 2 scatters off proton 1 is:
\begin{eqnarray}
  \rho_f &=& M_{12} \otimes I_3 ~  \rho_{\text{in}} ~   (M_{12} \otimes I_3)^{\dagger}
   ~.
 \label{eq_ap2}
\end{eqnarray}

Taking the form $M_{12}=| \psi^- \rangle_{23}   ~_{23}\langle \psi^- |$,
valid at $E_{lab}=10$~MeV, and $| \psi_i \rangle_{23} = | \psi^- \rangle_{23}$, 
a direct calculation leads to  the polarization of proton 3 being identical
to that of proton 1:
\begin{eqnarray}
  \rho_f &=&  \frac {1} {4} | \psi^- \rangle_{12}   ~_{12}\langle \psi^- |
  \otimes  \frac {1} {2} ( I_3 + P_y \sigma_y^3 ) 
   ~.
 \label{eq_ap3}
\end{eqnarray}

For the form $M_{12}=| \psi^+ \rangle_{23}   ~_{23}\langle \phi^- |$,
valid at $E_{lab}=151$~MeV, and $| \psi_i \rangle_{23} = | \psi^- \rangle_{23}$,
 the polarization of proton 3 has the opposite sign to that of proton 1:
\begin{eqnarray}
  \rho_f &=&  \frac {1} {4} | \psi^+ \rangle_{12}   ~_{12}\langle \psi^+ |
  \otimes  \frac {1} {2} ( I_3 - P_y \sigma_y^3 ) 
   ~.
 \label{eq_ap4}
\end{eqnarray}

With the same form of $M_{12}$, but replacing $| \psi^- \rangle_{23} $
with $| \psi^+ \rangle_{23} $, one 
again finds that the polarization of proton 3 is the
same as that of proton 1:
\begin{eqnarray}
  \rho_f &=&  \frac {1} {4} | \psi^+ \rangle_{12}   ~_{12}\langle \psi^+ |
  \otimes  \frac {1} {2} ( I_3 + P_y \sigma_y^3 ) 
   ~.
 \label{eq_ap5}
\end{eqnarray}


\acknowledgements

This work was supported by the National Science Centre,
Poland under Grant
IMPRESS-U 2024/06/Y/ST2/00135.   
The numerical calculations were partly performed on the supercomputers of
the JSC, J\"ulich, Germany.




\clearpage

\clearpage

\clearpage

\clearpage

\clearpage

\clearpage

\clearpage

\clearpage

\clearpage

\clearpage

\clearpage

\clearpage

\clearpage

\clearpage

\clearpage

\clearpage

\clearpage

\clearpage

\clearpage

\clearpage

\clearpage

\clearpage

\clearpage

\clearpage

\clearpage

\clearpage

\begin{figure}
\begin{center}
\begin{tabular}{c}
\resizebox{120mm}{!}{\includegraphics[angle=270]{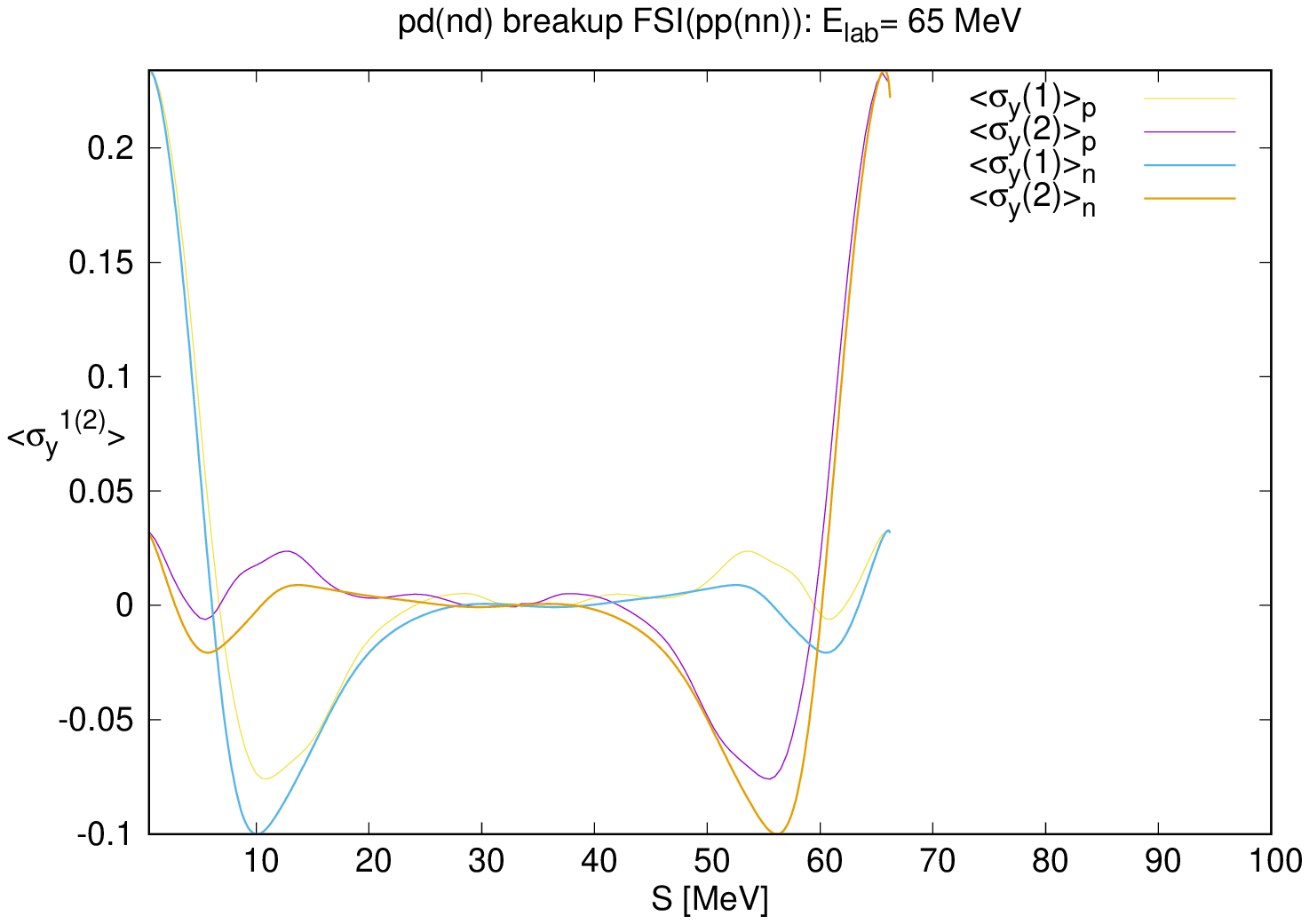}} \\
\resizebox{120mm}{!}{\includegraphics[angle=270]{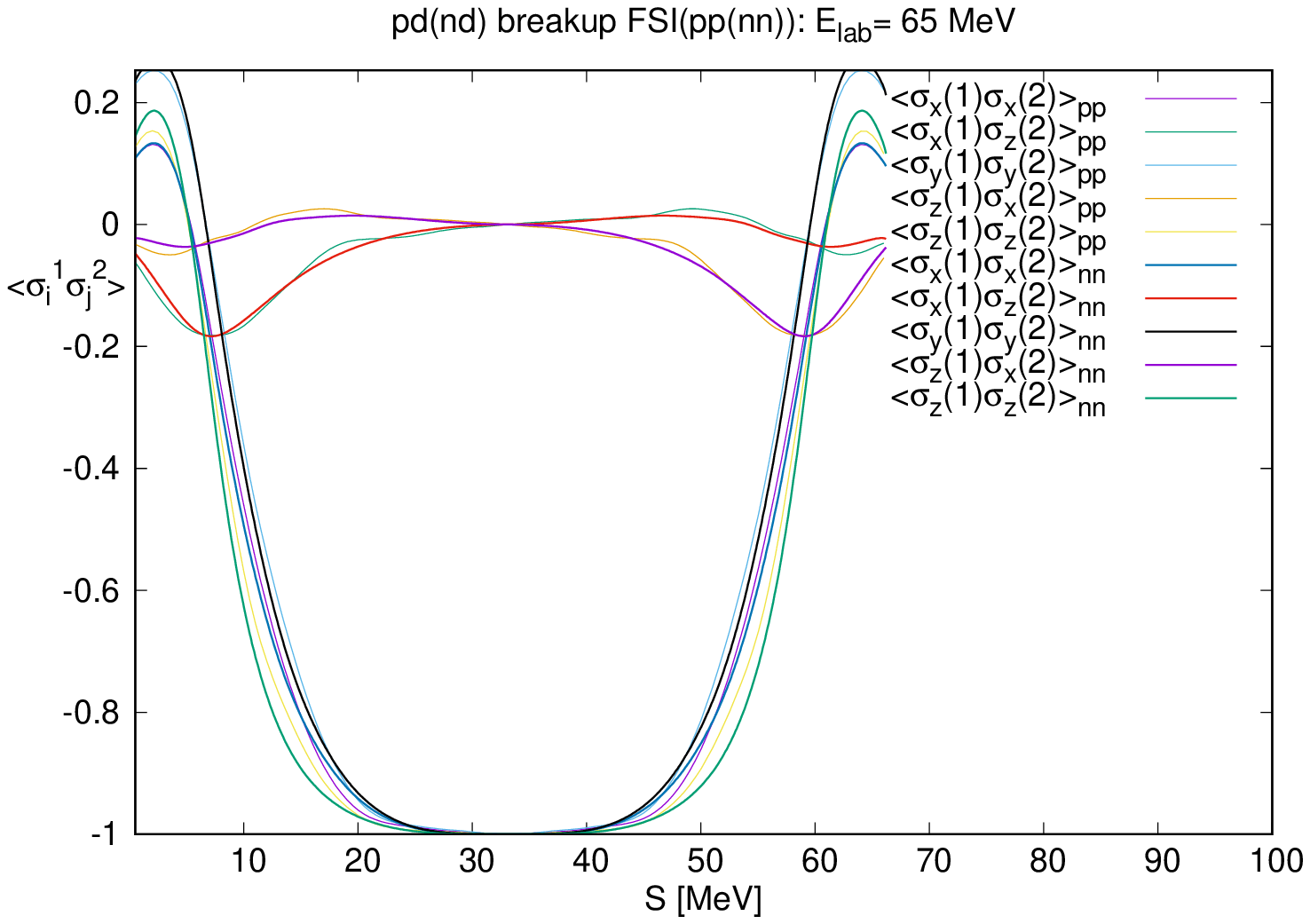}} \\
\end{tabular}%
\caption{
  (color online) Induced polarizations $P_y^{1(2)}=< \sigma_y^{1(2)} >
  $ and induced spin correlations
  $< \sigma_i^1 \sigma_j^2> $ of the outgoing nucleons  in unpolarized pd and
  nd  breakup  $d(n,N_1N_2)N_3$ at $E_{lab}=65$~MeV for a kinematically
  complete geometry
with $\Theta_1^{lab}=\Theta_2^{lab}=30^o$, and $\Phi_{12}=0^o$.
 They are shown as a function of the arc length S of the S-curve, with 
  the exact FSI(pp(nn)) condition occuring at $S=33$~MeV.
Calculations were performed using the AV18 NN potential, with and
    without the inclusion of the Coulomb force \cite{wit_coul}.
 }
\label{fig55}
\end{center}
\end{figure}

\clearpage

\begin{figure}
\includegraphics[bb=0 167 418 512,scale=1.0]{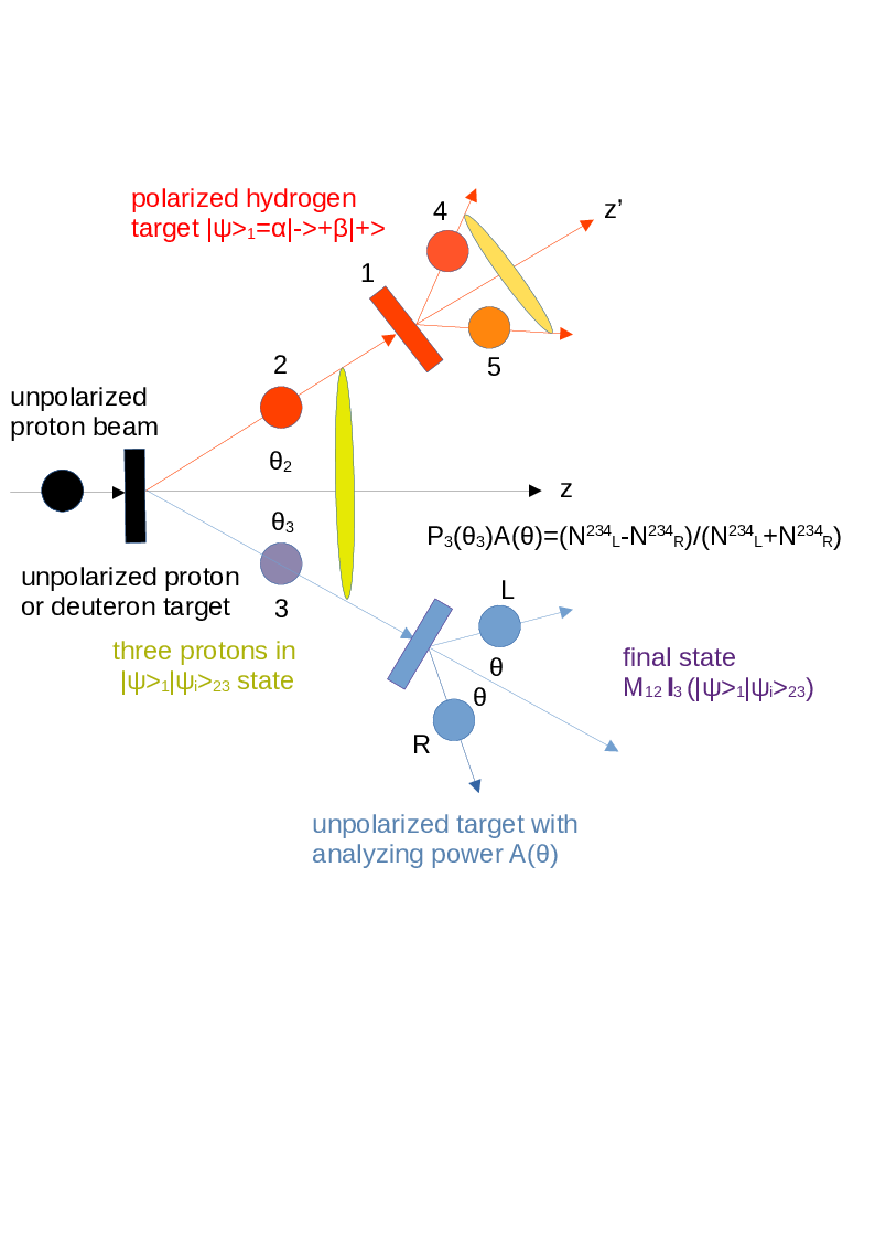}  
\caption{
  (color online) Setup of the quantum teleportation experiment.
  }
\label{exp_scheme}
\end{figure}

\end{document}